\documentclass[11pt]{article}
\newif\iffigs\figstrue

\usepackage{epsfig,latexsym}
\usepackage{amsmath}
\usepackage{verbatim}
\usepackage{mathrsfs}
\usepackage{amssymb}

\DeclareMathAlphabet{\mathpzc}{OT1}{pzc}{m}{it}

 \csname
@addtoreset\endcsname{equation}{section}

\def\gz0{\gamma^{0}}

\def\g{\gamma}

\def\vf{\varphi}

\def\be{\begin{equation}}
\def\ee{\end{equation}}
\def\bea{\begin{eqnarray}}
\def\eea{\end{eqnarray}}
\def\ba{\begin{array}}
\def\ea{\end{array}}
\def\bec{\begin{center}}
\def\ec{\end{center}}
\def\ba{\begin{align}}
\def\ena{\end{align}}

\def\12{\frac{1}{2}}

\begin{document}

\begin{flushright}
{\today}
\end{flushright}

\vspace{10pt}

\begin{center}


{\Large\sc Pre -- Inflationary Clues from String Theory ?}\\


\vspace{25pt}
{\sc N.~Kitazawa${}^{\; a}$ \ and \ A.~Sagnotti${}^{\; b}$}\\[15pt]

{${}^a$\sl\small Department of Physics, Tokyo Metropolitan University\\
Hachioji, Tokyo \\ 192-0397 JAPAN
\\ }e-mail: {\small \it
kitazawa@phys.se.tmu.ac.jp}\vspace{8pt}

{${}^b$\sl\small
Scuola Normale Superiore and INFN\\
Piazza dei Cavalieri, 7\\ 56126 Pisa \ ITALY \\
e-mail: {\small \it sagnotti@sns.it}}\vspace{10pt}

\vspace{24pt} {\sc\large Abstract}\end{center}
\noindent
``Brane supersymmetry breaking'' occurs in String Theory when the only available combinations of D--branes and orientifolds are not mutually BPS and yet do not introduce tree--level tachyon instabilities. It is characterized by the emergence of a steep exponential potential, and thus by the absence of maximally symmetric vacua. The corresponding low--energy supergravity admits intriguing spatially--flat cosmological solutions where a scalar field \emph{is forced to climb up} toward the steep potential after an initial singularity, and additional milder terms can inject an inflationary phase during the ensuing descent. We show that, in the resulting power spectra of scalar perturbations, an infrared suppression is typically followed by a pre--inflationary peak that reflects the end of the climbing phase and can lie well apart from the approximately scale invariant profile. A first look at WMAP9 raw data shows that, while the $\chi^2$ fits for the low--$\ell$ CMB angular power spectrum are clearly compatible with an almost scale invariant behavior, they display nonetheless an eye--catching preference for this type of setting within a perturbative string regime.

\setcounter{page}{1}

\pagebreak

\newpage
\section{\sc  Introduction}\label{sec:intro}

\vskip 12pt

It would be bizarre if Supersymmetry \cite{SUSY} were not to play a role in the Fundamental Interactions, since its local realization in Supergravity \cite{SUGRA} and its completion in String Theory \cite{strings} contain profound lessons on the links between gravity and the other forces. Yet, the apparent lack of super--partners of the known particles has already survived the first round of experiments at LHC \cite{LHC}. Under these circumstances, the standard recipe to keep Supersymmetry alive is to try and make all putative partners heavy enough, appealing to some mechanism of supersymmetry breaking \cite{SUSY_breaking}. However, arriving at a fully satisfactory scenario of this type remains a key challenge in current attempts to combine the Standard Model of Particle Physics with gravity, despite decades of intense effort and a number of important results.

A proper understanding of supersymmetry breaking cannot forego a detailed grasp of general matter couplings in Supergravity, which has long been available for models with ${\cal N}=1$ supersymmetry in four dimensions \cite{cfgv} but is not nearly as complete in the higher--dimensional settings that are so important in String Theory. As a result, perhaps, non--supersymmetric string compactifications are generally expected to spell trouble for the vacuum state and have been explored only to a limited extent. Still, there is a higher--dimensional setting that was identified long ago and stands out for its relative simplicity and rigidity. This is ``brane supersymmetry breaking'' (BSB) \cite{bsb}, which presents itself in some orientifold vacua \cite{orientifolds} where Ramond--Ramond (RR) charge neutrality requires the simultaneous presence, in the vacuum, of combinations of branes and orientifolds that are not mutually BPS and yet do not introduce tree--level tachyon instabilities. These extended objects leave behind a distinctive mark, a steep exponential potential proportional to their overall tension, whose lack of local minima excludes from the outset maximally symmetric geometries, and flat space in particular, for these systems. The phenomenon finds its simplest manifestation in the ten--dimensional Sugimoto model in \cite{bsb}, where supersymmetry appears exact, at tree level, insofar as the closed spectrum is concerned, but is actually non-linearly realized due to the non-supersymmetric brane configuration, whose modes include a singlet spinor that plays the role of a goldstino \cite{10d_bsb_couplings}. The nine dimensional vacuum geometry realizing the largest symmetry allowed for this system was presented by Dudas and Mourad in \cite{dm_vacuum}.

The ten--dimensional Sugimoto model in \cite{bsb} admits an intriguing spatially flat cosmological solution where the dilaton exhibits a striking behavior. The solution has a long history \cite{lm,exp_sol,dm_vacuum}, but its main lesson was appreciated only recently \cite{dks}. A minimally coupled scalar field ought to possess the two distinct options of emerging from an initial singularity while \emph{descending} or \emph{climbing} mild exponential potentials of the type
\be
{V}(\varphi) \ = \ {V}_0 \, e^{\,2\,\gamma\, \varphi} \ , \label{potonexp_intro}
\ee
and this is indeed the case. However, \emph{for $\gamma$ large enough only the climbing behavior becomes possible}, and this what we shall refer to as the climbing phenomenon. In particular, with the convenient non--canonical normalization for $\varphi$ described in the following sections the climbing behavior sets in at $\gamma=1$ for all space--time dimensions $d$.

A striking feature of ``brane supersymmetry breaking'' is that its potential lies precisely at the critical point $\gamma=1$ \cite{dks}. Moreover, this property continues to hold for $d<10$ for a special combination of the dilaton and the breathing mode, and under the assumption that this field dominates the early dynamics other branes present in String Theory \cite{branescan,branesugimoto} can contribute milder exponential terms \cite{dks,fss,as13} that could have injected inflation \cite{inflation} after the initial climbing phase. One is thus led to consider the class of potentials \cite{dks,dkps}
\be
{V}(\varphi) \ = \ {V}_0 \left( e^{\,2\,\varphi} \ + \ e^{\,2\,\gamma\, \varphi} \right) \  \label{potwoexp_intro}
\ee
in four dimensions, where $\gamma< 1/\sqrt{3}$ in order to allow for the onset of inflation following an initial climbing phase, with the eventual goal of comparing their implications with the low--$\ell$ end of the CMB angular power spectrum.

An enticing feature of the climbing phenomenon is that it links two apparently unrelated problems, the breaking of Supersymmetry and the onset of inflation. We are well aware of the limitations of the simple potentials of eq.~\eqref{potwoexp_intro}, which result for one matter in tensor--to--scalar ratios that are too large \cite{inflation}, but we can anticipate that the key phenomenon that we shall come to can be traced to the behavior of the scalar field near the ``hard'' exponential, the main datum that String Theory and ``brane supersymmetry breaking'' contribute to this discussion, so that it occurs in a variety of more realistic potentials, some of which will be touched upon in the following.

Some features of the scalar dynamics in the potential \eqref{potwoexp_intro} can be readily anticipated. To begin with, the ``hard'' potential forces the field to emerge from an initial singularity while climbing up from large negative values of $\varphi$, and this early phase is essentially driven by the mild exponential. Notice that the climbing phenomenon constrains the choice of initial conditions, a nice feature for a theory of inflation, so that a single parameter is left in this case, a constant $\varphi_0$ that determines to which extent the scalar feels the ``hard'' exponential. Depending on the value of $\varphi_0$, the reversal that opens the descent can be more or less abrupt, but lo and behold one would expect it to bring along a spurt of exponential expansion for the Universe, before the actual inflationary phase sets in. This generic feature was readily recognized and actually served as a motivation for the analysis in \cite{dks}, but turning into a quantitative prediction for the spectrum of scalar perturbations proved rather difficult, so much so that it failed to emerge in \cite{dkps}. As we shall see, a pre--inflationary peak does show up in the power spectrum of scalar perturbations that, as discussed in \cite{dkps}, experiences in general a wide infrared depression before merging with an almost scale invariant profile ${\cal P}_\zeta(k) \sim k^{n_s-1}$ as the dynamical evolution finally approaches the Lucchin--Matarrese (LM) attractor \cite{lm}. The reversal becomes less abrupt for lower values of $\varphi_0$, while the peak grows in size, until it eventually turns into the typical feature well described in \cite{destri}. This signals the approach to slow roll in more conventional models, and actually shows up, in the presence of a mild exponential alone, in the whole range of values for $\varphi_0$ that we have explored. As expected, however, the reversal of the scalar motion leaves no signs in the corresponding spectrum of tensor perturbations, consistently with the analysis in \cite{dkps}.

It is natural to inquire whether this type of behavior could have some bearing on our current understanding of the CMB angular power spectrum. We shall qualify under which assumptions this might well be the case, and we shall also manage to vindicate, to some extent at least, our expectation via a detailed, if preliminary, comparison with CMB data. As we have anticipated, the region of interest is the low--$\ell$ tail of the angular power spectrum, where some anomalies with respect to the $\Lambda$CDM setting have long been spotted but where cosmic variance, which reflects our very special observation site for the Universe, adds more than one word of caution to any attempt to interpret their actual meaning.

The main low--$\ell$ anomaly in the CMB angular power spectrum is quadrupole reduction, and it is large enough to go unnoticed. Indeed, a number of works recently touched upon the subject from different viewpoints \cite{dkps,quadrupole_red}, and they include a detailed re--analysis of the cosmic mask \cite{gruppuso}. Interestingly one can argue, on rather general grounds, that \emph{quadrupole reduction accompanies naturally the emergence from an initial singularity}. Moreover, for low multipoles $\ell \lesssim 35$ the actual CMB observable, the angular power spectrum, is determined essentially by a Bessel--like transform \cite{mukhanov_slow} that follows closely the power spectrum, so that in principle features of the former can reflect themselves in similar features of the latter, and vice versa.

In practice, however, this type of correspondence requires an additional assumption that, if correct, would make, by itself, the whole story quite interesting.
The assumption, which is not implausible numerically, posits that the largest wavelengths entering the horizon at the present epoch are essentially those that exited around the onset of the inflationary phase. Or, if you will, that the low--$\ell$ CMB data are opening in front of our instruments a small window on the onset of inflation, the very phenomenon that is usually advocated to explain the apparent flatness and homogeneity of our Universe and also explains naturally the slight tilt of the CMB power spectrum \cite{cm} that was recently confirmed to high precision by PLANCK \cite{PLANCK}. Working within this assumption, we shall begin to explore how far one can go in relating the available WMAP9 raw data \cite{wmap9} to the models at stake. We shall explore, to this end, the first 31 multipoles starting from the quadrupole, for a range of values of $\varphi_0$ that encompasses the emergence of the pre--inflationary peak, its growth and its eventual coalescence into the attractor profile.

As we shall see, the data are apparently not insensitive to the pre--inflationary peak, since centering it around $\ell = 5$ brings about a noticeable reduction of the $\chi^2$ value by two or three units. Amusingly the agreement improves, as we shall see, for values of $\gamma$ that lie below $0.08$, the choice that would tune the large -- $k$ behavior of ${\cal P}_\zeta(k)$ with the observed spectral index $n_s \approx 0.96$. This result resonates with a key indication of the PLANCK experiment, which favors generically concave inflationary potentials \cite{concave}. As we have anticipated, the pre--inflationary peak draws its origin from the region where the ``hard'' exponential begins to dominate and only the nearby behavior, which is naturally flatter in a concave potential, should play a role. We shall also vindicate this claim by displaying some power spectra computed directly in Starobinsky--like potentials \cite{starobinsky}
\be
{V}_S(\varphi) \ = \ {V}_0\,\left[ \left(1- e^{\,-\,\gamma\,(\varphi+\Delta)}\right)^2 \ + \ e^{\,2\,\varphi}\right] \label{starobinsky_intro}
\ee
that terminate on the same hard exponential, which possess very similar qualitative features. The Starobinsky potentials have aroused some interest lately in connection with Supergravity \cite{starobinsky_supergravity}, and are not foreign, in principle, to ``brane supersymmetry breaking'', if quantum corrections are taken into account. Interestingly, as we shall see, the comparison with CMB raw data favors scenarios of this type that appear to fit well within perturbative String Theory. The string coupling is in fact sized by $e^\varphi$, a quantity that stays well below one for the choices of $\varphi_0$ that result in better fits. We plan to return soon to a more detailed comparison with the CMB, modifying the standard $\Lambda$CDM setup to allow for quadrupole reduction and a pre--inflationary peak, the key features suggested by this class of models \cite{gnks}.

To reiterate, among a multitude of available vacua, String Theory suggests some peculiar options related to orientifold models where supersymmetry is broken at the string scale \cite{bsb}. Orientifold models generally allow a wide range of values for the string scale \cite{aadd}, all compatible with the standard values of Newton's constant and of gauge couplings, which includes the scales typically associated with inflation. And, as we have stressed, these two scales are linked in the simplest realizations of ``brane supersymmetry breaking'' with a ``climbing scalar''. A common origin for the two phenomena of supersymmetry breaking and inflation would represent an economy of principles, and our results can perhaps serve as a motivation in this respect, although they do not arise generically but only within a specific class of string vacua.

The structure of the paper is as follows. In Section \ref{sec:climbing} we review the climbing phenomenon starting from the one--exponential case, stressing its generality and illustrating its realization in the relatively simple class of potentials of eq.~\eqref{potwoexp_intro} and in the richer class of potentials of eq.~\eqref{starobinsky_intro}. In Section \ref{sec:powerspectrum}, which focusses on power spectra of scalar perturbations, we pinpoint the origin of the pre--inflationary climbing peak and we illustrate its dependence on $\varphi_0$. We also discuss briefly, for the sake of comparison, corresponding spectra of tensor perturbations for the same range of values for $\varphi_0$. In Section \ref{sec:observables} we move some first steps toward a quantitative comparison with the CMB, insofar as the first 30 multipoles or so are concerned. Finally, in Section \ref{sec:conclusions} we briefly summarize our main conclusions and some perspectives for future research along these lines, while the Appendix elaborates on the links between the two--exponential potentials \eqref{potwoexp_intro} and String Theory.
\vskip 24pt

\section{\sc  Climbing Scalars and String Theory}\label{sec:climbing}
\vskip 12pt

The spatially flat cosmologies of interest here correspond to a slight generalization of the Friedmann--Lemaitre--Robertson--Walker setting obtained considering the class of four--dimensional metrics
\be
ds^{\,2} \ = \ e^{\,2\,{\cal B}(t)} \, dt^2 \ - \ e^{\,\frac{2\,{\cal A}(t)}{3}}\ d{\bf x} \,\cdot \, d{\bf x} \ , \label{FLRW_gen}
\ee
where for later convenience we wrote the scale factor $a(t)$ in the form
\be
a(t) \ = \ e^{\,\frac{{\cal A}(t)}{3}} \ .
\ee
These types of cosmologies emerge naturally when Einstein gravity is minimally coupled to a scalar field subject to a potential $V(\phi)$, so that in a ``mostly negative'' signature,
\be
  \mathcal{S} \,= \, \int \, d^{\,4} \,x \, \sqrt{- \, \mbox{det} \,g} \left[ \, \frac{1}{2\, k_N^2} \ R \, + \ \frac{1}{2}\ g^{\mu\nu} \, \partial_\mu \,\phi \ \partial_\nu \, \phi \, - \, V(\phi) \right] \ .
  \label{scalanorma}
\ee

The introduction of the gauge function ${\cal B}$ is very convenient, since it allows to obtain analytic solutions, albeit not in terms of the actual cosmic time measured by comoving observes, in a single--exponential potential \cite{exp_sol} and in a number of similar instructive cases \cite{fss,bo_integrable}.
Here
\be
{R^\mu}_{\nu\rho\sigma} \ = \ \partial_\sigma \Gamma^{\mu}_{\nu\rho} \ - \ \partial_\rho \Gamma^{\mu}_{\nu\sigma} \ + \ \Gamma^{\mu}_{\sigma\tau} \, \Gamma^{\tau}_{\nu\rho} \ - \ \Gamma^{\mu}_{\rho\tau} \, \Gamma^{\tau}_{\nu\sigma}
\ee
and $R={\delta_\mu}^\rho \, g^{\,\nu\sigma}\, {R^\mu}_{\nu\rho\sigma}$, and with the convenient redefinition
\be
\varphi \ = \ k_N \, \sqrt{\frac{3}{2}} \ \phi \ , \label{redef}
\ee
for the class of metrics of eq.~\eqref{FLRW_gen} and with $\varphi$ only depending on $t$ the Lagrangian reduces, up to an overall constant, to
\be
{\cal L} \ = \ e^{\,{\cal A} \, -\, {\cal B}} \left[ \, \frac{1}{2} \, \left( \frac{d\varphi}{dt}\right)^2 \ - \ \frac{1}{2} \, \left(\frac{d{\cal A}}{dt} \right)^2 \ - \ \frac{3}{2} \ k_N^2 \ e^{\,2\, {\cal B}} \, V \, \right] \ .
\ee
If the potential $V$ is always \emph{positive}, one can work in the very convenient gauge determined by the condition
\be
V \, e^{\,2\,{\cal B}} \ = \ V_0 \ , \label{gauge}
\ee
where $V_0$ denotes its overall scale. In terms of the dimensionless ``parametric time''
\be
\tau \ = \ t\, \sqrt{3\, V_0\, k_N^2} \ ,
\ee
if $\dot{\cal A}>0$ the equations of motion take the convenient form \cite{dks}
\bea
&& \dot{\cal A}^{\,2} \ - \ \dot{\varphi}^{\,2} \ = \ 1 \ , \\
&& \ddot{\varphi} \ + \ \dot{\varphi} \sqrt{1 \ + \ \dot{\varphi}^{\,2}} \ + \ \frac{V_{\varphi}}{2\, V} \, \left( 1 \ + \ \dot{\varphi}^{\,2} \right) \ = \ 0 \ , \label{eqsgaugeB}
\eea
where ``dots'' denote derivatives with respect to $\tau$ and $V_{\varphi}$ denotes the derivative of the potential with respect to $\varphi$. Note that in this gauge the driving force originates from the logarithm of the potential. Therefore, it is exactly constant for an exponential potential and remains essentially piecewise constant in the presence of positive combinations of exponentials.

An interesting class of exact solutions exists, in terms of the parametric time $\tau$, for an exponential potential
\be
{V}(\varphi) \ = \ {V}_0 \, e^{\,2\,\gamma\, \varphi}\ . \label{potonexp}
\ee
For $0< \gamma<1$ there are actually \emph{two} classes of such solutions, which describe respectively a scalar that emerges from the initial singularity while \emph{climbing} or \emph{descending} the potential. To begin with, the \emph{climbing} solutions for the $\tau$--derivatives of $\vf$ and ${\cal A}$ are
\bea
\dot{\vf} &=& \frac{1}{2} \left[ \sqrt{\frac{1\,-\, \g}{1\,+\, \g}}\, \coth \left( \frac{\tau}{2}\ \sqrt{1\,-\, \g^{\,2}}\, \right) \ - \ \sqrt{\frac{1\,+\, \g}{1\,-\, \g}}\, \tanh
\left( \frac{\tau}{2}\ \sqrt{1\,-\, \g^{\,2}}\, \right)\right]\ , \nonumber \\
\dot{\cal A} &=& \frac{1}{2} \left[ \sqrt{\frac{1\,-\, \g}{1\,+\, \g}}\, \coth \left( \frac{\tau}{2}\ \sqrt{1\,-\, \g^{\,2}}\, \right) \ + \ \sqrt{\frac{1\,+\, \g}{1\,-\, \g}}\, \tanh
\left( \frac{\tau}{2}\ \sqrt{1\,-\, \g^{\,2}}\, \right)\right]
 \ , \label{speeds}
\eea
and the reader should appreciate that these expressions \emph{do not involve any initial--value constants} other than the Big--Bang time, here set for convenience at $\tau=0$. On the other hand, the corresponding fields read
\bea
\varphi  &=& \varphi_0 \ + \ \frac{1}{1+\gamma} \ \log \sinh  \left( \frac{\tau}{2}\ \sqrt{1\,-\, \g^{\,2}}\, \right) \ - \ \frac{1}{1-\gamma} \ \log \cosh  \left( \frac{\tau}{2}\ \sqrt{1\,-\, \g^{\,2}}\, \right)\ , \nonumber \\
{\cal A} &=& \ \frac{1}{1+\gamma} \ \log \sinh  \left( \frac{\tau}{2}\ \sqrt{1\,-\, \g^{\,2}}\, \right) \ + \ \frac{1}{1-\gamma} \ \log \cosh  \left( \frac{\tau}{2}\ \sqrt{1\,-\, \g^{\,2}}\, \right) \ ,\label{fields_1exp}
\eea
and do involve an important \emph{integration constant}, $\varphi_0$. This determines the value of $\varphi$ at a reference ``parametric time'' $\tau>0$ or, what is more interesting for us, bounds from above the largest value that it can attain during the cosmological evolution. Strictly speaking, ${\cal A}$ would also involve an additive constant, but one can set it to zero up to a rescaling of the spatial coordinates. On the other hand, $\varphi_0$ has interesting effects on the dynamics that become particularly pronounced in the two--exponential potentials
\be
{V}(\varphi) \ = \ {V}_0 \left( e^{\,2\,\varphi} \ + \ e^{\,2\,\gamma\, \varphi} \right) \ . \label{potwoexp}
\ee
Much information on these systems can be extracted from the preceding special case even though a general exact solution is not available, and the one--exponential solutions provide accurate accounts of the behavior close to the initial singularity and at late epochs, where one of the two terms dominates.

As we have anticipated, for $\gamma<1$ another class of solutions exists in the potential of eq.~\eqref{potonexp}, which describes a scalar that emerges from the initial singularity \emph{descending} the potential. The corresponding expressions can be simply obtained from eqs.~\eqref{speeds} and \eqref{fields_1exp} with the two replacements $\gamma \to - \gamma$ and $\varphi \to - \varphi$, which are a symmetry of the action of eq.~\eqref{scalanorma} with this potential, and eventually both classes of solutions converge for large $\tau$ on the LM attractor \cite{lm}, for which
\be
\dot{\varphi} \ = \ - \ \frac{\gamma}{\sqrt{1-\gamma^{\,2}}} \ , \qquad \dot{\cal A} \ = \ \frac{1}{\sqrt{1-\gamma^{\,2}}} \ . \label{lm}
\ee
However, only climbing solutions exist for $\gamma \geq 1$, and we should stress that the example has general implications: in any potential that for $\varphi\to +\infty$ is dominated by the first term in eq.~\eqref{potwoexp} the scalar field cannot emerge from an initial singularity descending that end.

The two--exponential potentials of eq.~\eqref{potwoexp} find a key motivation in String Theory, in a link between string scale and supersymmetry breaking scale that manifests itself in a class of orientifold models \cite{orientifolds}. The effect is brought about by \emph{classically stable} and yet \emph{non--mutually BPS} combinations of branes and orientifolds that are to be present simultaneously in some orientifold vacua to guarantee RR charge neutrality. It is usually called ``brane supersymmetry breaking'' (BSB) \cite{bsb} and finds its simplest manifestation in the ten--dimensional Sugimoto model in \cite{bsb}. This mechanism is directly responsible for the first contribution present in eq.~\eqref{potwoexp}, a ``hard'' term  with exponent $2\,\varphi$ that is left over at the (projective--)disk level by the ${\overline D}9$ branes (anti--BPS objects with tension $T>0$ and RR charge $Q<0$) and the $O9_+$ planes (BPS objects with $T>0$ and $Q>0$) that are present in the vacuum, whose opposite charges cancel one another but whose identical tensions add up. The Polyakov expansion of String Theory \cite{polyakov} \emph{predicts} that in ten dimensions this exponent lies precisely at the ``critical value'' where descending solutions disappear.

Let us stress that the uncanceled $D9$--$O9_+$ tension introduces conceptual and technical difficulties in the applications of BSB to Particle Physics, since for one matter flat space is not a solution of the equations of motion in the low--energy effective field theory, and consequently in dealing with this type of models one is inevitably confronted with the presence of uncanceled tadpoles. This leads readily to the need for resummations, which are complicated in Field Theory and prohibitive at the string level \cite{resummations}. For this reason in \cite{dks} we started to explore the possible role of these types of models in Cosmology. After all, the typical scales of inflationary models and the typical ranges for the string scale in type-I orientifold models can be close to one another \cite{aadd}, while in Cosmology the issue is not vacua but rather evolving states. Moreover, the nature of the classical solutions of the field equations that we have described gives us some confidence that a full--fledged string embedding might be found eventually, at least in special cases.

All in all, in the class of potentials \eqref{potwoexp} the scalar field can only emerge from the initial singularity \emph{climbing up} their left portion, which is essentially determined by the second, ``mild'' exponential. Therefore, it should not come as a surprise that the exact solutions for the one--exponential potential \eqref{potonexp} provide an effective way of setting initial conditions close to the initial singularity when solving numerically for the dynamics in the two--exponential potential of eq.~\eqref{potwoexp}, according to \footnote{The corresponding expressions in the cosmic time $t_c$ are independent of $\gamma$, and are both asymptotic to $1/t_c$.}
\bea
\varphi  &\ \ \ \thicksim\!\!\!\!\!\!\!\!\!_{{}_{{\tau \to 0}}}& \varphi_0 \ + \ \frac{1}{1+\gamma} \ \log \left( \frac{\tau}{2}\ \sqrt{1\,-\, \g^{\,2}}\, \right) \ , \nonumber \\
{\cal A} &\ \ \ \thicksim\!\!\!\!\!\!\!\!\!_{{}_{{\tau \to 0}}}& \ \frac{1}{1+\gamma} \ \log \left( \frac{\tau}{2}\ \sqrt{1\,-\, \g^{\,2}}\, \right)  \ . \label{earlytimes}
\eea
The climbing phase ends at a turning point whose location, sensitive to $\varphi_0$, determines to which extent the scalar feels the first, ``hard'' exponential while reverting to a descending phase. Eventually, if $\gamma< {1}/{\sqrt{3}}$ the Universe will attain an accelerated expansion, again largely under the spell of the mild exponential alone. Clearly, what makes this phenomenon interesting is that \emph{the climbing phase can provide a rationale for the very onset of inflation within perturbative String Theory}. In the ten--dimensional Sugimoto model of \cite{bsb} $\varphi$ is in fact the dilaton $\phi_{10}$, whose expectation value determines the string coupling in terms of the dilaton vacuum value according to
\be
g_s \ = \ e^{\, \langle \phi_{10} \rangle}
\ee
so that, with $\varphi$ bounded from above during the whole cosmic evolution, the available initial conditions leave naturally room for models with $g_s<1$. Moreover, for $d<10$ the performer changes and yet the music somehow does not: the (non--canonically normalized) field $\varphi$ becomes a $d$--dependent linear combination of dilaton and internal breathing mode \cite{fss,as13}, but the ``hard'' exponential term retains in all cases the ``critical'' form $e^{\,2\,\varphi}$. Insofar as the orthogonal combination of the two fields is somehow stabilized, climbing thus remains an inevitable fate. In terms of a canonically normalized scalar $\phi$, however, the potential changes in a definite fashion with the number of space--time dimensions, becoming in particular $e^{\,\sqrt{6} k_N \phi}$ in four dimensions.
\begin{figure}[h]
\begin{center}$
\begin{array}{ccc}
\epsfig{file=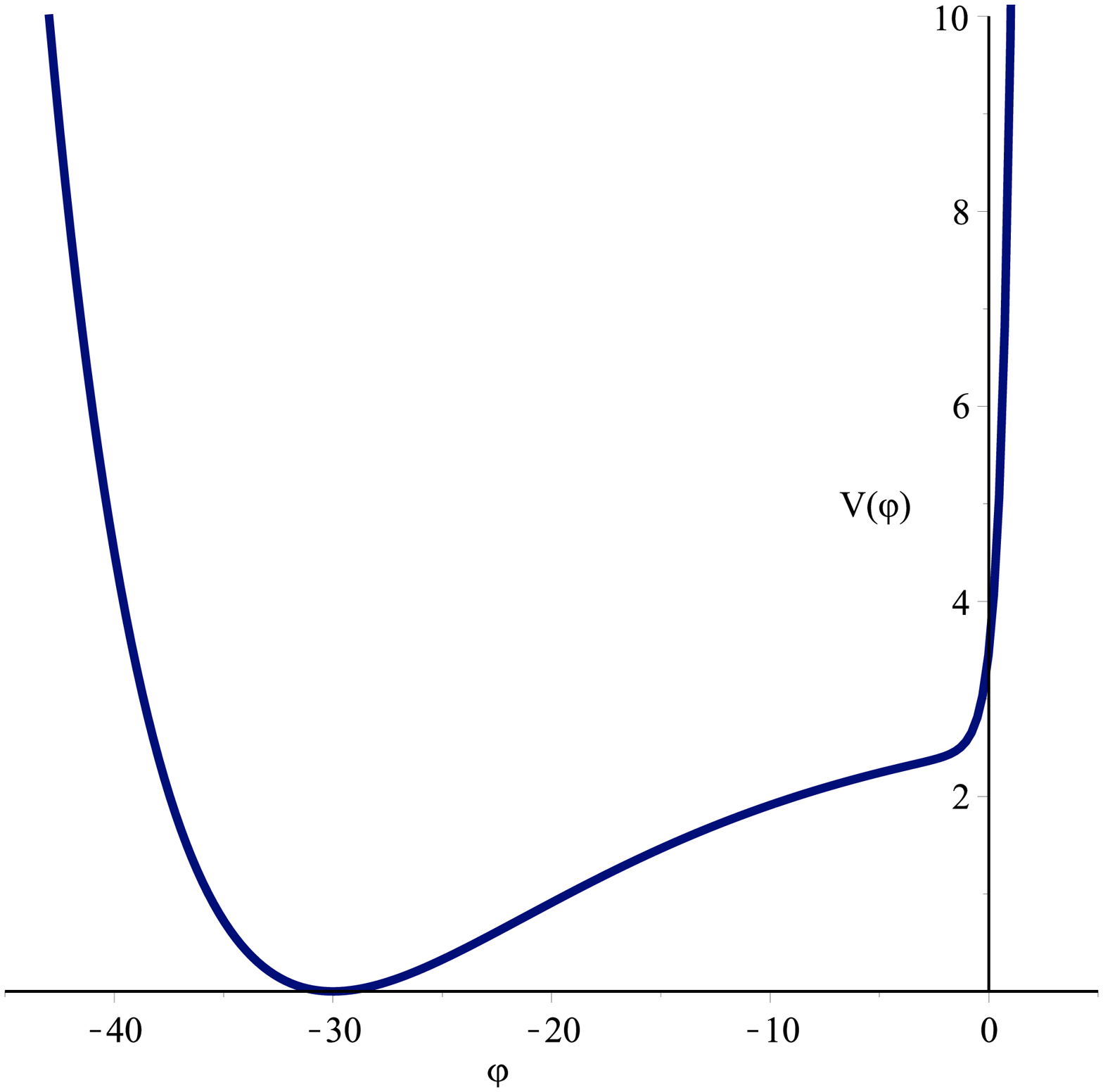, height=1.2in, width=1.2in} & \qquad\quad
\epsfig{file=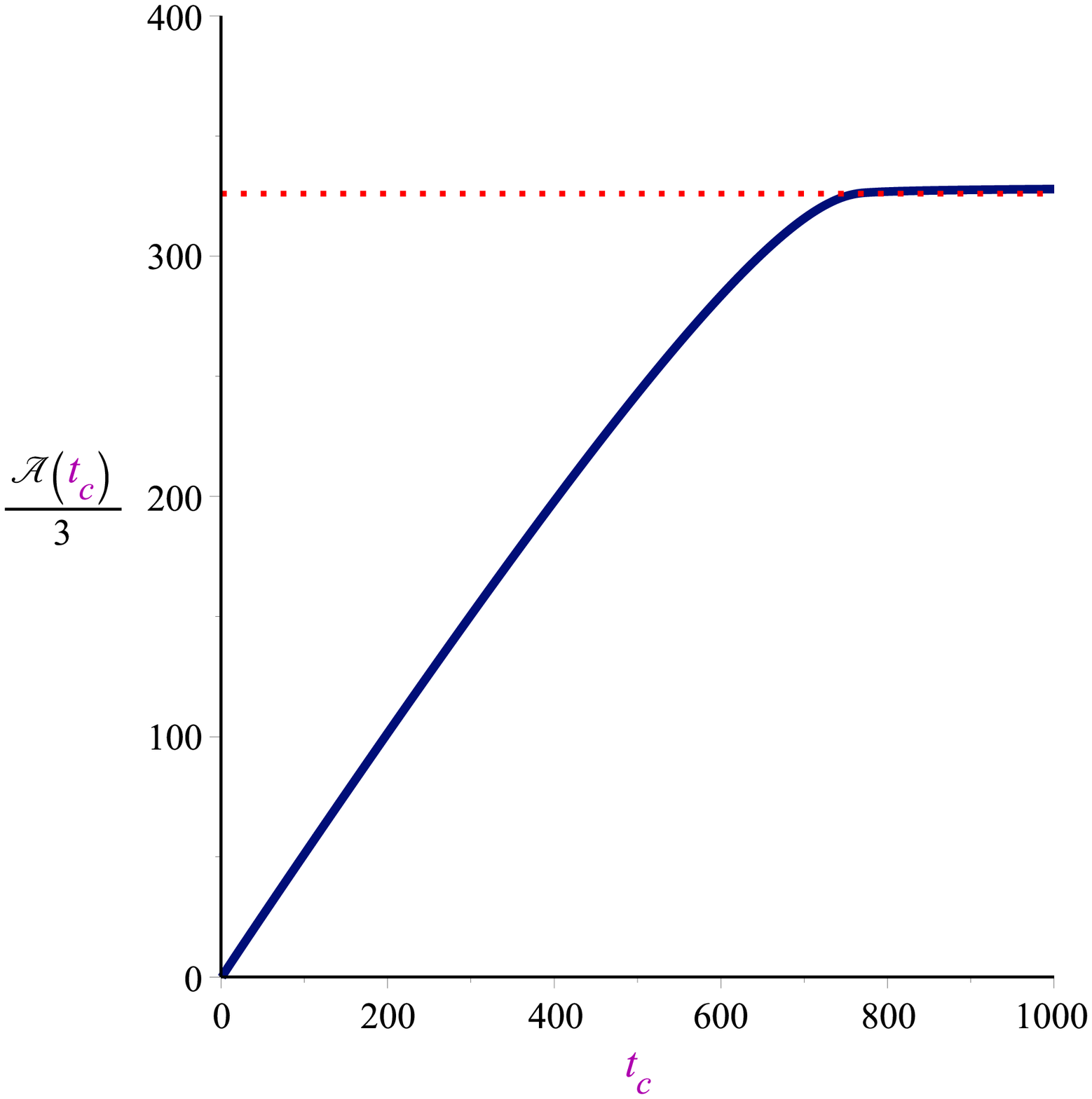, height=1.2in, width=1.2in} & \qquad\quad
\epsfig{file=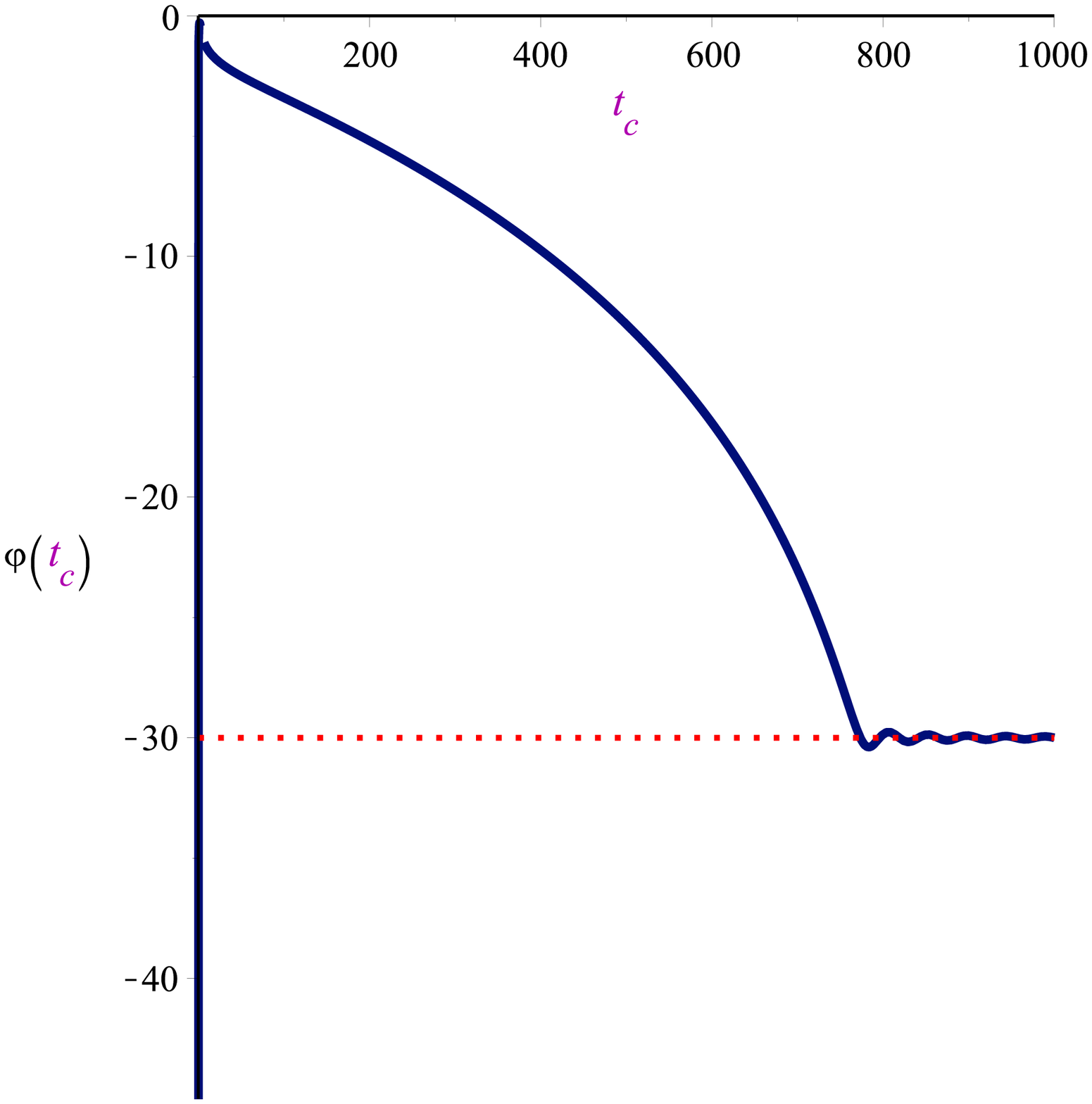, height=1.2in, width=1.2in}
\end{array}$
\end{center}
\caption{\small A Starobinsky-like potential whose right end terminates on a ``hard'' exponential (left) and the corresponding evolutions of ${\cal A}/3$ (center) and $\varphi$ (right) in cosmic time $t_c$. After an initial fast--roll descent of the left end, the scalar climbs up to a point, reverts its motion, attains a slow--roll regime for a while and eventually reaches the bottom of the potential well, where it comes to rest after some oscillations. An enlarged early--time view is provided in fig.~\ref{fig:twoexp}.}
\label{fig:starobinsky}
\end{figure}
\begin{figure}[h]
\begin{center}$
\begin{array}{ccc}
\epsfig{file=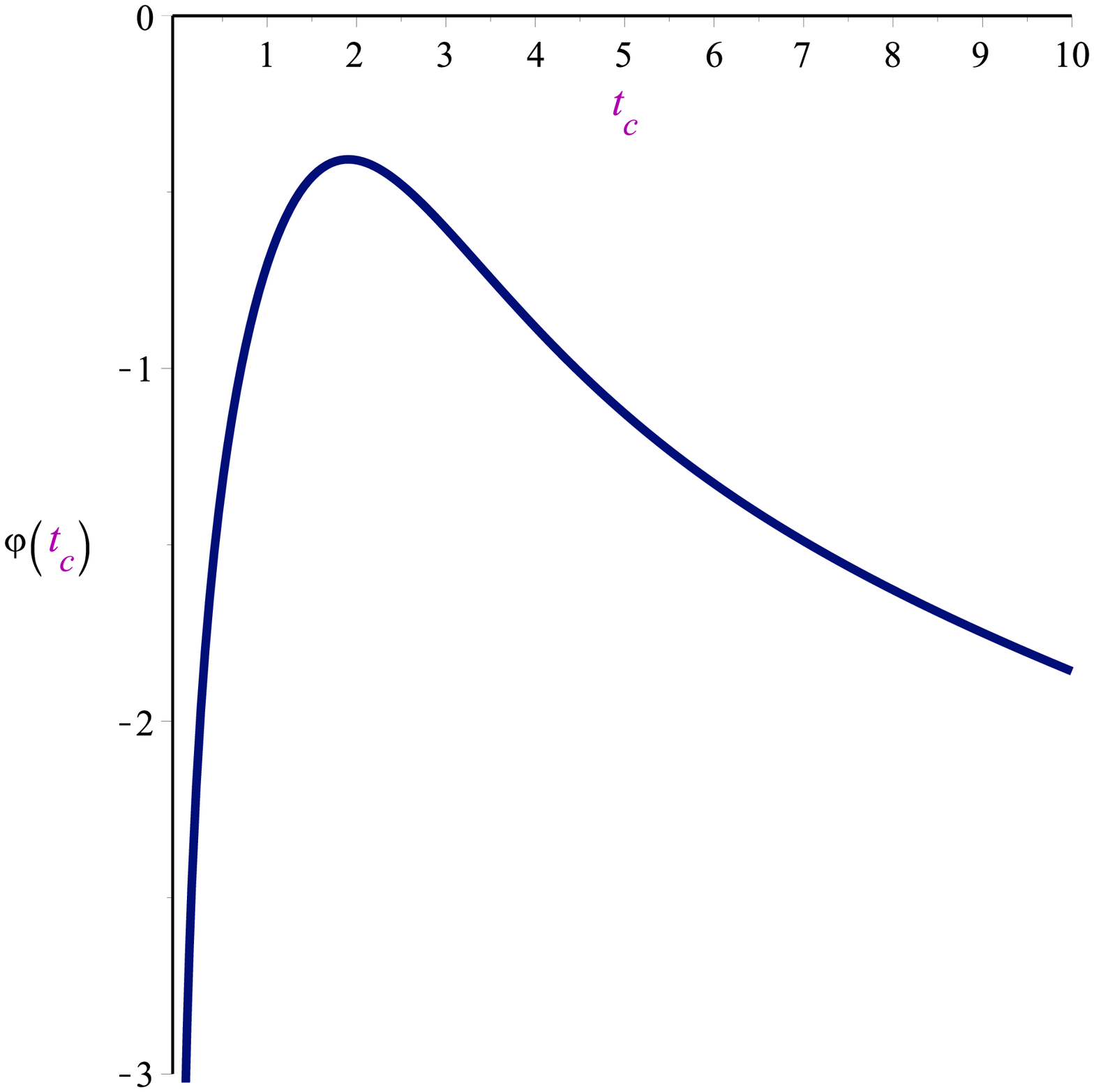, height=1.2in, width=1.2in} & \qquad\qquad
\epsfig{file=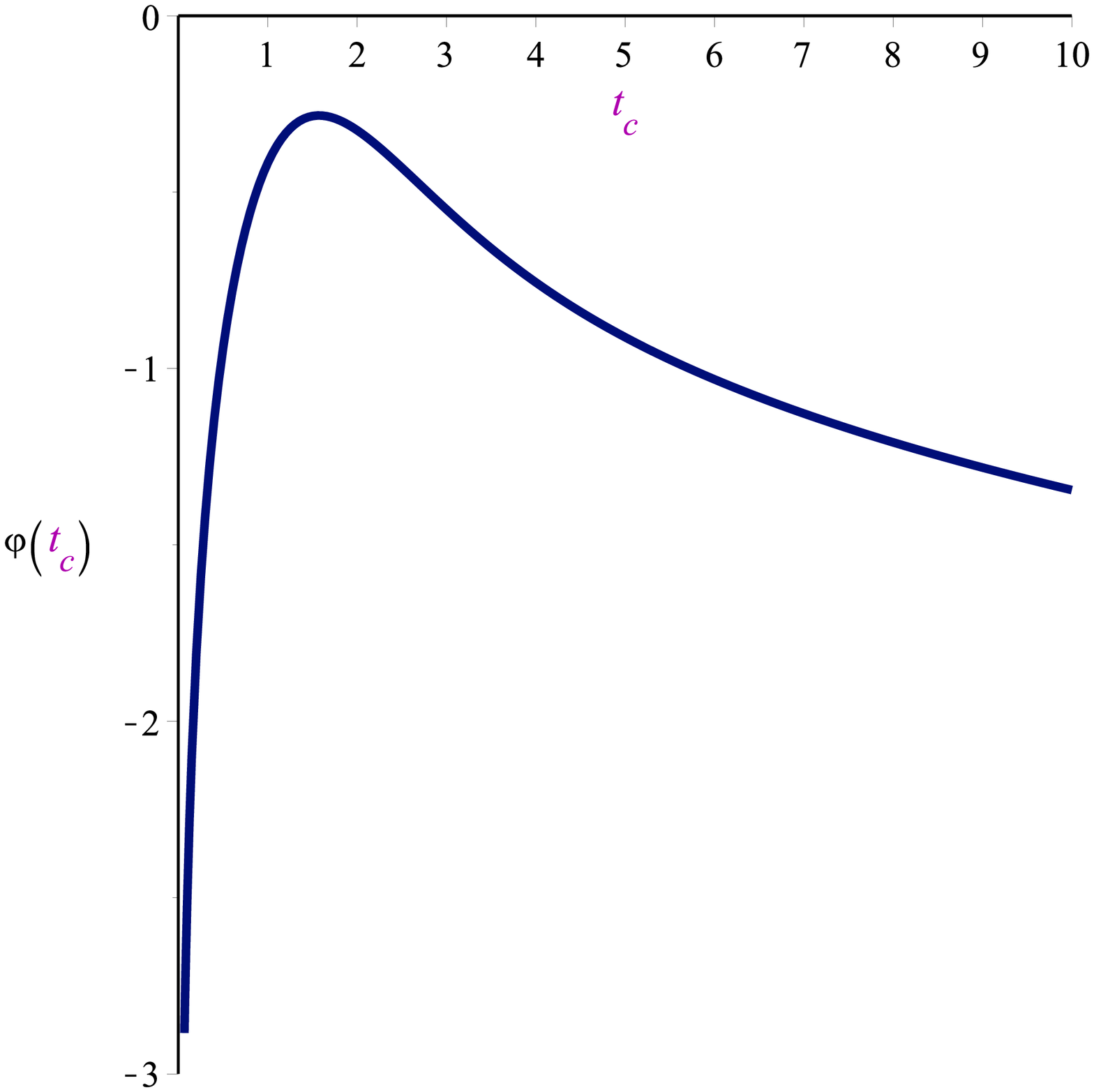, height=1.2in, width=1.2in}
\end{array}$
\end{center}
\caption{\small The scalar behaves in qualitatively similar ways (displayed here in cosmic time) near the inversion points of the two--exponential potential of eq.~\eqref{potwoexp} (left) and of the Starobinsky--like potential of fig.~\ref{fig:starobinsky} (right), if the inversion occurs close enough to the ``hard'' exponential.}
\label{fig:twoexp}
\end{figure}

As we have stressed, the second term in eq.~\eqref{potwoexp} plays an essential role both in the early ascent and in the final descent. Yet, its origin is admittedly less compelling. It can be traced, to some extent, to other string $p$--branes, which under some assumptions spelled out in \cite{fss,as13} give rise for space--filling branes to the discrete set of values
\be
\gamma \ = \ \frac{1}{12} \left( p \ + \ 9 \ - \ 6\, \alpha \right) \ ,
\ee
if the dilaton enters their world--volume actions in the string frame via the exponential $e^{\,-\,\alpha \phi_{10}}$. This set includes the non--BPS D3 brane found long ago in \cite{branesugimoto} following the approach of Sen \cite{sen}, but there are clearly two familiar types of branes: $\alpha$ would be one for $D$--branes and two for the Neveu--Schwarz (NS) fivebrane, but the formula has in principle a wider range of applicability since a zoo of exotic branes with higher values of $\alpha$ that are present for $d<10$ was also identified in \cite{branescan}. As argued in \cite{fss,as13}, all these branes ought to have been generically present in the vacuum at very early epochs, close to the initial singularity. In particular, an NS fivebrane wrapped on a small internal cycle, corresponding to $p=4$ and $\alpha=2$, would yield a ``mild'' exponential term with $\gamma=\frac{1}{12}$, while its instability in orientifold models and its consequent decay could perhaps account for the eventual graceful exit of the Universe from the inflationary phase.  A brief discussion of the role of these branes is presented in the Appendix. On the other hand, $\gamma=\frac{1}{12}$ would naively translate, in the potentials of eq.~\eqref{potwoexp}, into a spectral index equal to $0.957$, which lies intriguingly within the experimentally allowed range for the CMB, $n_s = 0.9603 \pm 0.0073$.

All in all, the two--exponential model of eq.~\eqref{potwoexp} is admittedly somewhat naive, but nonetheless in the next sections we shall hopefully convince the reader that it can convey interesting dynamical lessons, possibly with some bearing on the CMB power spectrum. Let us therefore concentrate on very early epochs in these cosmologies, leaving aside for the moment a detailed account of an eventual graceful exit following a typical inflationary epoch with about 60 $e$--folds or so. As we have already stressed, we are drawing some motivation from the striking fact that climbing can make inflation an inevitable feat while also linking it to another phenomenon, the breaking of Supersymmetry, which would also occur at very high scales in this context. We would like to stress that, in this whole class of systems, the scalar field spans twice a given region of its configuration space during the cosmic evolution, first moving toward the steep potential in a regime of fast roll and then reverting from it. An eventual slow--roll regime can be attained for suitable completions of the potential, and the scalar can even be stabilized in potential wells that left no tangible signs on its fast ascent.

For instance, the Starobinsky--like potentials \cite{starobinsky}
\be
{V}_S(\varphi) \ = \ {V}_0\,\left[ \left(1- e^{\,-\,\gamma\,(\varphi+\Delta)}\right)^2 \ + \ e^{\,2\,\varphi}\right]\ , \label{starobinsky}
\ee
\noindent which have received some attention lately in connection with Supergravity \cite{starobinsky_supergravity}, can be combined with the ``hard exponential'' in eq.~\eqref{potwoexp} to yield this type of dynamics. A typical solution for $\varphi(t)$ in this context is displayed in fig.~\ref{fig:starobinsky}, which vindicates some of the preceding claims since the scalar field:
\begin{itemize}
\item[1. ] emerges in fast roll from the left end of the potential (dominated by a ``mild'' exponential that actually grows rapidly as $\varphi \to -\infty$), moves to the right and climbs up, leaving behind a potential well;

\item[2. ] reverts its motion, more or less abruptly depending on how close it comes to the ``hard wall'', before attaining during the ensuing descent a slow--roll regime driven by the milder terms in the potential;

\item[3. ] eventually settles at the bottom of the potential well after some oscillations.
\end{itemize}

As we have anticipated, our aim here is to elucidate how the effects of the transition between the early fast--roll ascent driven by the ``hard'' exponential of BSB and the subsequent descent depend on $\varphi_0$. Transitions to slow roll in conventional inflationary potentials were nicely investigated in \cite{destri}, and were shown to leave a distinctive mark in power spectra of scalar perturbations: a quick growth from a deep infra--red depression followed by a sharp peak and a few damped oscillations before a rapid approach to an almost scale invariant spectrum. Both this result and the far wider infrared depression described in detail in \cite{dkps} will emerge again from our analysis for special choices of $\varphi_0$, but taking a closer look will unveil the generic emergence of a new type of spectral distortion.
\begin{figure}[h]
\begin{center}$
\begin{array}{ccc}
\epsfig{file=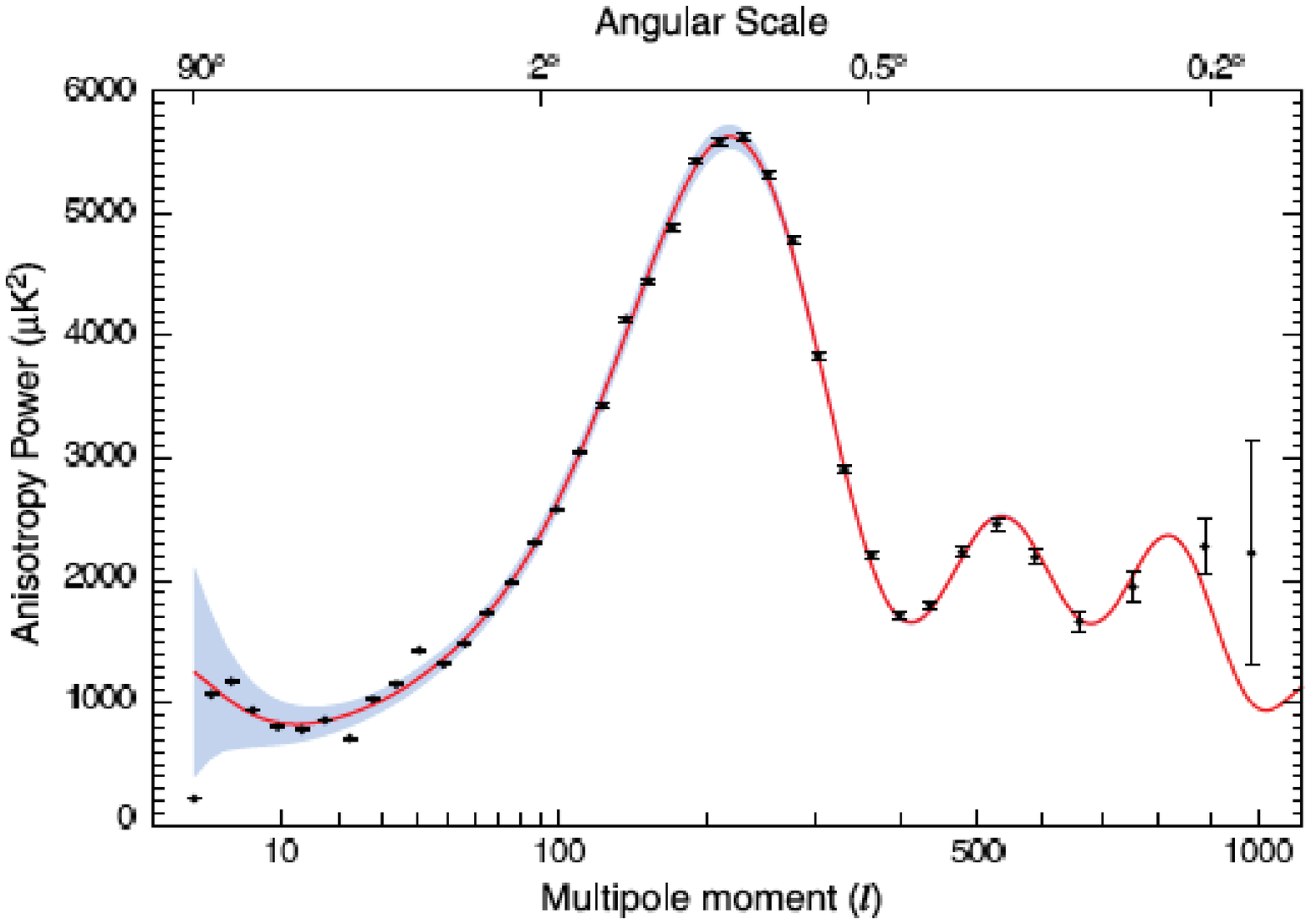, height=1.5in, width=1.5in} & \qquad\quad
\epsfig{file=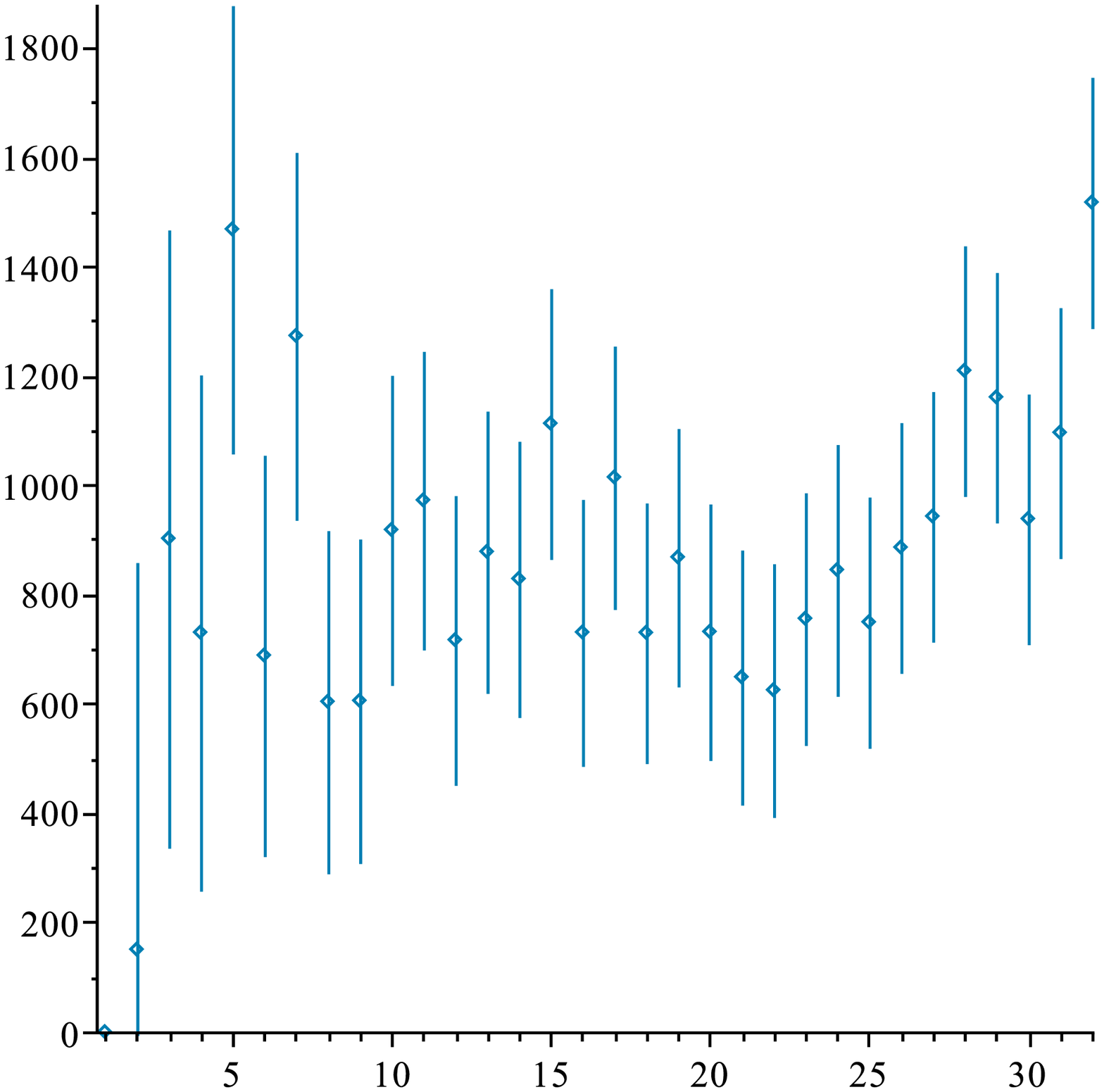, height=1.5in, width=1.5in} & \qquad\quad
\epsfig{file=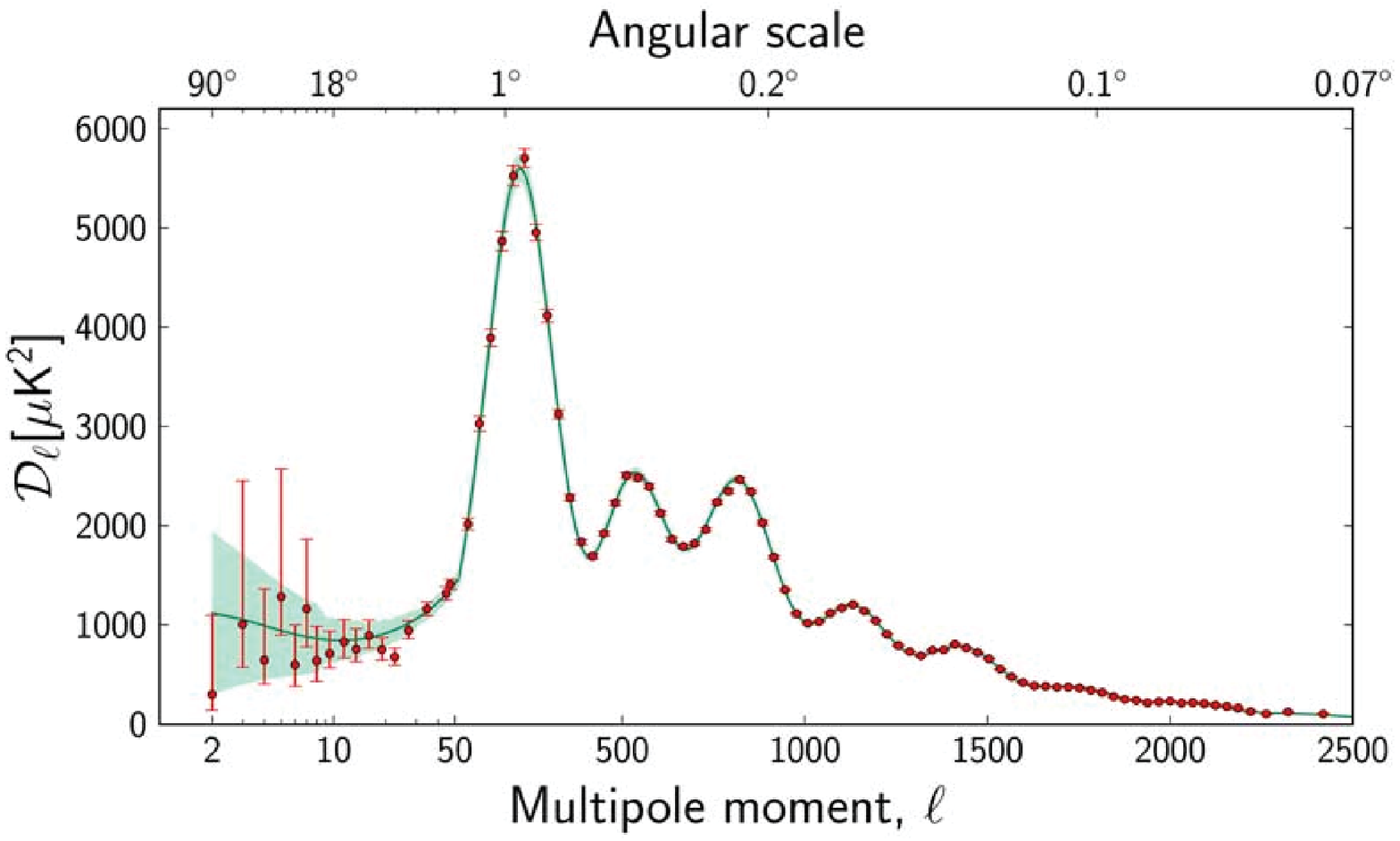, height=1.5in, width=1.5in}
\end{array}$
\end{center}
\caption{\small WMAP9 determination \cite{wmap9} of the CMB angular power spectrum (left), the raw data for its low--$\ell$ portion (center) and the corresponding PLANCK determination \cite{PLANCK} (right). The anomalies of interest in this paper concern the low-$\ell$ region detailed in the central portion of the figure, while the shadows in the outer portions are meant emphasize the role of ``cosmic variance''.}
\label{fig:WMAP9-PLANCK}
\end{figure}

A proper characterization of these phenomena cannot forego some reference to the behavior of the Hubble parameter
\be
H \ = \ k_N\ \sqrt{\frac{V(\varphi)}{3} \ \left(1 \ + \ \dot{\varphi}^2\right)} \ . \label{hubble}
\ee
and of two familiar slow--roll parameters
\bea
\epsilon_\phi &\equiv& - \ \frac{1}{H^2}\ \frac{dH}{dt_c} \ = \ 3 \ \frac{\dot{\varphi}^2}{1 \ + \ \dot{\varphi}^2} \ ,\nonumber \\
\eta_\phi &\equiv& \frac{1}{k_N^2\, V} \ V_{\phi\phi} \ = \ \frac{3}{2\,V} \  V_{\varphi\varphi} \ , \label{eps_eta}
\eea
where we have used a shorthand notation for the second derivative of $V$ with respect to the scalar field and $t_c$ denotes the cosmic time measured by comoving observers, defined in the two--exponential models according to
\be
dt_c\ = \ e^{\,{\cal B}}\, dt\ .
\ee

Restricting the attention to the two--exponential potentials $\eqref{potwoexp}$ brings about a number of technical simplifications and yet, we believe, can still capture the essential new features that can be largely traced to the epoch when the scalar terminates its ascent. The ensuing analysis refines and corrects to some extent the results in \cite{dkps}, and hopefully it can also convey an overall picture of the potential imprints of BSB on the CMB power spectrum of scalar perturbations. The first of these imprints is a \emph{reduction of power at low--frequencies} within a window that, as we shall see, becomes \emph{significantly wider for larger values of $\varphi_0$}, as the scalar feels more intensely the ``hard wall''. This is the scenario that was elaborated upon at length in \cite{dkps}, but as we shall see the available CMB data do not seem to favor it. Moreover, \emph{a sizable lack of power in wide--angle correlations is a generic feature in cosmologies emerging from an initial singularity}, while the climbing phenomenon can leave behind a more distinctive mark. Low and behold, current measurements may be confronting us with some pre--inflationary information, since for one matter the low--$\ell$ tails of WMAP9 or PLANCK angular power spectra, if taken at face value, point to a reduction of the quadrupole, and refined and well motivated alternative choices for the cosmic mask enhance the effect \cite{gruppuso} rather than reducing it. However, in the same spirit one can perhaps spot in fig.~\ref{fig:WMAP9-PLANCK} a rather pronounced peak for $\ell \simeq 5$ and some more oscillations. Following the suggestion of \cite{as13}, in this paper we would like to elaborate in detail on the possible lessons that a climbing scalar can provide in this respect, and conversely on how these types of spectra can select preferred values for $\varphi_0$. The resulting picture will comply to the intuitive idea that a ``hard'' reflecting wall can make the reversal of the scalar motion more or less abrupt depending on how close it gets to it, and thus on $\varphi_0$.

We can conclude this section adding to this positive note some cautionary remarks on the role of climbing in String Theory. To begin with, climbing is not a definite prediction of String Theory, although it occurs in a wide class of cosmologies related to orientifold models with BSB, where the supersymmetry breaking scale is tied to the scale of inflation. Moreover, even in this context the phenomenon is inevitable only in one--field reductions, which are at any rate a familiar choice in Cosmology. Remarkably, as we have already stressed, even after compactification there is always a one--field reduction that leaves behind a ``critical'' scalar $\varphi$, which is purely the dilaton only in ten dimensions \cite{fss,as13}. All in all, larger values of $\varphi$ bring about larger values of the string coupling, so that as we have stressed a dynamics where these are subject to an upper bound possesses the attractive feature of being naturally captured by string perturbation theory. Still, one should not forget that, in taking these models seriously, one is pushing rather far our current grasp of String Theory. Curvature corrections become in fact important near the initial singularity, and are naively expected to dominate precisely at epochs where the early climbing would occur. These intricacies were examined in a number of cases in \cite{cd}, with due attention to possible ways of bypassing them, at least insofar as quadratic curvature corrections go, but no definite conclusion was reached in this sense. However, with no better way to proceed at present, one may well explore the possible consequences of this intriguing dynamics while keeping well in mind this important proviso. This is what we shall do in the following sections.
\vskip 24pt

\section{\sc  A New Look at the Power Spectra of Climbing Scalars}\label{sec:powerspectrum}
\vskip 12pt

Following \cite{maldacena}, one can study scalar perturbations in the class of cosmological backgrounds of the preceding section, even at the non--linear level, starting from the ADM decomposition
\be
ds^{\,2} \ = \ N^2\, e^{\,2\,{\cal B}(\tau)}\ d\tau^2 \ - \ h_{ij} \, \left(dx^i \,+\, N^i\, e^{\,{\cal B}(\tau)}\ d\tau \right)\left(dx^j \,+\, N^j\, e^{\,{\cal B}(\tau)}\ d\tau \right)
\ee
and working with the gauge choice
\be
h_{ij} \ = \ e^{\,\frac{2\,{\cal A}(\tau)}{3}} \, e^{\,2\,\zeta} \ \delta_{ij} \ , \qquad \delta \phi \ = \ 0 \ .
\ee
The perturbations of the scalar field then disappear and $\zeta({\bf x},\tau)$ becomes the key variable, both for the power spectrum and for the bi-spectrum, and in particular its two--point function at the large ``parametric times'' $\tau_F$ that correspond to the end of inflation determines the power spectrum according to
\be
\langle \zeta({\bf x},\tau_F)\,  \zeta({\bf x},\tau_F) \rangle \ = \ \int_0^\infty \frac{dk}{k} \ P_\zeta(k) \ . \label{power_zeta}
\ee
$\zeta({\bf x},\tau_F)$ possesses the striking property of being conserved outside the horizon: its very existence opens a window on the Early Universe, since information stored on super--horizon scales during an early inflationary phase is ready to reemerge unabridged, in front of our detection instruments, after a decelerated phase.

The Mukhanov--Sasaki (MS) variable
\be
v({\bf x},\tau) \ = z(\tau) \, \zeta({\bf x},\tau) \ ,
\ee
where
\be
{z}(\tau) \ = \ \frac{1}{k_N}\ \sqrt{6}\ e^{\,\frac{{\cal A}(\tau)}{3}} \ \frac{d\varphi(\tau)}{d{\cal A}(\tau)} \ , \label{zeta}
\ee
does not share the property of $\zeta({\bf x},\tau)$, but has the virtue of leading to a very instructive formulation of the quadratic problem. Actually, the difference between $\zeta$ and $v$ is not marginal in our case, since $z(\tau)$ vanishes at the end of the climbing phase, so that $\zeta$ develops a pole there \cite{kodama}. Since the power spectrum depends only on the large--$\tau_F$ behavior of $\zeta$, this fact does not introduce serious difficulties, even in numerical studies, which can be effected in terms of other related quantities as in \cite{dkps}, but special care should be exercised  in studying the bi--spectrum, which depends on the detailed behavior of $\zeta$ over the whole range of $\tau$ that is traced out during the cosmological evolution. We hope to reconsider elsewhere the Schwinger--Keldysh formalism \cite{sk} in models exhibiting the climbing phenomenon.

Returning to the power spectrum, let us recall that expanding the quantum MS field as
\be
v({\bf x},\eta) \ = \ \int \frac{d^3 {\bf k}}{(2\pi)^3} \left[ v_k (\eta)\, \alpha({\bf k})\,e^{\,i{\bf k}\cdot {\bf x}} \ + \  v_k^\star (\eta)\, \alpha({\bf k})^\dagger\,e^{\,-\,i{\bf k}\cdot {\bf x}}\right]
\ee
and working in terms of the (dimensionless) conformal time $\eta$ defined according to
\be
ds^{\,2} \ = e^{\,\frac{2\,{\cal A}(\tau)}{3}} \left(  d \eta^2 \ - \ d{\bf x} \,\cdot \, d{\bf x}\right) \ ,
\ee
so that
\be
d \eta \ = \ e^{\,-\,\frac{{\cal A}}{3}} \, \sqrt{\frac{V_0}{V}} \, d \tau \ ,
\ee
the Fourier coefficients $v_k$ that play the role of the flat--space exponentials $e^{\pm i \omega t}$
satisfy the Schr\"odinger--like equation
\be
\left(\frac{d^2}{d \eta^2} \ + \ k^2 \ - \ W_s(\eta) \right) v_k(\eta) \ = \ 0
\ee
with the Bunch--Davies condition
\be
v_k(\eta) \ \ \ \thicksim\!\!\!\!\!\!\!\!\!\!_{{}_{{k \to \infty}}} \frac{1}{\sqrt{2k}}\ e^{\,-\,i\,k\,\eta}\
\ee
and the Wronskian constraint
\be
v_k \, \frac{\partial}{\partial \, \eta} \ v_k^\star \ - \ v_k^\star \, \frac{\partial}{\partial \, \eta} \ v_k \ = \ i \ .
\ee

$W_s$ is the MS potential, which is determined by the background cosmology via the relation
\be
W_s \ = \ \frac{z^{\prime\prime}(\eta)}{z(\eta)} \ , \label{MS_potential}
\ee
where ``primes'' denote derivatives with respect to the conformal time $\eta$ and $z$ is defined in eq.~\eqref{zeta}.
The power spectrum is then
\be
P_\zeta(k) \ = \ \frac{k^3}{2\,\pi^2} \ \left| \frac{v_k(-\epsilon)}{z(-\epsilon) }\right|^2 \ , \label{power_spec}
\ee
where the quantities involved are computed at the end of inflation, for small positive values of $\epsilon$, or equivalently for large values of $\tau_F$, when the ratio becomes independent of $\epsilon$ and reduces to a well--defined function of $k$.

From the limiting behavior displayed in eqs.~\eqref{earlytimes} one can deduce that
\be
W_s \ \ \ \thicksim\!\!\!\!\!\!\!\!\!\!_{{}_{{\eta \to - \eta_0}}} \ - \ \frac{1}{4}\ \frac{1}{(\eta+\eta_0)^2} \ , \label{early_MS}
\ee
where the conformal time $\eta\,= \,-\,\eta_0$ corresponds to the initial singularity, and in a similar fashion from the late--time behavior of eqs.~\eqref{lm} one can deduce that
\be
W_s \ \ \ \thicksim\!\!\!\!\!\!\!\!\!\!_{{}_{{\eta \to - 0^-}}} \  \ \frac{\nu^2 - \frac{1}{4}}{\eta^2} \ , \label{attractor_MS}
\ee
with
\be
\nu \ = \ \frac{3}{2} \ \frac{1\,-\,\gamma^2}{1\,-\,3\,\gamma^2} \ . \label{nugamma}
\ee

The initial singularity thus translates into a singular attractive behavior for the MS potential, while the final inflationary epoch builds up a ``centrifugal'' barrier. The dynamical properties of different models leave their signature in the intermediate region, which encodes the distinctive features of their power spectra. We shall therefore begin our analysis from $W_s$, following \cite{dkps} in establishing a dictionary between some of its key features and the power spectrum, before extending the lessons to the scalar dynamics itself. In fact, one can apply to this problem much of the machinery that is familiar from one--dimensional Quantum Mechanics, but for a key difference that should not be overlooked, since here one is solving an \emph{initial--value} problem, rather than a typical \emph{boundary--value} problem. As a result the inflationary barrier gives rise to a large \emph{amplification} of (generically ill--tuned) initial signals, which lies at the heart of a slight tilt of the power spectrum of the form
\be
P_\zeta(k) \ \sim \ k^{3-2\,\nu} \ \equiv \ k^{n_s-1}  \label{cibmukh}
\ee
for slow--roll cosmologies that proceed on an attractor curve (on the LM attractor of eq.~\eqref{lm}, in this type of systems), and thus of the whole $\Lambda$CDM setup. This early prediction of \cite{cm} was finally confirmed to high precision by PLANCK \cite{PLANCK}, with $n_s = 0.9603 \pm 0.0073$, and is currently regarded as a main lesson of inflation.

As we have anticipated in Section \ref{sec:climbing}, our aim here is to depart slightly from this canonical setting under the spell of String Theory and BSB, and to examine in detail the predictions of the two--exponential model of eq.~\eqref{potwoexp}, whose attractor curve is only approached after a pre--inflationary climbing phase. The dynamical problem, as we shall see, has some interest of its own since it reveals novel effects, but our analysis was clearly motivated by the probable discrepancies between the low--$\ell$ CMB tail of fig.~\ref{fig:WMAP9-PLANCK} and the predictions of the $\Lambda$CDM model, which rest after all on the attractor power spectrum of eq.~\eqref{cibmukh}. Some first steps toward a quantitative comparison between the CMB and the refined spectra that we are about to describe will be the subject of Section \ref{sec:observables}.

Some of the results of \cite{dkps} provide a convenient starting point for our current discussion. To begin with, combining the limiting behaviors in eqs.~\eqref{early_MS} and \eqref{attractor_MS} one can readily conclude that \emph{an initial singularity finitely back in the past forces $W_s(\eta)$ to cross the horizontal axis}. The area bounded by its upper portion, which determines the WKB amplification factor, saturates as $k \to 0$, and drawing from standard facts of Quantum Mechanics one can conclude that, in view of the $k^3$ pre--factor in eq.~\eqref{power_spec}, the resulting power spectrum is bound to decrease for low $k$. The lack of power in large--angle correlations that both WMAP9 \cite{wmap9} and PLANCK \cite{PLANCK} apparently see, if their data are taken at face value despite cosmic variance, could thus seemingly translate into an indication that our instruments are capturing some glimpses of an initial singularity.

It is actually simple to exhibit a class of MS potentials displaying exactly this type of effect. To this end, it suffices to displace the attractor MS potential \eqref{attractor_MS} according to
\be
\frac{\nu^2 - 1/4}{\eta^2} \ \longrightarrow \ \frac{\nu^2 - 1/4}{\eta^2} \ - \ \frac{\nu^2 - 1/4}{\eta_0^2}  \ , \label{lower_Ws}
\ee
so that it meets the horizontal axis at $\eta\,=\,-\,\eta_0$, which is tantamount to effecting the replacement
\be
k \ \longrightarrow \sqrt{k^2\ + \ \frac{1}{\eta_0^2}\ \left(\nu^2 - \frac{1}{4}\right)} \ ,
\ee
but only in the \emph{second} factor of eq.~\eqref{power_spec}. The resulting power spectrum is then exactly
\be
P_\zeta(k) \ \sim \ \frac{\left(k\, \eta_0\right)^{3}}{\sqrt{\left(k\, \eta_0\right)^2\ + \  \left(\nu^2 - \frac{1}{4}\right)}} \ , \label{cibmukh_BB}
\ee
and exhibits clearly the type of low--frequency depression that we had anticipated.
\begin{figure}[h]
\begin{center}$
\begin{array}{ccc}
\epsfig{file=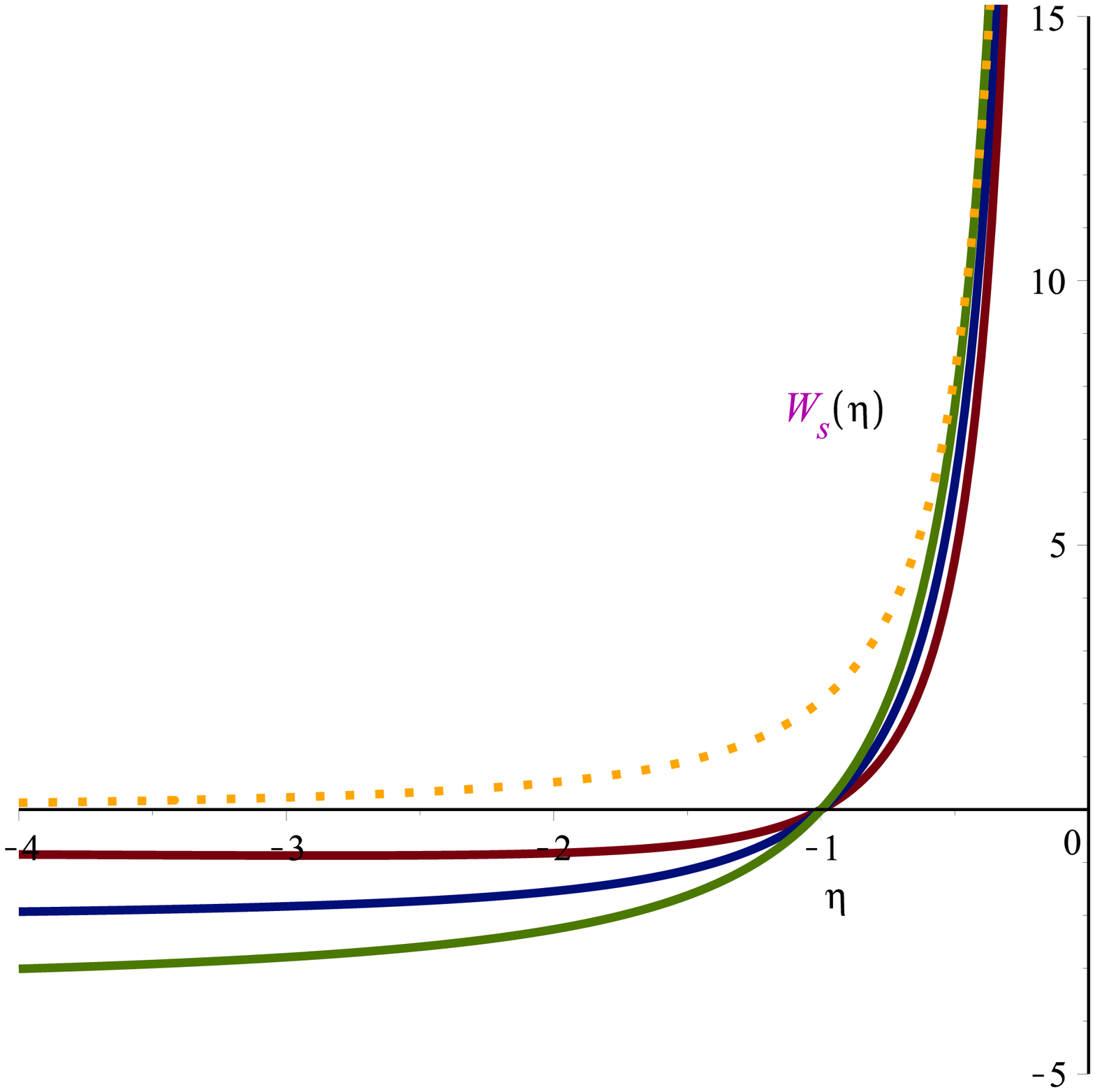, height=1.5in, width=1.5in} & \qquad\quad
\epsfig{file=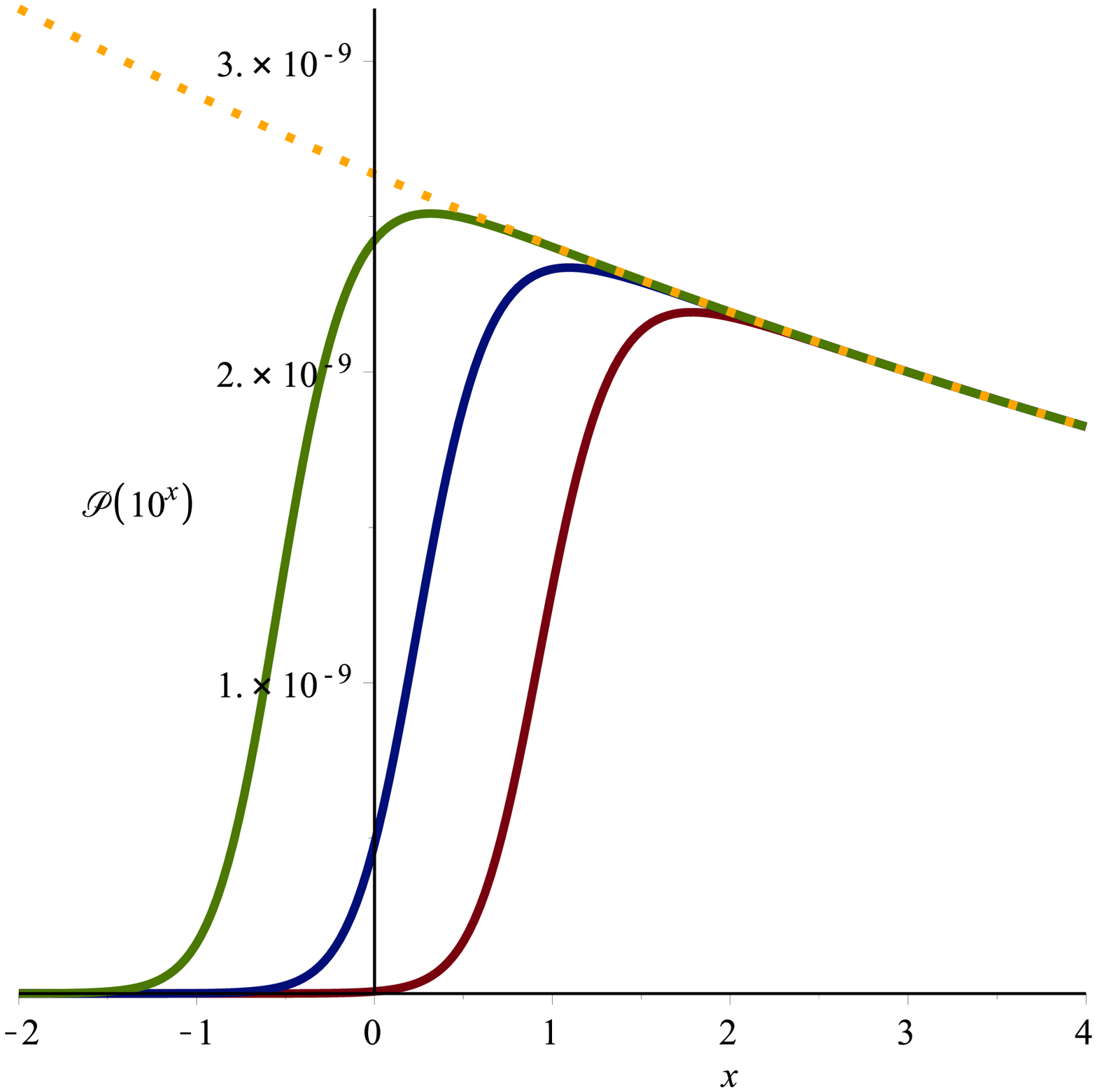, height=1.5in, width=1.5in} & \qquad\quad
\epsfig{file=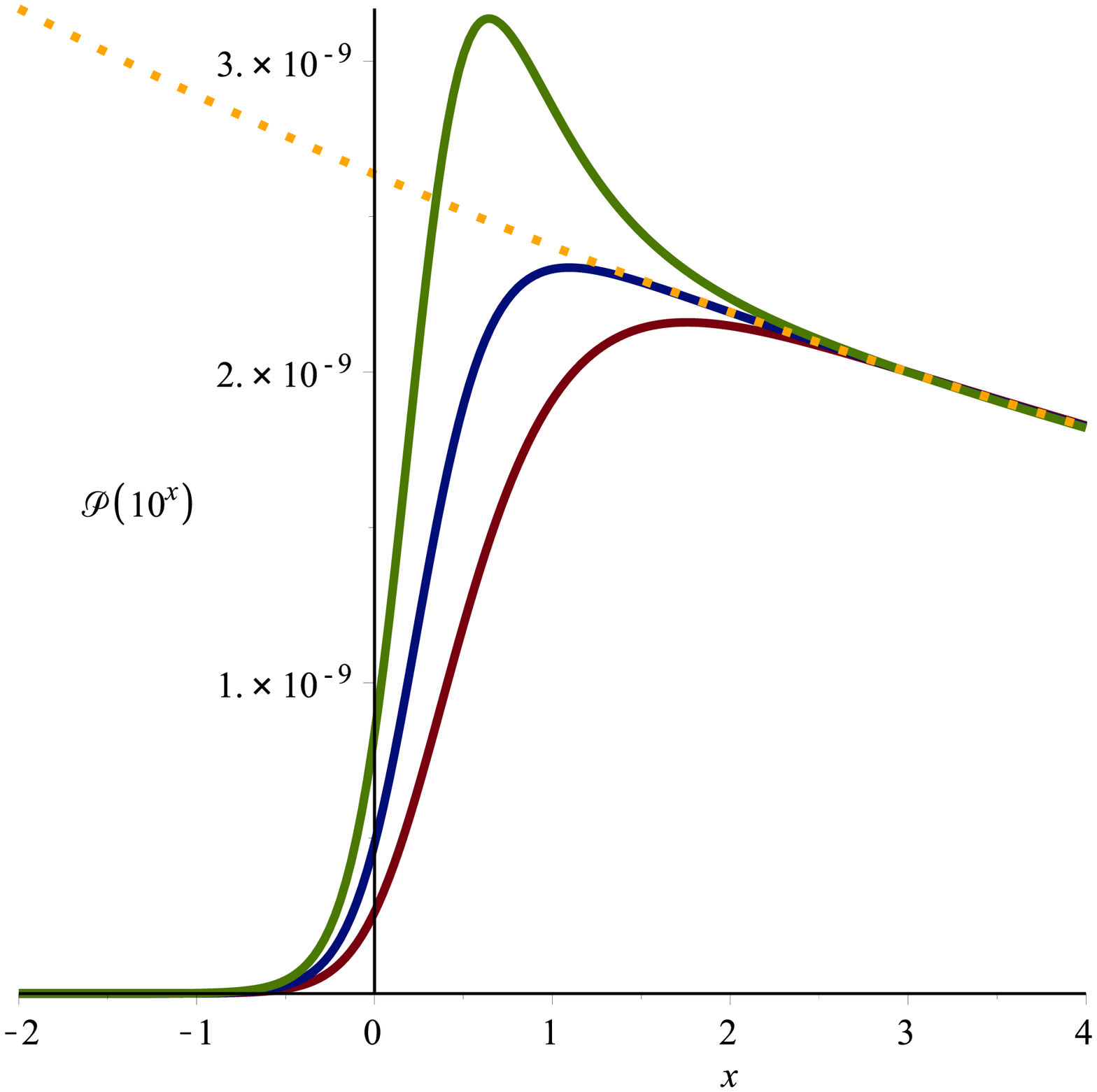, height=1.5in, width=1.5in}
\end{array}$
\end{center}
\caption{\small Attractor (orange, dotted) and Coulomb--like $W_s$ (left) \emph{vs} conformal time $\eta$ for $\eta_0=1$ and $c=1$ (red), $2$ (blue), $3$ (green). Coulomb--like spectra (center) for $c=2$ and $\eta_0=0.2$ (red), $1$ (blue), $6$ (green), or (right) with $\eta_0=1$ and $c=1.3$ (red), $2$ (blue), $2.7$ (green). In the last two cases on the horizontal axis we display $x$, where $k=10^x$.}
\label{fig:Coulomb}
\end{figure}

Actually, in an Appendix of \cite{dkps} we presented an exact solution for the family of Coulomb--like MS potentials
\be
W_s(\eta) \ = \ \frac{\nu^2 - \frac{1}{4}}{\eta^2} \ \left[ c\left(1\, +\, \frac{\eta}{\eta_0} \right) \ + \ (1-c)\left( 1\, +\, \frac{\eta}{\eta_0} \right)^2\right] \ ,
\ee
which include the attractor MS potential \eqref{attractor_MS} and, for $c \geq 1$, also displacements as in eq.~\eqref{lower_Ws} and rotations with respect to it. The corresponding power spectra read
\be
P(k) \ \sim \ \frac{\left({k\, \eta_0}\right)^{3}\, \exp\left[{{\frac {\pi \, \left( \frac{c}{2}-1
 \right)  \left( {\nu}^{2}-\frac{1}{4} \right) }{\sqrt {\left({k\, \eta_0}\right)^{2}+ (c-1)({\nu}^{2}-\frac{1}{4})}}}}\right]}{ \left| \Gamma\left( \nu+\frac{1}{2}+{\frac {i \left( \frac{c}{2}-1 \right)
 \left( {\nu}^{2}-\frac{1}{4} \right) }{\sqrt {\left({k\, \eta_0}\right)^{2}+ (c-1)({\nu}^{2}-\frac{1}{4})}}} \right)\right|^2\, \big[\left({k\, \eta_0}\right)^{2}+ \left( c-1 \right)  \left( {\nu}^{2}-\frac{1}{4} \right) \big]^{\nu}} \ , \label{power_coulomb}
\ee
and contain as a special case, for $c=2$, the deformed power spectra of eq.~\eqref{cibmukh_BB}. However, modifying $c$ one can also affect the growth rate, lowering it, enhancing it and even introducing an overshoot with respect to eq.~\eqref{cibmukh}, as in fig.~\ref{fig:Coulomb}.
Moreover, as described in \cite{dkps}, an additional type of imprint can be associated to local departures of $W_s$ from its attractor shape. First--order perturbation theory \emph{\`a la} Schwinger--Keldysh implies that this superposes an oscillatory behavior, in general, to the preceding effects, but the Coulomb--like potentials deviate from the attractor $W_s$ in an infinite domain, so that these oscillations are somehow washed out in eq.~\eqref{power_coulomb}. Oscillations of this type were clearly seen to accompany the transition from fast roll to slow roll in more conventional inflationary potentials in \cite{destri}, and are also responsible for part of the behavior displayed in \cite{dkps}.

Summarizing, an initial singularity thus translates, almost verbatim, into two interesting types of imprints:
\begin{itemize}
\item[a. ] an \emph{inevitable suppression of the power spectrum for low frequencies with respect to the attractor form}, which is not necessarily ${\cal O}(k^3)$ but can be milder, as in the cases that were the focus of \cite{dkps}, or more pronounced, as in the cases analyzed in \cite{destri};
\item[b. ] a \emph{possible overshoot}, which can present itself when the actual $W_s$ happens to emerge from the horizontal axis more steeply than the attractor curve. We saw clear signs of the overshoot in the tensor spectra analyzed in \cite{dkps}.
\end{itemize}
\begin{figure}[h]
\begin{center}$
\begin{array}{ccc}
\epsfig{file=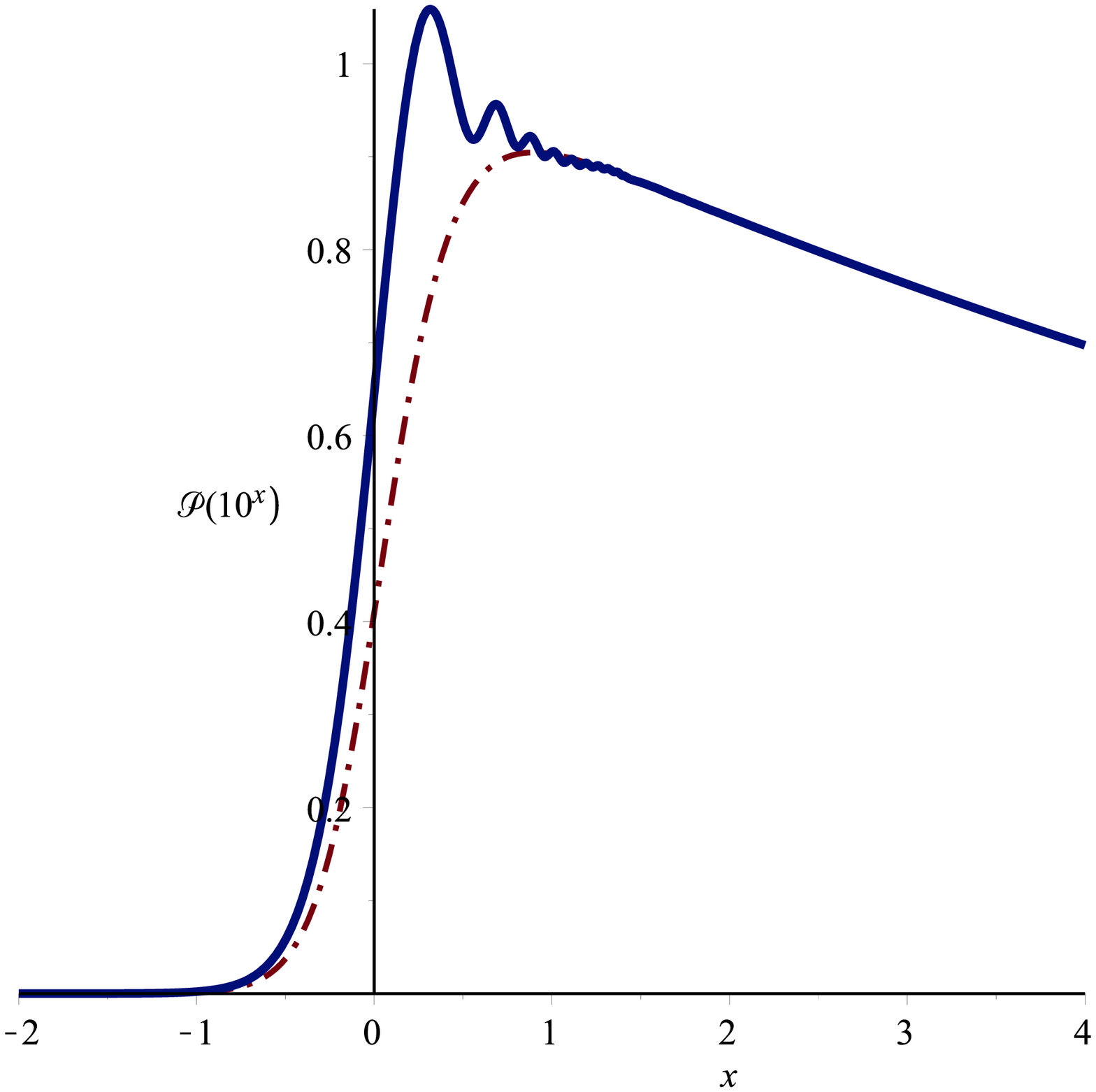, height=1.5in, width=1.5in} & \qquad\qquad
\epsfig{file=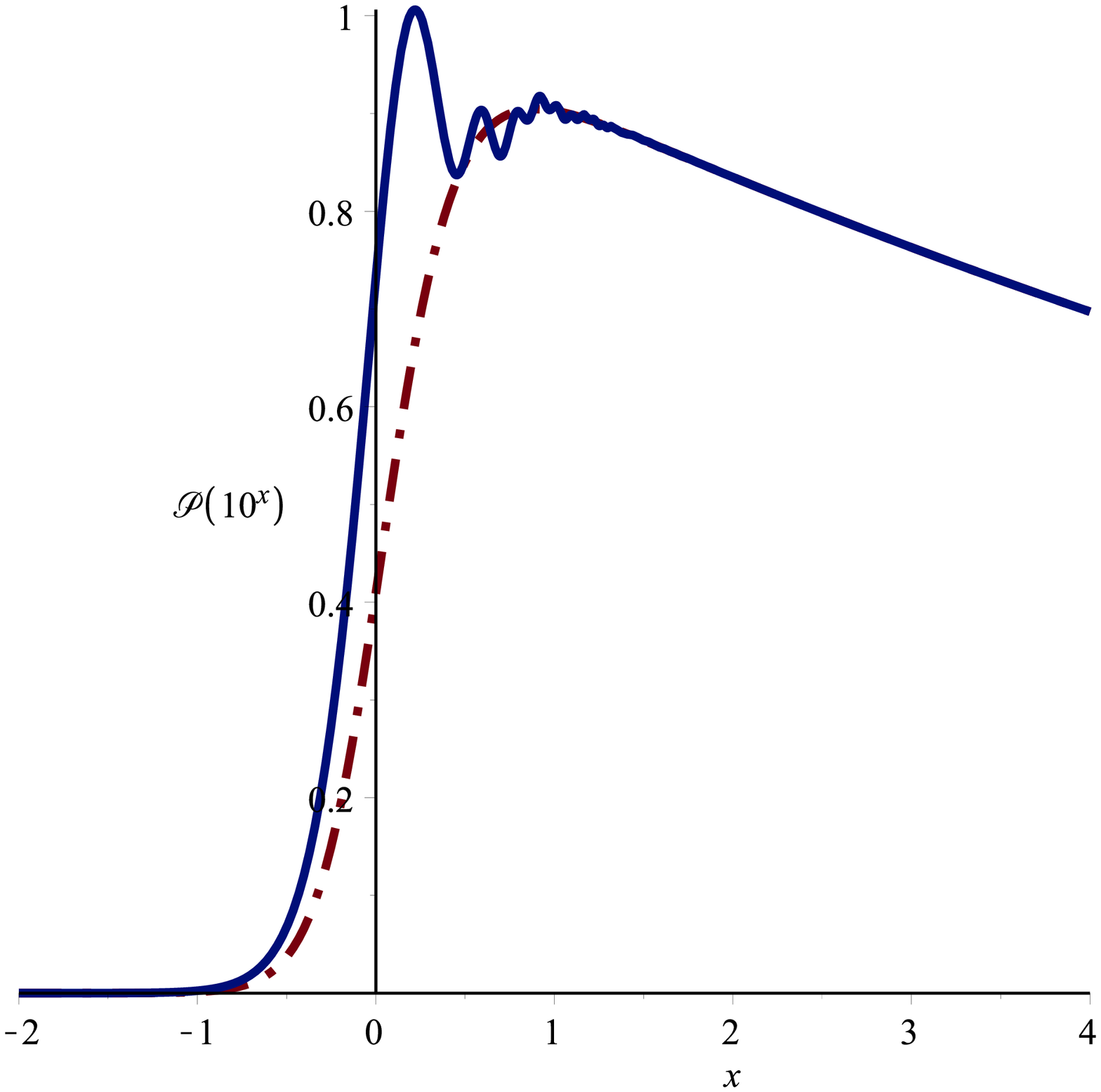, height=1.5in, width=1.5in} & \qquad\qquad
\epsfig{file=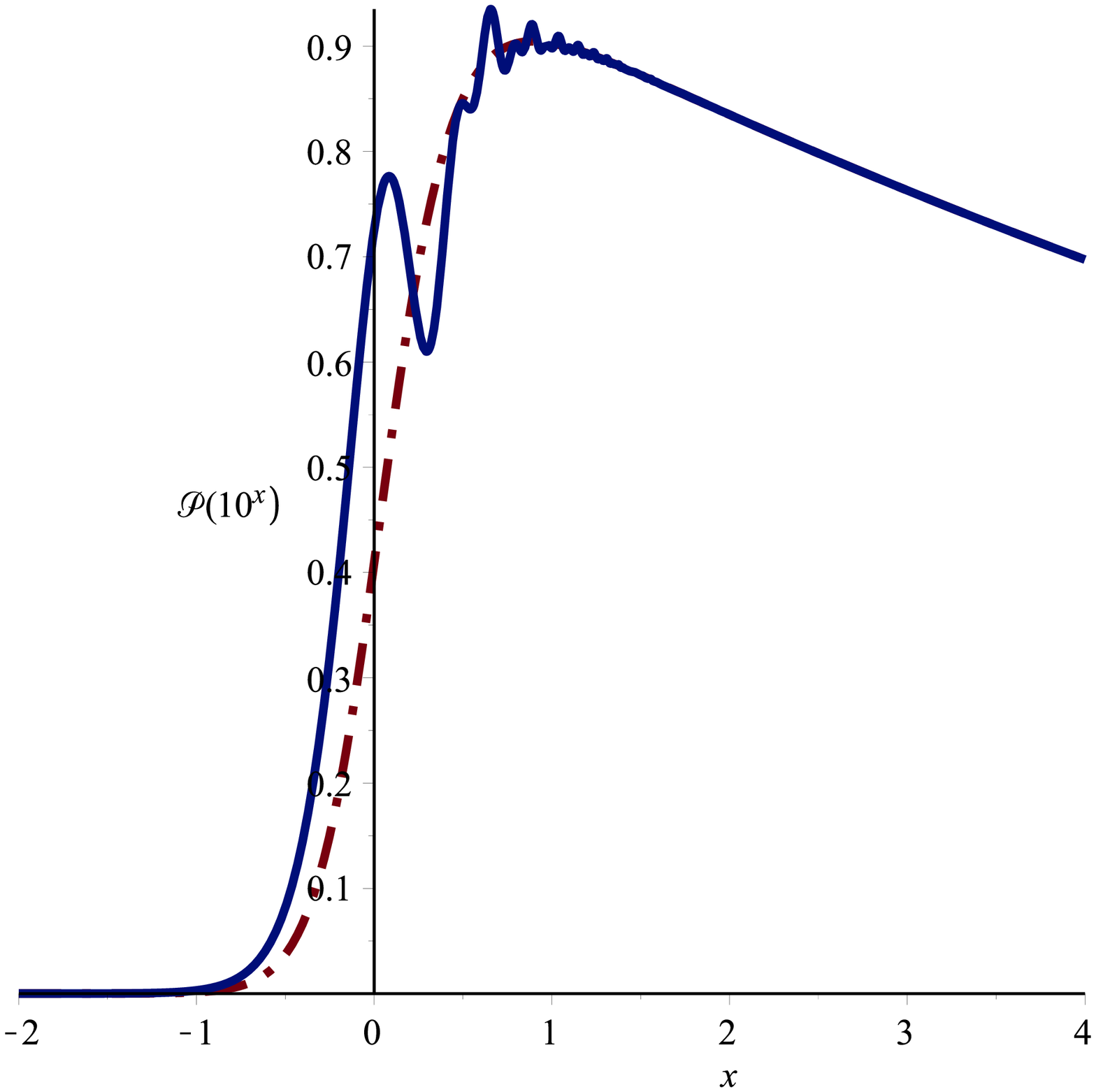, height=1.5in, width=1.5in}
\end{array}$
\end{center}
\caption{\small Oscillations induced by square--well perturbations of a Coulomb--like $W_s$.  These plots have an arbitrary normalization and correspond to perturbations acting in intervals of the same length but centered around decreasing values of $\eta$, and on the horizontal axis we display $x$, where $k=10^x$.}
\label{fig:local_perturbations}
\end{figure}
Moreover:
\begin{itemize}
\item[c. ] localized perturbations of the MS potential translate into \emph{localized oscillations} in $k$--space, as those that were seen in \cite{destri} to accompany transitions from fast roll to slow roll.
\end{itemize}
The emergence of this type of behavior in the presence of a climbing scalar was already discussed in \cite{dkps} \footnote{The relevant example of square--well perturbation displayed in eq.~(3.20) contains however a typo, since the overall $k$ should be replaced with an overall $k^{-2}$.} and is also illustrated qualitatively in fig.~\ref{fig:local_perturbations} with reference to square--well perturbations, but here we can be rely on more accurate numerical calculations, while drawing also a better comparison with similar phenomena that were discussed in the recent literature.
\begin{figure}[h]
\begin{center}$
\begin{array}{cc}
\epsfig{file=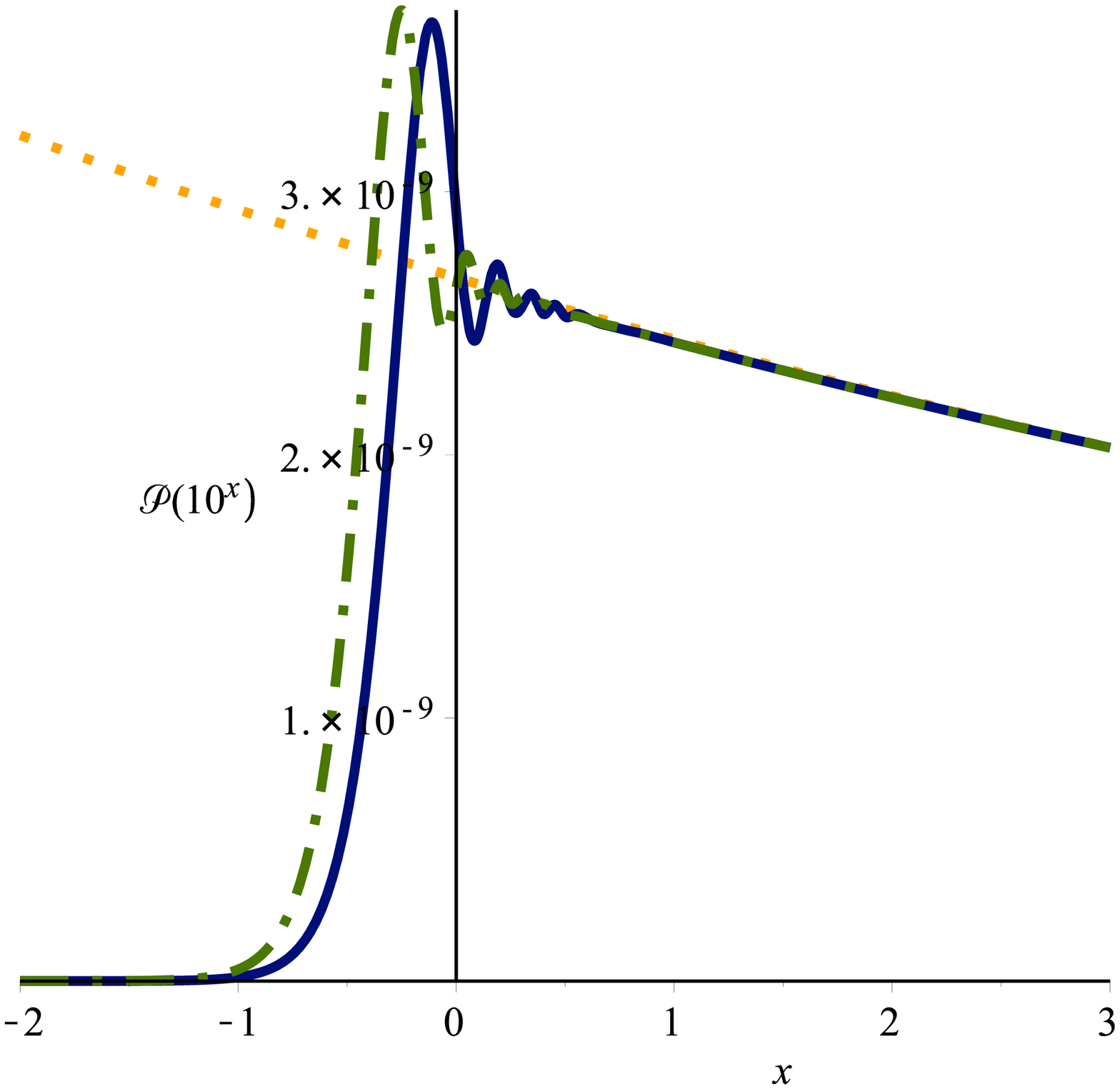, height=1.5in, width=1.5in} & \qquad\qquad
\epsfig{file=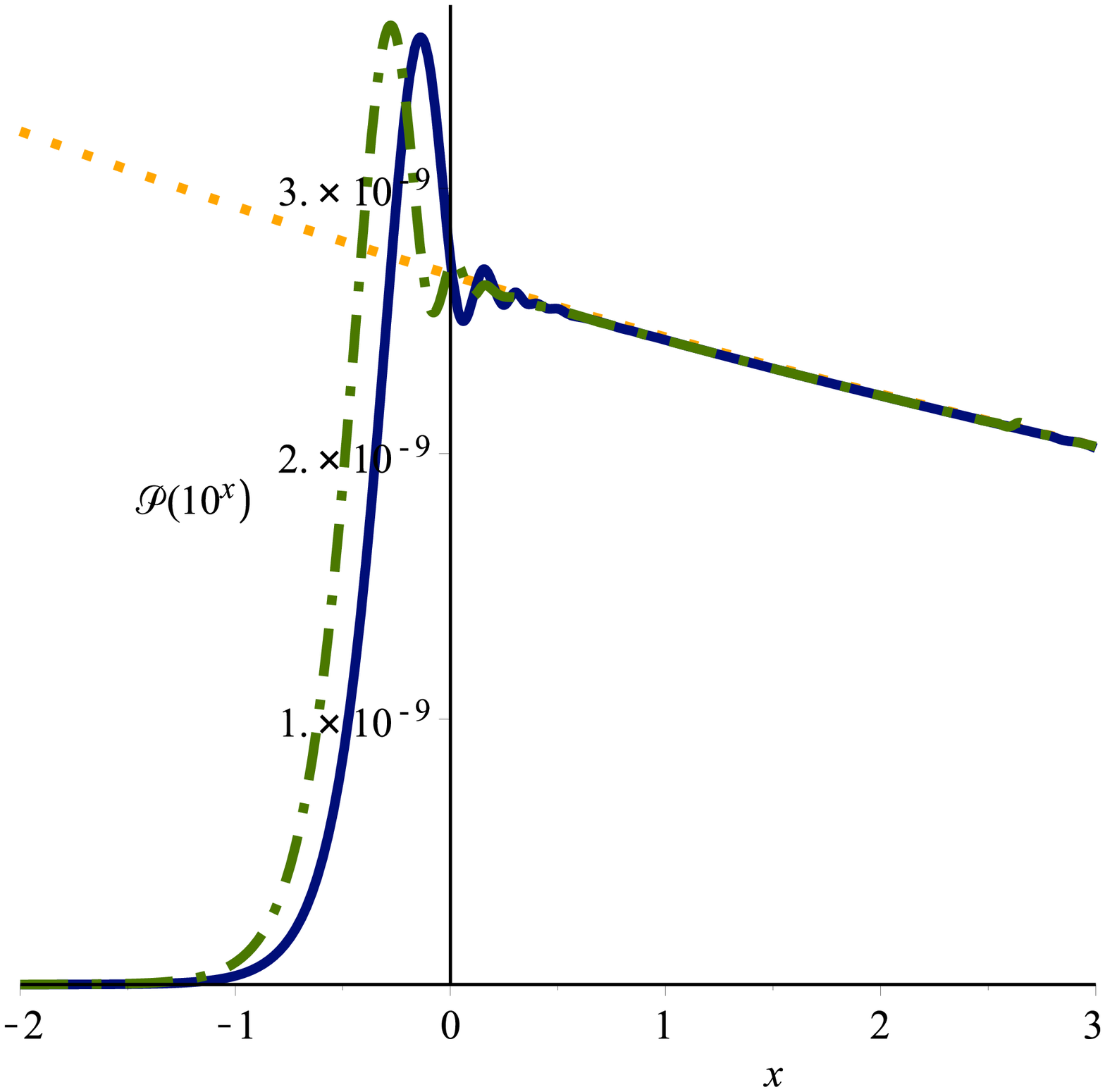, height=1.5in, width=1.5in}
\end{array}$
\end{center}
\caption{\small Power spectra of scalar perturbations for the single--exponential potential of eq.~\eqref{potonexp}, for a \emph{descending} scalar, with $\varphi_0=0$ (left, continuous, red) and $\varphi_0=-4$ (left, dashed, blue) and for a \emph{climbing} scalar, with $\varphi_0=0$ (right, continuous, red) and $\varphi_0=-4$ (right, dashed, blue). The two backgrounds leave very similar imprints, which are recurrent in transitions from fast roll to slow roll in other potentials and were discussed at length in \cite{destri}. In both cases, increasing $\varphi_0$ moves the transitions to slightly larger values of $k$. These plots are COBE normalized at $k=38$, and on the horizontal axis we display $x$, where $k=10^x$.}
\label{fig:single}
\end{figure}

The models typified by the two--exponential potentials of eq.~\eqref{potwoexp} and emerging somehow from BSB are the site, in general, of new type of phenomenon, \emph{incomplete transitions to slow-roll}. This feature was not clearly recognized in \cite{dkps}, although in retrospect it could be regarded as their own distinctive signature. Characterizing its effects was a main motivation for the present work, and to this end let us begin by stressing that, intuitively, the scalar has a tendency to climb up too fast the two--exponential potential since, as we have already stressed in Section \ref{sec:climbing}, as it emerges from the initial singularity it is largely driven by the ``mild'' exponential alone. As a result, the encounter with the ``hard'' exponential typically occurs somewhat abruptly and brings about a consequent tendency to bounce against it. Unless of course the parameter $\varphi_0$ is too large and negative for the scalar to ever feel the first term in eq.~\eqref{potwoexp}. In other words, for $\varphi_0$ sufficiently large and negative the two potentials of eqs.~\eqref{potonexp} and \eqref{potwoexp} should lead to power spectra that are essentially identical.
\begin{figure}[h]
\begin{center}$
\begin{array}{ccc}
\epsfig{file=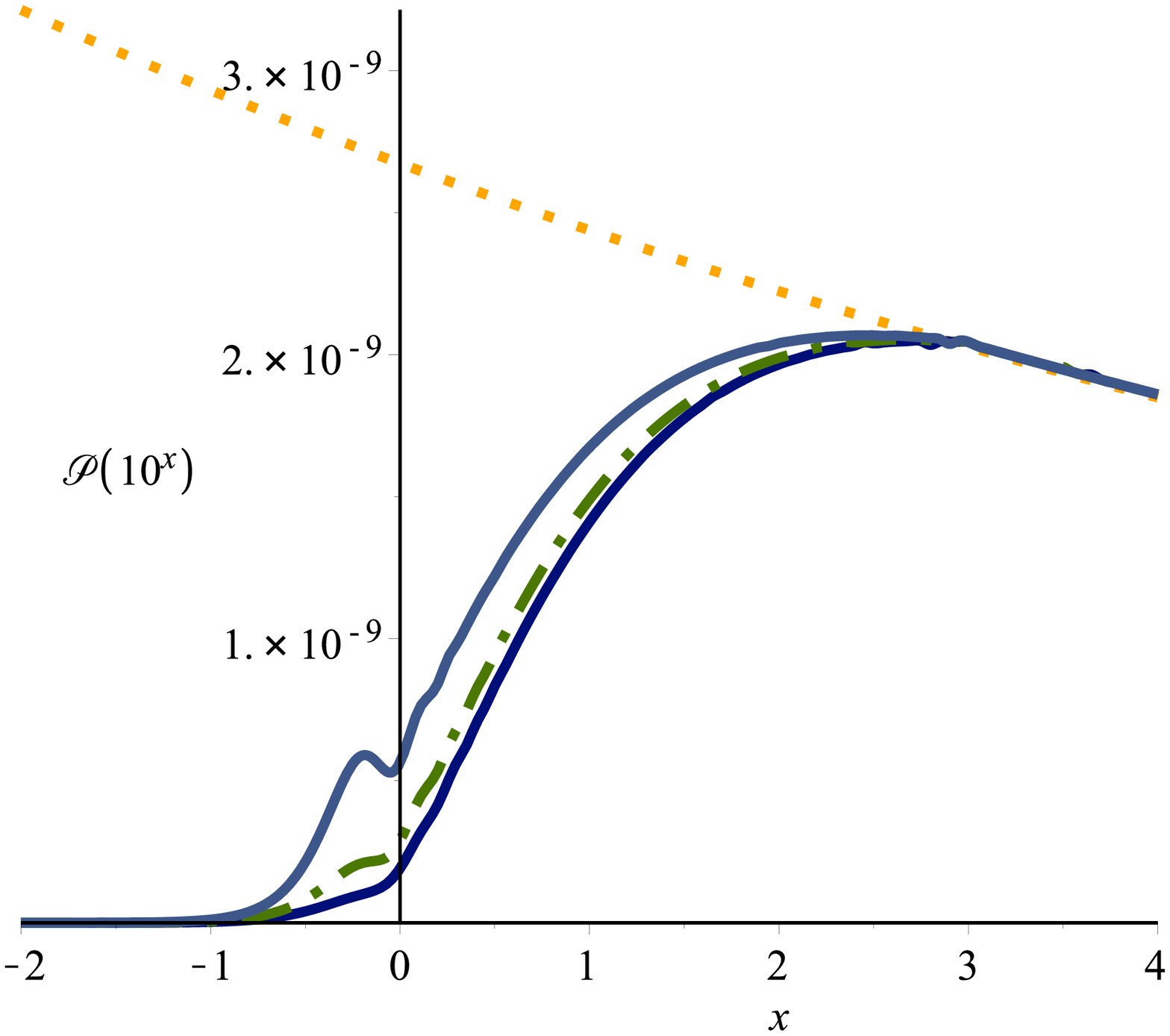, height=1.5in, width=1.5in} & \qquad\qquad
\epsfig{file=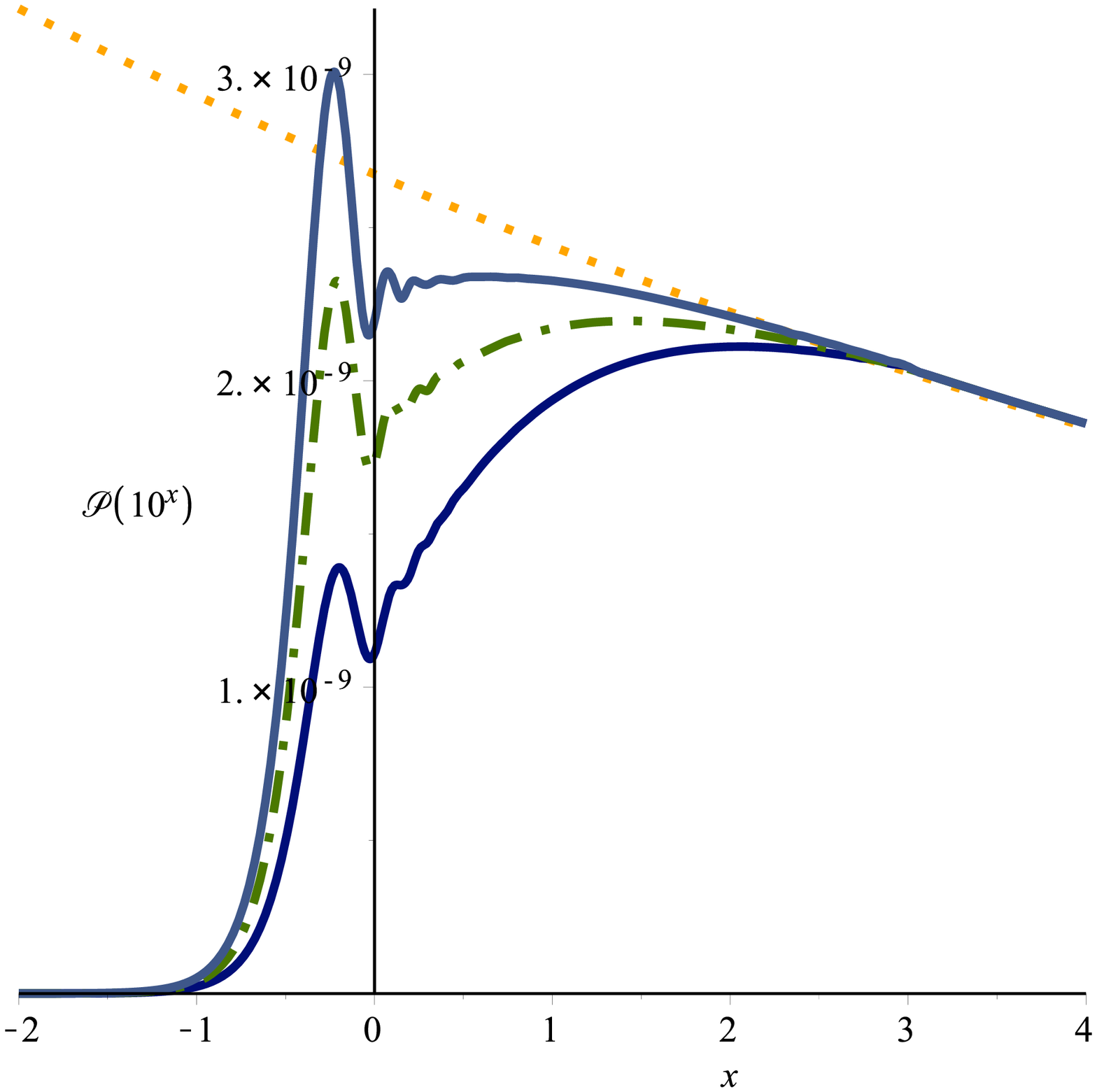, height=1.5in, width=1.5in} & \qquad\qquad
\epsfig{file=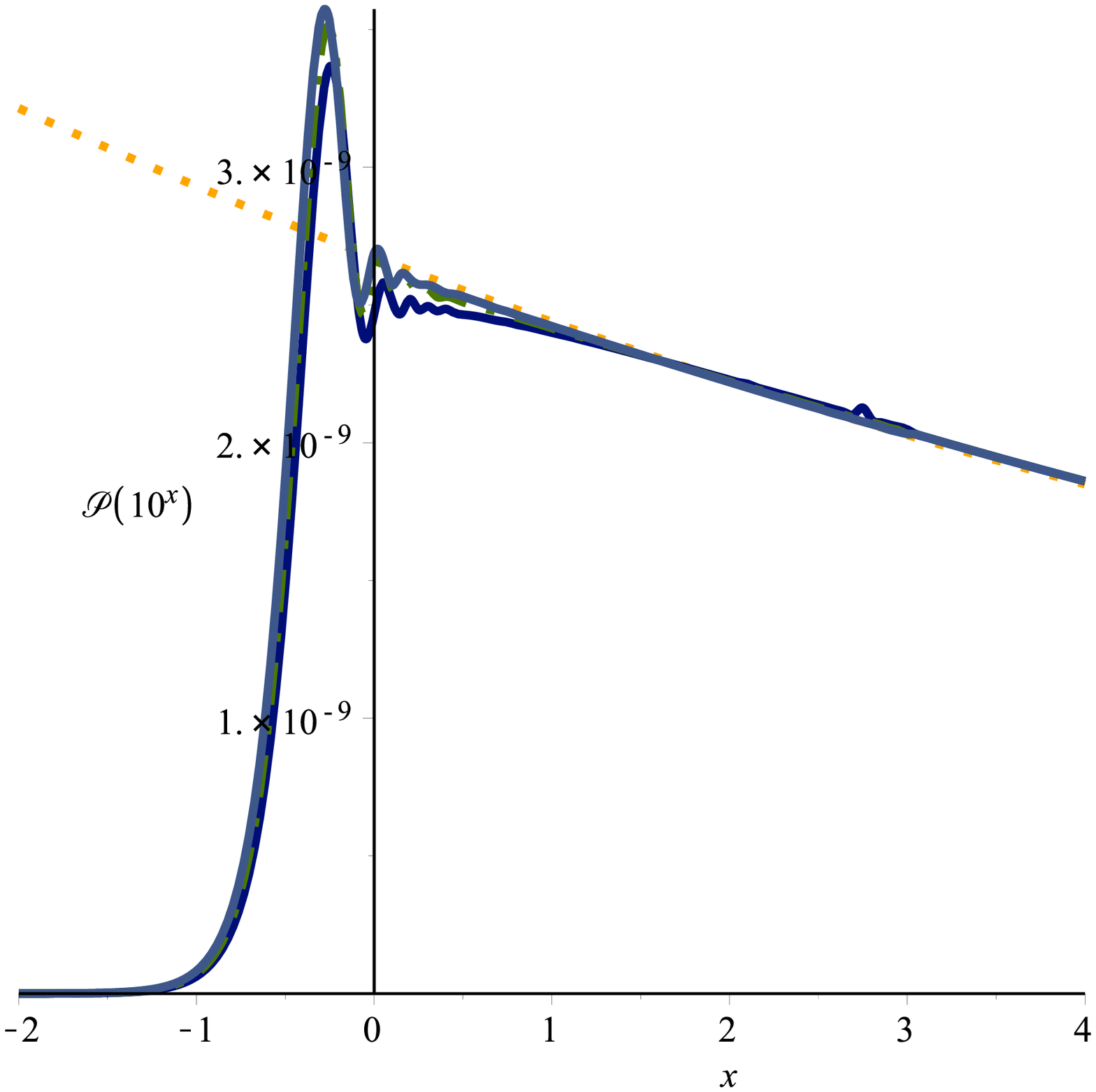, height=1.5in, width=1.5in}
\end{array}$
\end{center}
\caption{\small Power spectra of scalar perturbations for the double--exponential potential of eq.~\eqref{potwoexp} (in all cases the dotted line is the attractor curve). For $\varphi_0=0$ a featureless power spectrum approaches the attractor curve after four decades in $k$ (left), while for $\varphi_0=-0.5,-1$ a small pre--inflationary peak starts to build up (left). For $\varphi_0=-1.5,-2,-2.5$ the pre--inflationary peak becomes more and more pronounced, but remains well separated from the attractor curve (center). For $\varphi_0=-3,-3.5,-4$ the power spectrum is essentially the same as in fig.~\ref{fig:single}: it rises steeply, and a narrow peak overshoots the attractor curve that is readily reached after a few oscillations (right). In all cases on the horizontal axis we display $x$, where $k=10^x$, and these power spectra are normalized, in an arbitrary but convenient fashion, so that they all meet at the end of the explored range.}
\label{fig:double}
\end{figure}

The left portion of fig.~\ref{fig:single} displays some typical power spectra for a \emph{descending} scalar in the one--exponential potential of eq.~\eqref{potonexp}. Repeating the exercise for a \emph{climbing} scalar entails, surprisingly, some numerical subtleties, but the reader will not fail to recognize that the end result, in the right portion of fig.~\ref{fig:single}, is almost identical and is again almost independent of $\varphi_0$. Moreover, these spectra are also strikingly similar to those found in \cite{destri}. This universality is very interesting: their structure typifies conventional transitions from fast roll to slow roll, and emerges again, nicely enough, for $\varphi_0$ sufficiently large and negative in the double--exponential potentials, as shown in the last portion of fig.~\ref{fig:double}.

Turning to the two--exponential potential of eq.~\eqref{potwoexp}, the opposite limit of a scalar impinging on a ``hard wall'' formed the core of the numerical analysis in \cite{dkps}, but the neater results displayed there actually correspond to the upper limit of the perturbative regime. Indeed, for sufficiently large values of $\varphi_0$ the scalar experiences a hard bounce and attains a slow--roll regime far later than in the single--exponential potential, so that the power spectrum is widely depressed with respect to the attractor curve over several decades in $k$--space, as can be seen again in fig.~\ref{fig:double}.
\begin{figure}[h]
\begin{center}
\epsfig{file=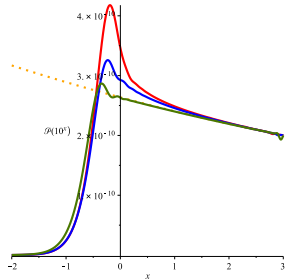, height=1.5in, width=1.5in}
\end{center}
\caption{\small Power spectra of tensor perturbations for the double--exponential potential of eq.~\eqref{potwoexp} for $\varphi_0=0$ (red), $\varphi_0=-1$ (blue) and $\varphi_0=-4$ (green). The only feature is the overshoot that was already discussed in \cite{dkps}, which is only prominent for $\varphi_0=0$ but becomes rapidly less pronounced as $\varphi_0$ is reduced. The dotted line is the attractor curve and the normalization is fixed conventionally in order to grant a tensor--to--scalar ratio $r \simeq 0.1$ at the highest scale explored. As above, $k=10^x$.}
\label{fig:double_tensor}
\end{figure}

The intermediate values of $\varphi_0$ in fig.~\ref{fig:double} are most interesting, since they give rise to the new phenomenon that we are addressing here. Indeed, in reverting its motion, the scalar undergoes inevitably a short period of slow--roll, giving rise to a spurt of almost exponential expansion for the Universe. A period of this type is always present, but it merges into the slow--roll descent for $\varphi_0$ large and negative while it is too short to leave any tangible signs for sufficiently large values of $\varphi_0$. On the other hand, in the intermediate region the reversal does leave a sign although, after reverting its motion, the scalar returns to a fast--roll regime for a while before finally slowing down. This type of dynamics leaves a striking signature: for intermediate values of $\varphi_0$ the power spectra of the two--exponential model display a \emph{pre-inflationary peak} of variable size, which superposes to the slowly growing spectrum present for large values of $\varphi_0$. The peak is well separated in the vertical scale from the more standard feature signaling the onset of the eventual slow--roll phase and yet both occur essentially for the same range of frequencies, albeit for different values of $\varphi_0$. No similar phenomena show up in power spectra of tensor perturbations, which reflect the evolution of the scale factor alone, as can be seen in fig.~\ref{fig:double_tensor}.

One can be slightly more quantitative, since on general grounds the link between wave numbers of perturbations exiting the horizon and cosmological dynamics rests on the correspondence
\be
k \ \sim \ {{e^{\,\frac{\cal A}{3}} \ H} \over {k_N \ \sqrt{3V_0}}} \ = \ \dot{\cal A} \, e^{\,\frac{\cal A}{3}} \, \frac{\sqrt{V(\varphi)/V_0}}{3} \ .
\ee
In all cases displayed in fig.~\ref{fig:double} these numbers are very close to -0.3, in the log--scale of the plots, for the end of the climbing phase, where $\dot{\cal A}=1$. And indeed all pre--inflationary peaks lie in that region in $k$--space, which confirms the link to the reversal of the scalar motion that we have advocated.
\begin{figure}[h]
\begin{center}$
\begin{array}{ccc}
\epsfig{file=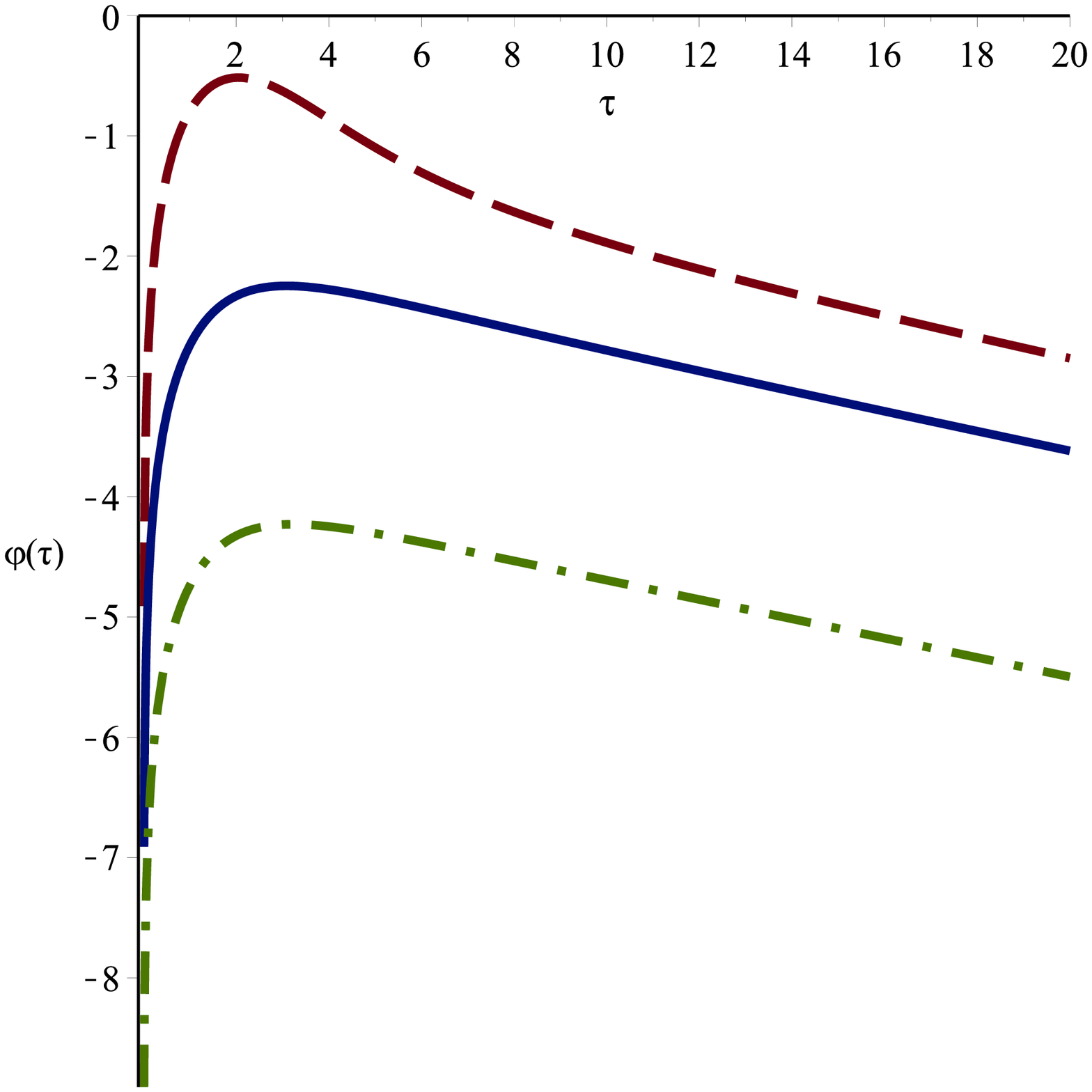, height=1.5in, width=1.5in} & \qquad\qquad
\epsfig{file=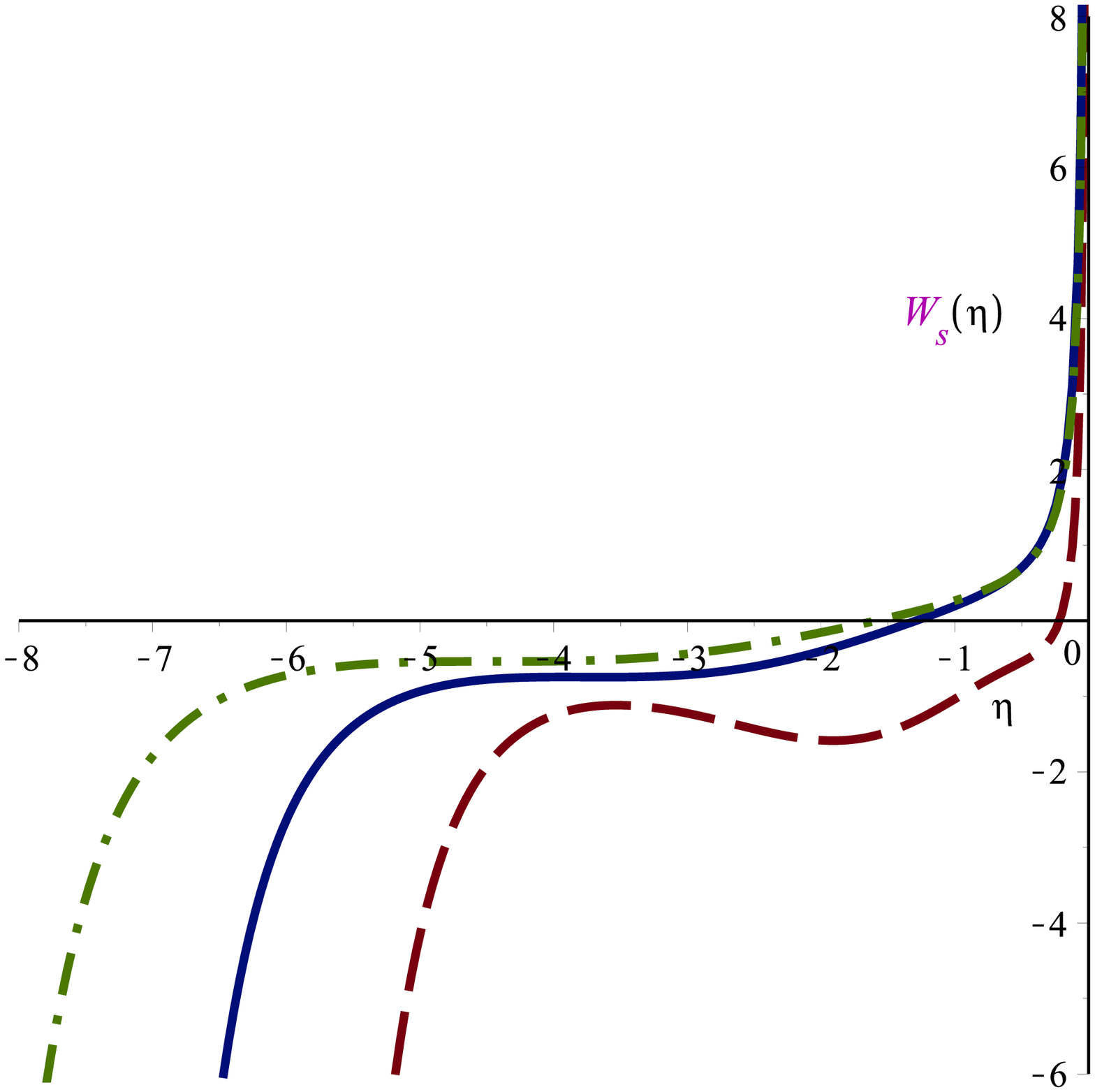, height=1.5in, width=1.5in} & \qquad\qquad
\epsfig{file=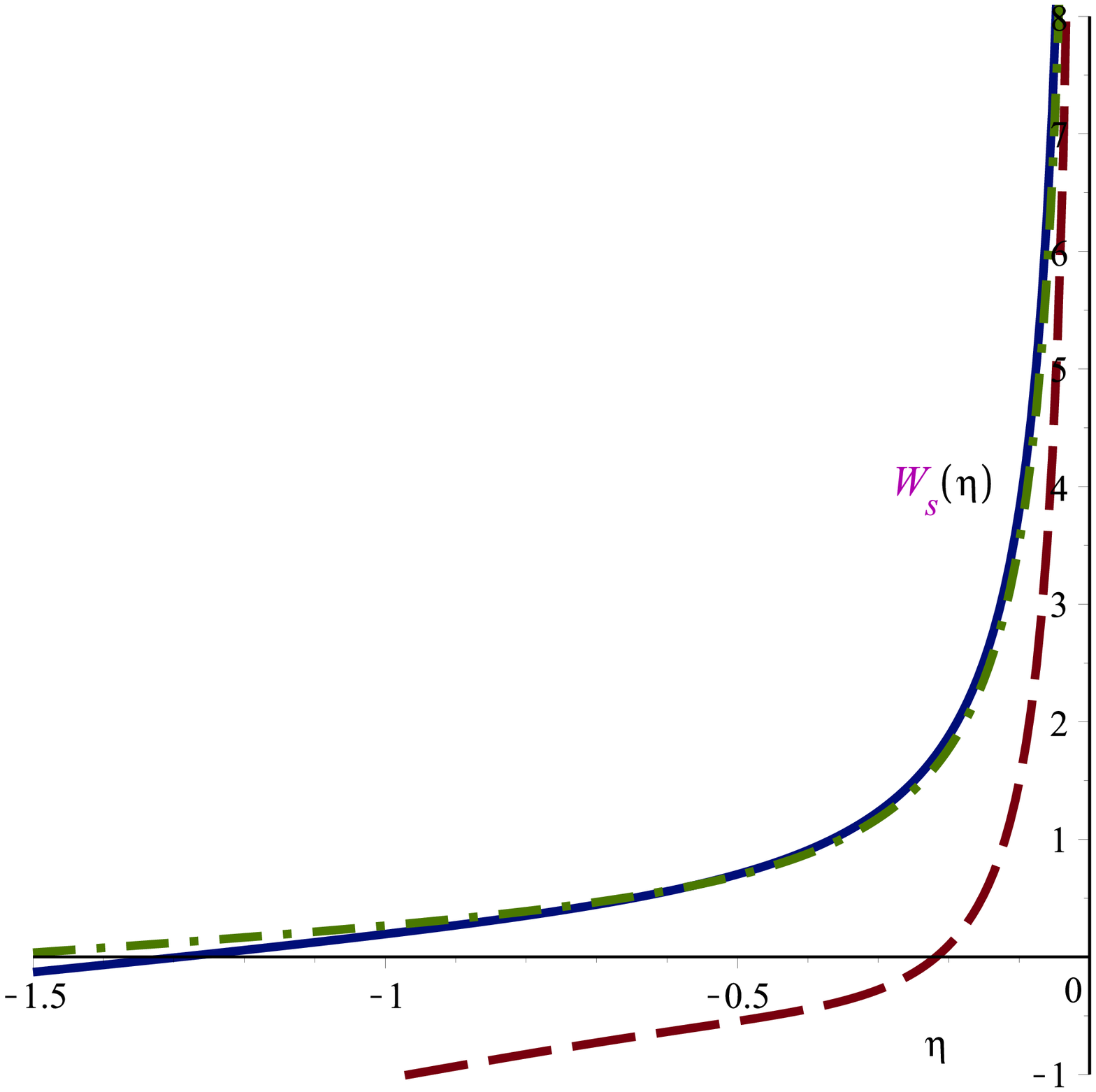, height=1.5in, width=1.5in}
\end{array}$
\end{center}
\caption{\small $\tau$--evolution of the scalar field near the turning point (left), the corresponding MS potentials $W_s$ in conformal time $\eta$ (center) and an enlarged view of the region where the $W_s$ curves cross the horizontal axis (right). The lines for $\varphi_0=0,-2,-4$ are dashed, continuous and dashed--dotted.}
\label{fig:phi_Ws}
\end{figure}

\begin{figure}[h]
\begin{center}$
\begin{array}{ccc}
\epsfig{file=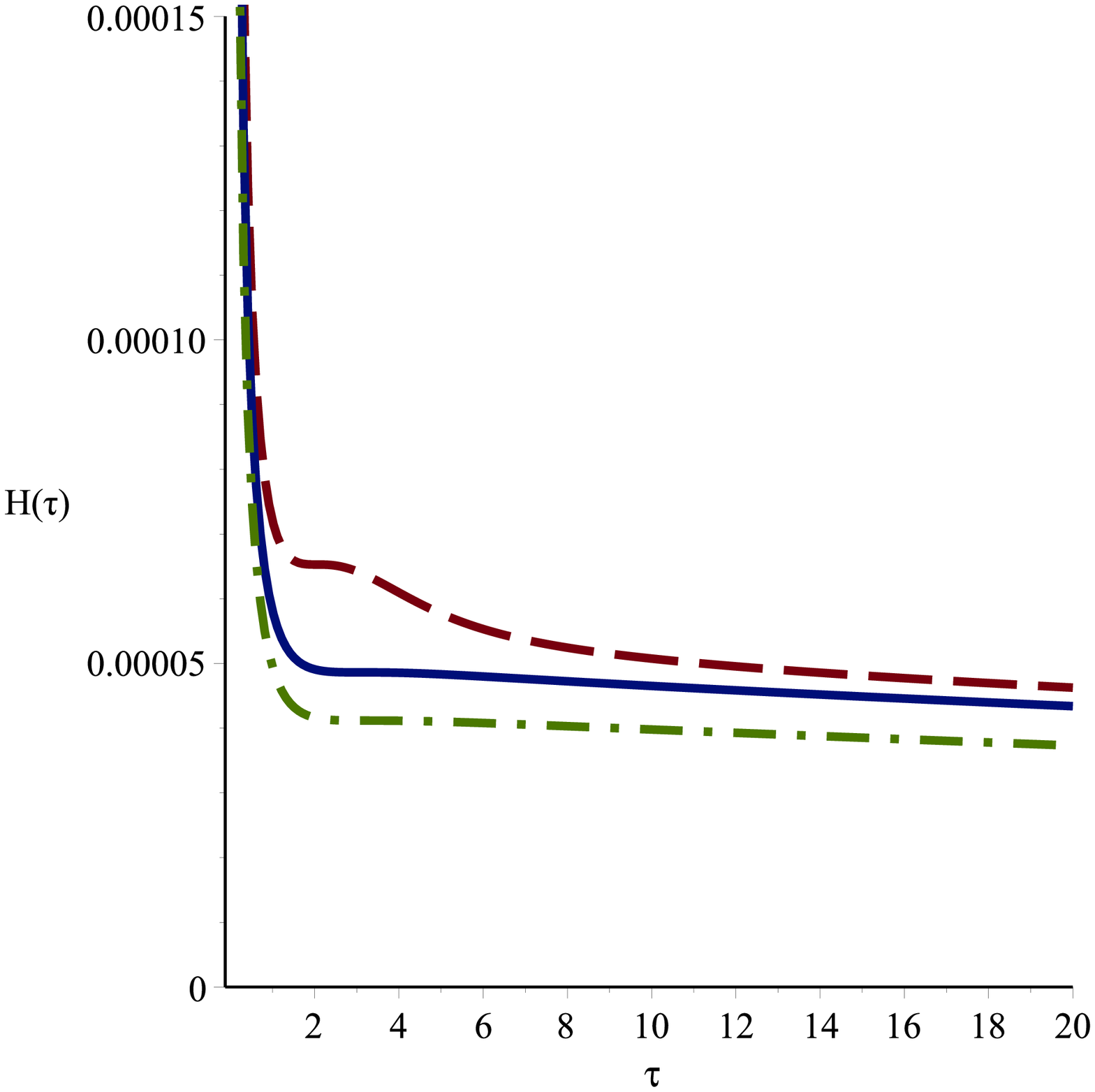, height=1.5in, width=1.5in} & \qquad\qquad
\epsfig{file=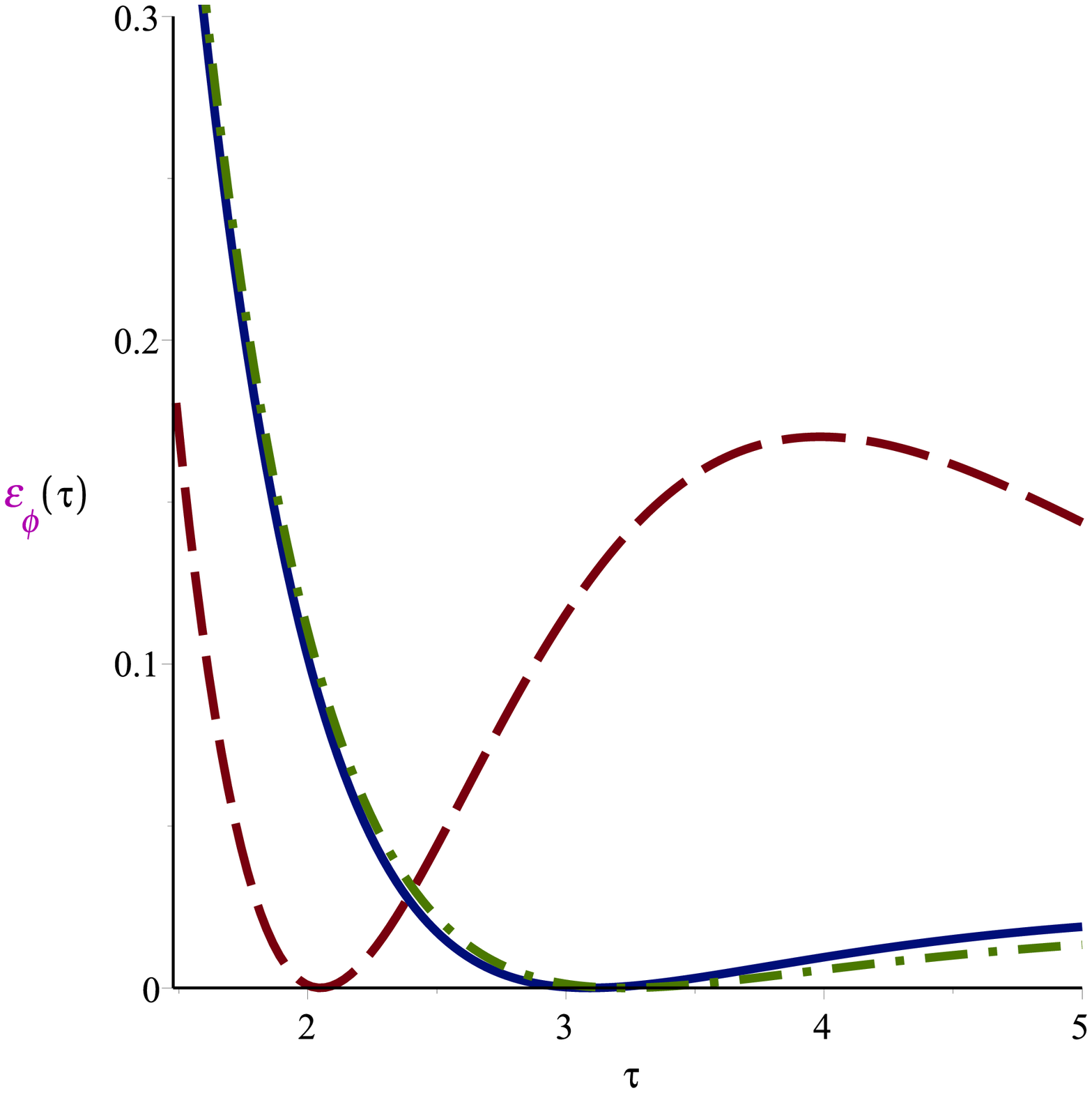, height=1.5in, width=1.5in} & \qquad\qquad
\epsfig{file=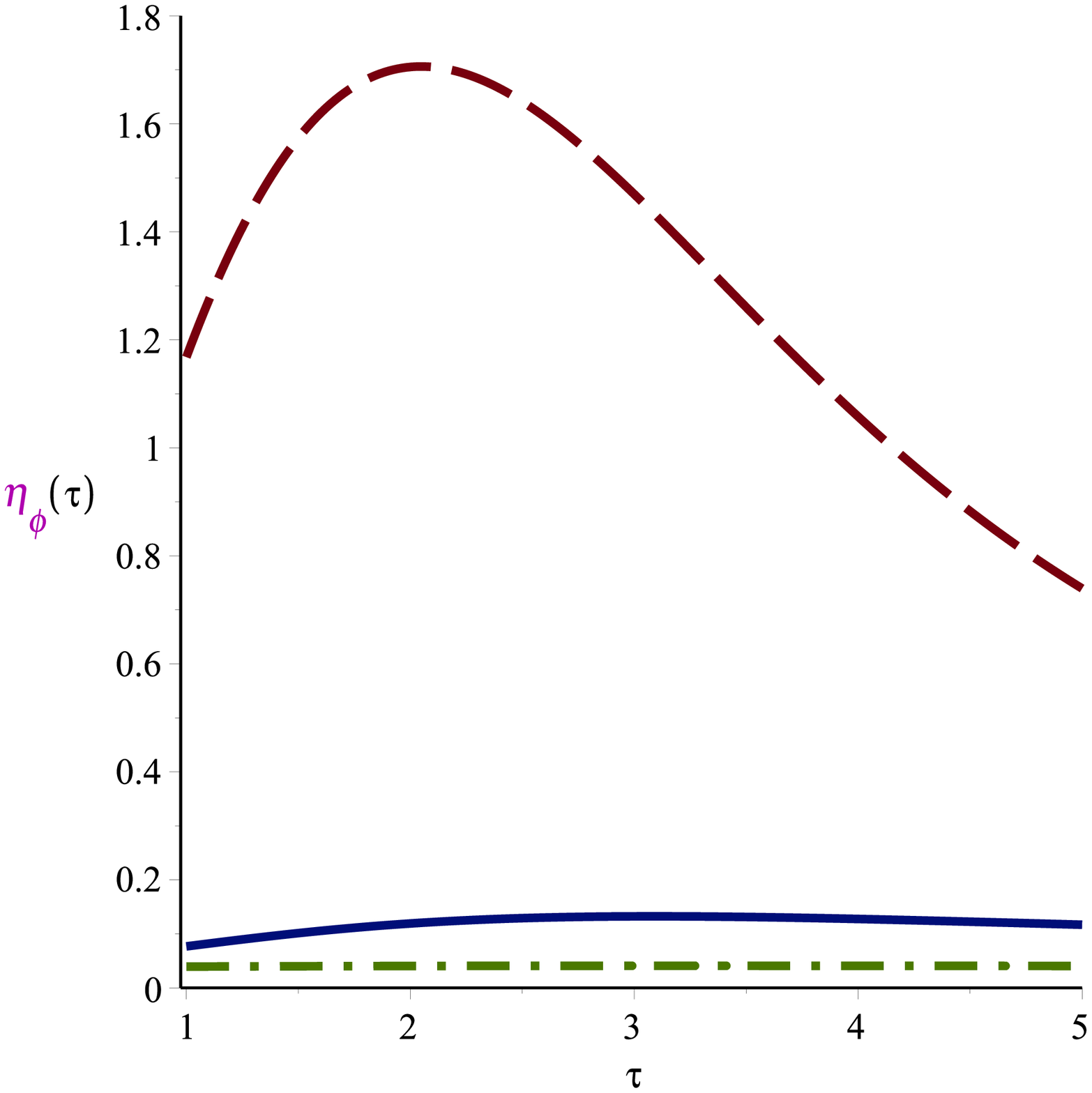, height=1.5in, width=1.5in}
\end{array}$
\end{center}
\caption{\small $\tau$--evolution of $k_N \, H$, or $H$ in reduced Planck units (left), and the corresponding $\tau$--evolutions of $\epsilon_\phi$ (center) and $\eta_\phi$ (right).  The lines for $\varphi_0=0,-2,-4$ are dashed, continuous and dashed--dotted.}
\label{fig:H_eps_eta}
\end{figure}

Some additional details can perhaps allow a better grasp of these dynamical effects.
A closer look at various dynamical quantities in the three significant cases with $\varphi_0=-4$ (dashed--dotted curves), $\varphi_0=-2$ (continuous curves) and $\varphi_0=0$ (dashed curves) in figs.~\ref{fig:phi_Ws} and \ref{fig:H_eps_eta} can provide a clearer picture of the phenomenon that is taking place in two--exponential systems. The $\tau$--evolutions of $\varphi$ collected in the left portion of fig.~\ref{fig:phi_Ws} exhibit clearly the sharp change of behavior experienced by the scalar: as it comes closer to the ``hard'' exponential the reversal of its motion becomes more abrupt, so that it is nearly a reflection for $\varphi_0=0$. On the other hand, there are no appreciable differences between the curves for $\varphi_0=-4$ and $\varphi_0=-2$, and yet as we have seen their spectra are qualitatively rather different. The differences between these two cases become more evident, however, in the rest of fig.~\ref{fig:phi_Ws}, which collects the corresponding $W_s(\eta)$. To begin with, the $W_s$ curve for $\varphi_0=0$ lies well to the right of the others, consistently with a wide reduction of the area below it and hence with the wide WKB suppression of power that, as we have seen, occurs in this case. Moreover, one can also see that the scalar reverts its motion well within the region where $W_s$ is negative, so that for $\varphi_0=0$ the phenomenon can leave no tangible signs in the power spectrum, consistently with fig.~\ref{fig:double}. On the other hand, in the other two cases the inversion occurs in regions where $W_s$ is positive, consistently with the presence of tangible imprints in their power spectra.

The differences among the three cases can be scrutinized in further detail from another vantage point, referring to three quantities whose definitions were recalled of Section \ref{sec:climbing}: the Hubble parameter $H$ and the $\epsilon_\phi$ and $\eta_\phi$ parameters. The corresponding curves for $\varphi_0=-4$ (dashed--dotted), for $\varphi_0=-2$ (continuous) and for $\varphi_0=0$ (dashed) are collected in fig.~\ref{fig:H_eps_eta}. In all cases, the dashed line lies well apart from the others, which are relatively close and yet exhibit appreciable differences. Even for $\varphi_0=0$ $H$ contains a noticeable horizontal portion that signals a quasi--exponential expansion around the inversion point, but the corresponding $\epsilon_\phi$ grows rapidly away from it, consistently with the picture of a fast bounce, while finally the corresponding $\eta_\varphi$ is also far larger than in the other cases, consistently with the lack of a sizable growth of perturbations accompanying the reversal of the scalar in this case. The comparison between the $\epsilon_\phi$ curves for $\varphi_0=-2$ and $\varphi_0=-4$ is also very instructive. The former lies in fact first below and then above the latter, since the reversal is more efficient close to the ``wall'', but this makes the intermediate case initially closer to slow roll than the other and then, after the reversal, farther from it, an information that is almost tantamount to drawing the pre--inflationary peak in fig.~\ref{fig:double}. For this case, however, a larger $\eta_\varphi$ makes it comparatively harder for the perturbations to grow. These details are all consistent with the fact that a pre--inflationary peak is not visible for $\varphi=0$, is visible for $\varphi_0=-2$ and is large for $\varphi_0=-4$.

Before coming to our first comparisons with the CMB power spectrum, let us pause to summarize some technical details of our computations. To begin with, the natural option to compute power spectra would seem to rest on the Fourier modes of the variable $\zeta$, whose differential equation in conformal time,
\be
\frac{d^2 \zeta_k}{d \eta^2} \ + \ 2 \ \frac{z^\prime}{z} \ \frac{d \zeta_k}{d\eta} + k^2 \, \zeta_k \ = 0 \ ,
\ee
also takes a nice form. Moreover, $\zeta_k$ is to approach a constant for large $\tau_F$, or equivalently for small negative $\eta$, which could be used as a strong test of the numerics. As we have already stressed, however, this variable develops a pole at the inversion point for the climbing scalar, where $z$ vanishes \cite{kodama}. This makes $\zeta_k$ inconvenient in numerical integrations, and therefore both here and in \cite{dkps} we have actually resorted to a different variable,
\be
Q_k \ = \ e^{\,-\, \frac{{\cal A}}{3}}\ {v_k} \ = \ e^{\,-\, \frac{{\cal A}}{3}}\ z \ {\zeta_k} \ ,
\ee
which combines the virtues of both $\zeta_k$ and $v_k$: it has no pole at the inversion point for the climbing scalar and yet it also attains ($k$--dependent) limiting values for large $\tau$, after several $e$--folds of inflation.

Working in terms of the parametric time $\tau$, which is particularly convenient for the two--exponential system as we have seen, we were led to the differential equation
\be
\ddot{Q}_k \, + \, \left( \dot{\cal A} \, + \, \frac{V_\varphi}{2\, V} \, \dot{\varphi} \right) \dot{Q}_k \, + \,  \left( \frac{k^2\,e^{\,-\, \frac{2}{3}\, {\cal A}}}{V/V_0}  \, + \, \frac{V_{\varphi\varphi}}{2\, V} \, + \,  \frac{V_\varphi}{2\, V} \ \frac{4 \dot{\varphi}}{\sqrt{1 \, + \, \dot{\varphi}^2}} \, + \, \frac{2\, \dot{\varphi}^2}{1 \, + \, \dot{\varphi}^2} \right) Q_k \, = \, 0 \ . \label{Qeq}
\ee
As in the preceding section, here ``dots'' denote $\tau$--derivatives. The reader should appreciate that the various terms in eq.~\eqref{Qeq} are indeed manifestly free of singularities at the inversion point for the scalar field $\varphi$.

The power spectra were computed in this fashion with Maple programs, exercising special care both with the numerical integration and with the choice of initial conditions. As a result, the present analysis is more precise than that reported in \cite{dkps}, where wide oscillations were present that, in retrospect, reflect in part an imprecise translation of the fixed initial conditions of Section \ref{sec:climbing} into cosmic time. The improved precision was essential to unveil the pre--inflationary peak, and to this end special care was exercised to deal with the oscillatory nature of the complex solutions of the Mukhanov--Sasaki equation. We actually found out, by trial and error, that the numerics is typically more stable if, rather than working with the complex differential equation for $Q_k$, one combines the linear \emph{second--order} differential equation \eqref{Qeq} for its real part $Q_{R,\,k}$ with the \emph{first--order} Wronskian condition
\be
Q_{I,\,k} \ \frac{dQ_{R,\,k}}{d\tau} \  - \ Q_{R,\,k} \ \frac{dQ_{I,\,k}}{d\tau} \  = \ \frac{1}{2\,\sqrt{V(\varphi)/V_0}} \ e^{\,-\, {\cal A}(\tau)}
\ee
to determine its imaginary part $Q_{I,k}$. All in all, we have gathered the impression that Maple handles numerical instabilities in a clear fashion, so that we have had all the way a good control on which spectra required more sophisticated methods for their determination. Most of the results were obtained with a high--order Runge--Kutta method working with large numbers of digits. As we have stressed, they satisfy the nice consistency condition of converging, for sufficiently large negative $\varphi_0$, to the single--exponential power spectra of fig.~\ref{fig:single}, which are also along the lines of \cite{destri}. In this fashion we reached relatively handily the border of present observations, which lies around $k=10^3$. All these results were obtained, as in \cite{dkps}, starting from a Bunch--Davies--like vacuum, setting initial conditions close to the singularity (at $\tau=0.01$, in terms of parametric time) as we have explained and working in terms of the parametric time $\tau$ in the gauge of eq.~\eqref{gauge}. Our results thus rest on the choice of an initial Bunch--Davies--like vacuum: as we saw in detail in \cite{dkps}, moving away from it blurs the reduction of power at low frequencies.
\vskip 24pt

\section{\sc  A First Look at the CMB}\label{sec:observables}
\vskip 12pt

We can now see how the power spectra that we have identified, and in particular their pre--inflationary peaks, translate into some features of angular power spectra that do not seem foreign to what WMAP9 \cite{wmap9} and PLANCK \cite{PLANCK} are observing. The potential relevance of pre--inflationary scenarios for the CMB rests, of course, on the key assumption that our Universe gives us access, via the seven or so observable $e$--folds, to the features present in the power spectra of Section \ref{sec:powerspectrum}. Or, if you will, that the CMB gives us access somehow to the onset of inflation. However, a back--of--the--envelope computation shows that this is not implausible, insofar as inflation did not last too long (not more than 60--70 $e$--folds, in a crude scenario), and given this assumption features present in the primordial power spectrum do translate into corresponding features of the angular power spectrum, on account of the neat relation discussed in \cite{mukhanov_slow}, which we shall present in the form
\be
A_\ell\;(\varphi_0,{\cal M},\delta) \ = \ {\cal M} \ \ell(\ell+1)\ \int_0^\infty \frac{dk}{k} \ {\cal P}_\zeta \big( k , \varphi_0 \big) \, {j_\ell}^2 \big( k \, 10^\delta \big) \, \label{bessel}
\ee
where we are emphasizing the dependence on $\varphi_0$, and is relatively accurate for $\ell \lesssim 35$.
Here $j_\ell$ denotes a spherical Bessel function, and nicely enough there is no dependence on complicated plasma effects after recombination. Actually, a further ``plus'' of eq.~\eqref{bessel} is that $j_\ell^{\,2}$ is sizably peaked when its argument is of order $\ell$, so that
\be
A_\ell \ \sim \ {\cal P}_\zeta \left( \ell \ 10^{-\delta} \right) \ . \label{Aelldelta}
\ee
In other words, the $A_\ell$ encode direct information on the primordial spectral function ${\cal P}_\zeta$, but this remarkable fact goes unfortunately on the par with a big ``minus'' of the whole setting. To begin with, very few independent data, $2\ell+1$ for each value of $\ell$, determine the first few multipoles and thus the large--scale structure of the CMB angular power spectrum, and this brings about correspondingly large error bars. In addition, we are observing the CMB from a very special vantage point, so that ``cosmic variance'' induces a properly conservative attitude, so much so that the sizable reduction of the CMB quadrupole is often signalled as a puzzle, but is not widely regarded, at present, as a critical problem for Cosmology. On the other hand, the discussion presented in Section \ref{sec:powerspectrum} has anticipated our idea that the reduced quadrupole might represent a natural shadow of an initial singularity, while the current estimates of cosmic variance might prove too conservative. For the time being, we shall concentrate on the low--$\ell$ tail of the CMB power spectrum and on eq.~\eqref{bessel}, but we plan to perform a more complete analysis in a future publication \cite{gnks}.

The reader should have noticed the two parameters that we have inserted in eq.~\eqref{bessel}. The first plays a more evident role: it is an overall normalization ${\cal M}$, which accounts for various constants entering the relation between the angular power spectrum and the primordial power spectrum of scalar perturbations and for the conversion to the proper units, $\mu K^2$, but ultimately reflects the scale of inflation. The second, the exponent $\delta$, is even more interesting. It can be moved, up a sign, into the argument of ${\cal P}_\zeta$, and controls the horizontal displacement of the features present in the power spectra of Section \ref{sec:powerspectrum} with respect to the main peaks of the Bessel functions. Alternatively, in more physical terms, $\delta$ allows a finer tuning between the largest wavelengths that are entering the cosmic horizon at the present epoch and those that were exiting it at the onset of the inflationary phase, which are assumed to be roughly identical in our setting.
\begin{figure}[h]
\begin{center}$
\begin{array}{ccc}
\epsfig{file=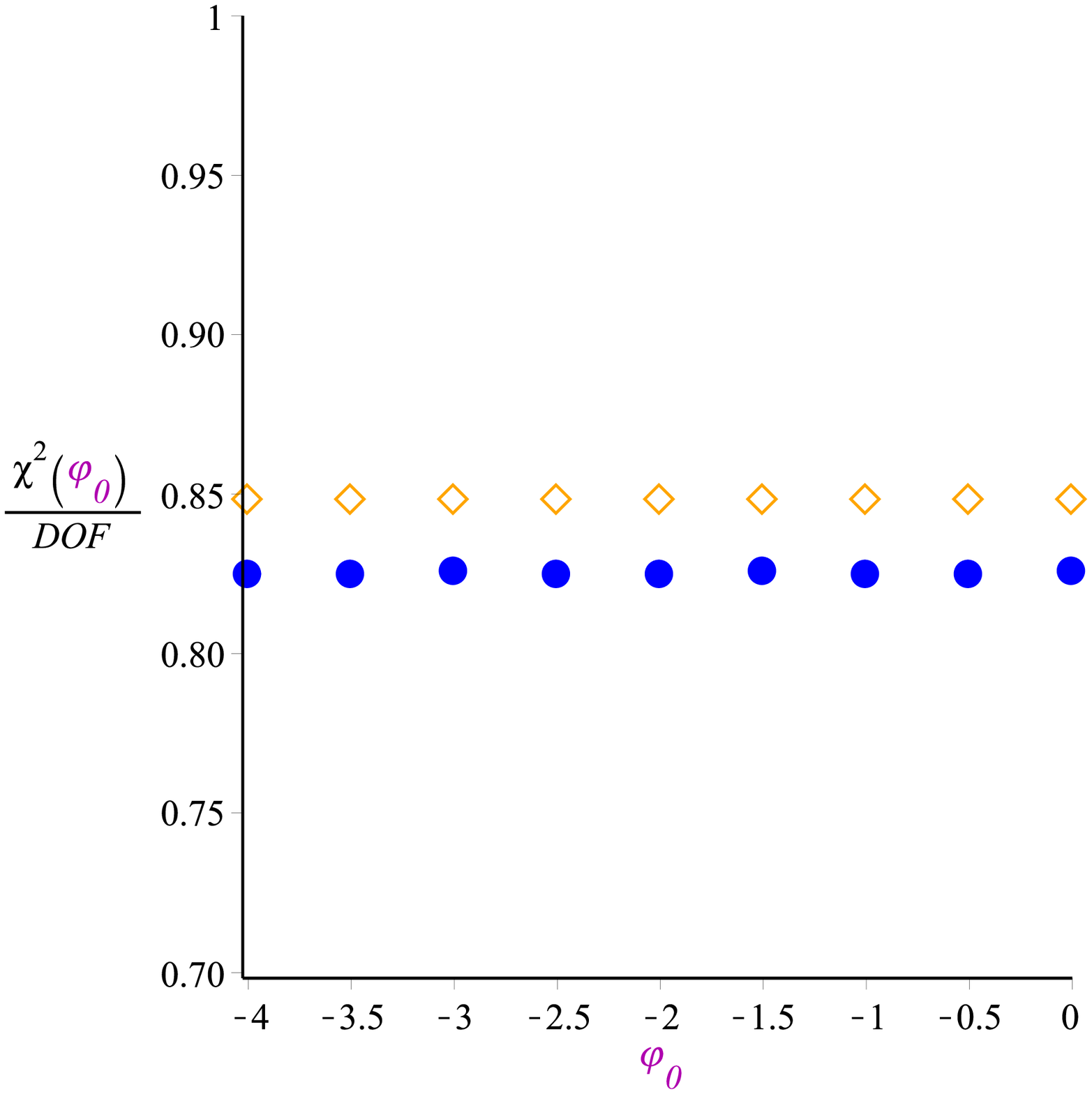, height=1.5in, width=1.5in}& \qquad\qquad
\epsfig{file=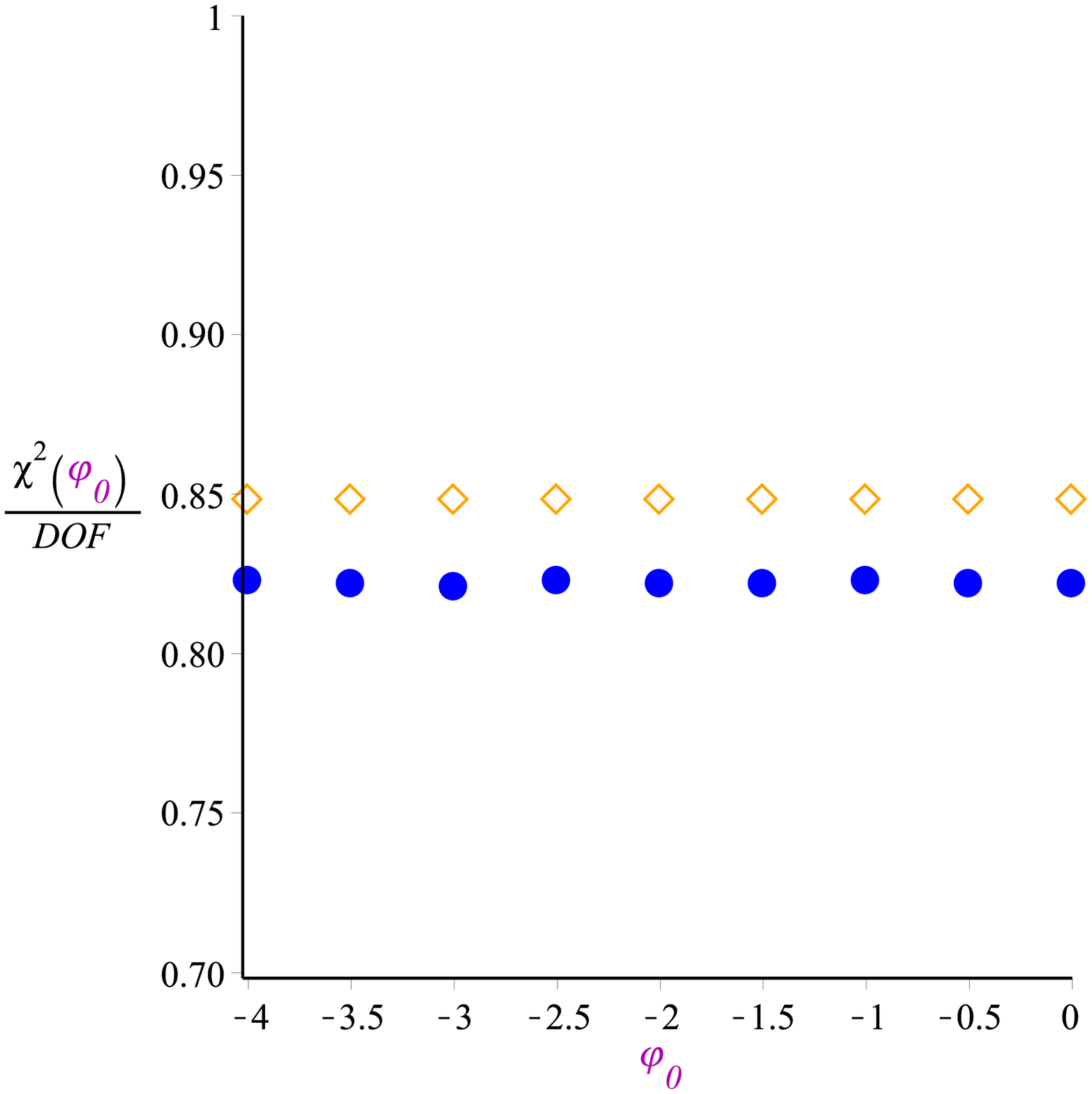, height=1.5in, width=1.5in}
\end{array}$
\end{center}
\caption{\small $\chi^2/DOF$ arising from comparisons between WMAP9 raw data and the angular power spectra predicted by the one--exponential potential of eq.~\eqref{potonexp} for a climbing scalar (left, blue dots), for a descending scalar (right, blue dots), and for the attractor (orange, diamonds).}
\label{fig:angular_fit_single}
\end{figure}

Let us begin our analysis from the attractor curve
\be
{\cal P}_\zeta \left( k \right) \ = \ k^{n_s-1} \ ,
\ee
noticing that in this case $\delta$ can be set to zero, since it plays \emph{no independent role} with respect to ${\cal M}$, and that the $A_\ell$ can be computed exactly, with the end result
\be
A_\ell^{\rm (attr)}({\cal M}) \ = \ {\cal M} \ \ \frac{\sqrt {\pi }\, \ell(\ell+1)}{4} \ {\frac { \Gamma  \left( \frac{3\,-\,{\it n_s}}{2} \right)
\Gamma  \left( \frac{2\,l\,-\,1\,+\,{\it n_s}}{2} \right) }{\Gamma  \left( \frac{4\,-\,{\it n_s}}{2} \right) \Gamma  \left( \frac{2\,l\,+\,5\,-\,{\it n_s}}{2} \right) }} \ . \label{bessel_attractor}
\ee
We can now compare this expression, computed for the preferred value of the spectral index $n_s \simeq 0.96$, with the first 31 raw WMAP9 data, adjusting the normalization ${\cal M}$ in such a way as to minimize
\be
\chi^2 \ = \ \ \sum_{\ell=2}^{32} \frac{\left(A_\ell\ - \ A_\ell^{\rm WMAP9}\right)^2}{\left(\Delta A_\ell^{\rm WMAP9}\right)^2} \ , \label{chi_squared}
\ee
where $A_\ell^{\rm WMAP9}$ are the WMAP9 central values and $\Delta A_\ell^{\rm WMAP9}$ are the corresponding errors.
In this fashion, after adjusting ${\cal M}$ one is effectively left with 30 independent degrees of freedom ($DOF$'s), and the end result for the least $\chi^2$ per $DOF$ is
\be
\frac{\chi^2_{\rm (attr,\, min)}}{DOF} \ = \ \frac{25.46095}{30} \ \simeq \ 0.849 \ . \label{chi2_attractor}
\ee

We can now proceed to the pre--inflationary models of the preceding sections, starting from the one--exponential climbing and descending systems. In the one--exponential systems, as we have seen, the power spectra depend little on $\varphi_0$, so that we contented ourselves with the nine values $\varphi_0=0,-0.5,-1,\ldots,-4$, first minimizing in all cases eq.~\eqref{chi_squared} analytically with respect to ${\cal M}$ and then exploring the result for more than 60 values of $\delta$ belonging to the interval $[-1.2,1.2]$, which encompasses all the features described in Section \ref{sec:powerspectrum}. Let us stress that the behavior of the spherical Bessel functions makes our knowledge of the primordial power spectra, which is limited to the range $10^{-2} < k < 10^3$, widely sufficient to obtain relatively accurate results in eq.~\eqref{bessel} for $\delta \in [-1.2,1.2]$, since we are only interested in values of $\ell \leq 32$. As expected, the lowest values of $\chi^2/DOF$ are essentially the same for the one--exponential climbing and descending cases, and are essentially independent of $\varphi_0$, as can be seen clearly in the point plots of fig.~\ref{fig:angular_fit_single}. Notice that both lie only slightly below the attractor points although they correspond to $\chi^2 \approx 24$, since in both cases one is left with $DOF=29$, after fixing both ${\cal M}$ and $\delta$, for any choice of $\varphi_0$.

For the two--exponential system of eq.~\eqref{potwoexp}, whose dependence on $\varphi_0$ is far richer, we performed a more detailed investigation as follows:
\begin{itemize}
\item[1. ] we explored a wider sequence of about 25 values including $\varphi_0=0,-0.25,-0.5,,\ldots,-3.75,-4$, in order to make the features of the transition region involving the pre--inflationary peak more transparent. Our initial choice for $\gamma$ was motivated by the naive correspondence between the mild exponential and the spectral index,
\be
n_s \ - \ 1 \ = \ 3 \ - \ 2\, \nu \ = \ - \ \frac{6\, \gamma^2}{1\,-\,3\,\gamma^2} \ .
\ee
This result would hold exactly for power--law inflation and gives $\gamma \simeq 0.08$ for $n_s \simeq 0.96$;
\item[2. ] we also repeated the analysis for the potentials of eq.~\eqref{potwoexp} with $\gamma=0.04$ and with $\gamma=0.02$, in order to take a first look at systems where an exponential ``hard wall'' accompanies \emph{concave} potentials like those in eq.~\eqref{starobinsky}. Further terms could complete in fact the two--exponential potentials for lower values of $\varphi_0$, turning them into concave functions capable of associating to later portions of the angular power spectrum the proper spectral index, as in fig.~\ref{fig:starobinsky}. This supplement of analysis may be regarded as a first attempt to take into account the well--known difficulty of power--law inflation with large tensor--to--scalar ratios, but as we anticipated we plan to subject the whole construction to more stringent statistical tests \cite{gnks} over a wider range of frequencies. With this proviso in mind, the $\chi^2$ analysis of the angular power spectra in figs.~\ref{fig:single} and \ref{fig:double}, which builds upon the expectation that the key features are well captured by the two--exponential models, represents a natural first step.
\end{itemize}

As in the one--exponential system, for any choice of $\varphi_0$ two parameters were determined in order to optimize the comparison with WMAP9 raw data:
\begin{itemize}
\item we minimized eq.~\eqref{chi_squared} analytically with respect to the normalization factor ${\cal M}$ present in eq.~\eqref{bessel};
\item we then identified optimal choices for the parameter $\delta$ in eq.~\eqref{bessel}, which allows a fine tuning between the largest wave numbers entering the horizon at the present epoch and those exiting it around the onset of the inflationary phase.
\end{itemize}
\begin{figure}[h]
\begin{center}$
\begin{array}{cc}
\epsfig{file=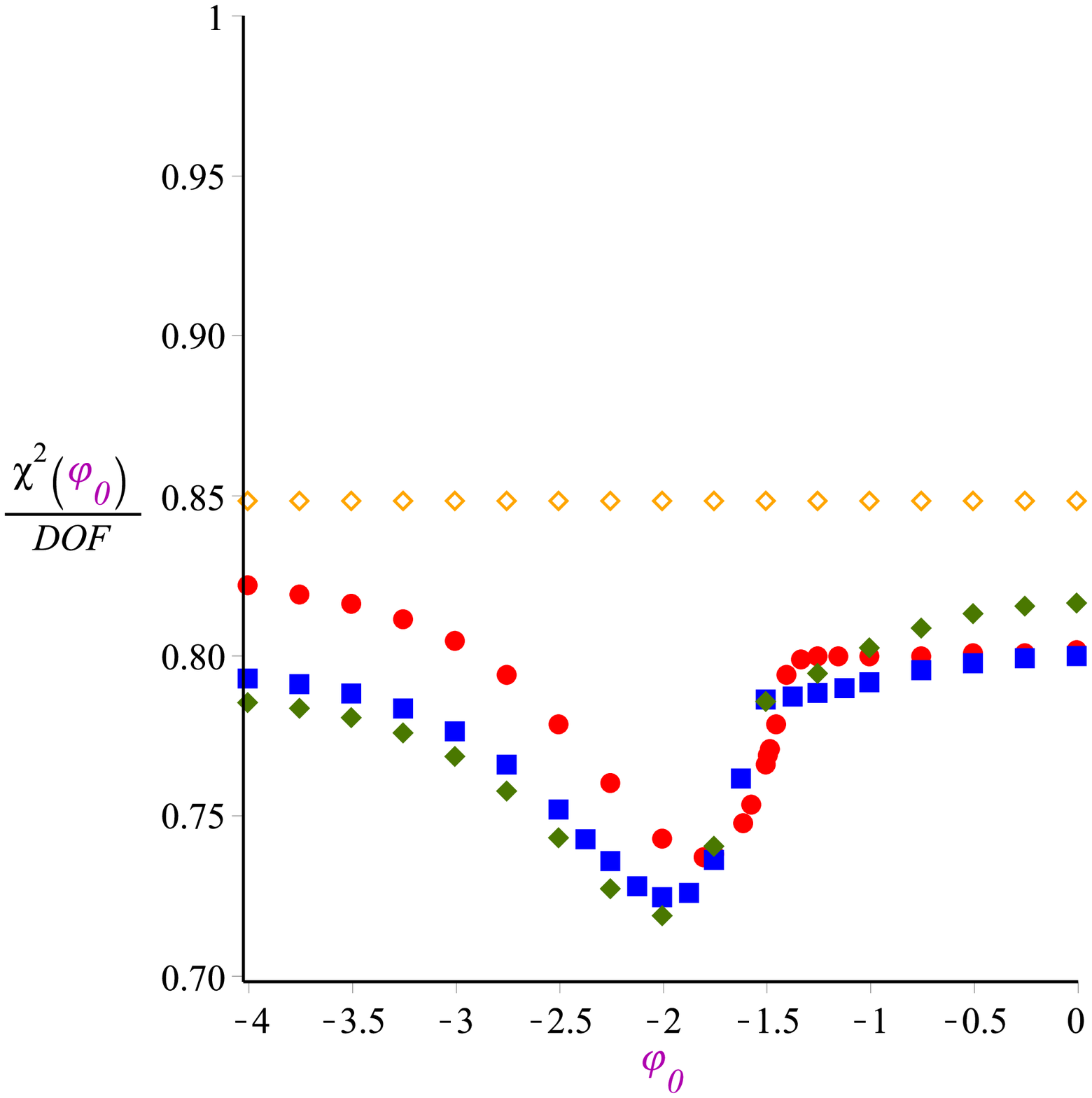, height=1.7in, width=1.7in}& \qquad\qquad
\epsfig{file=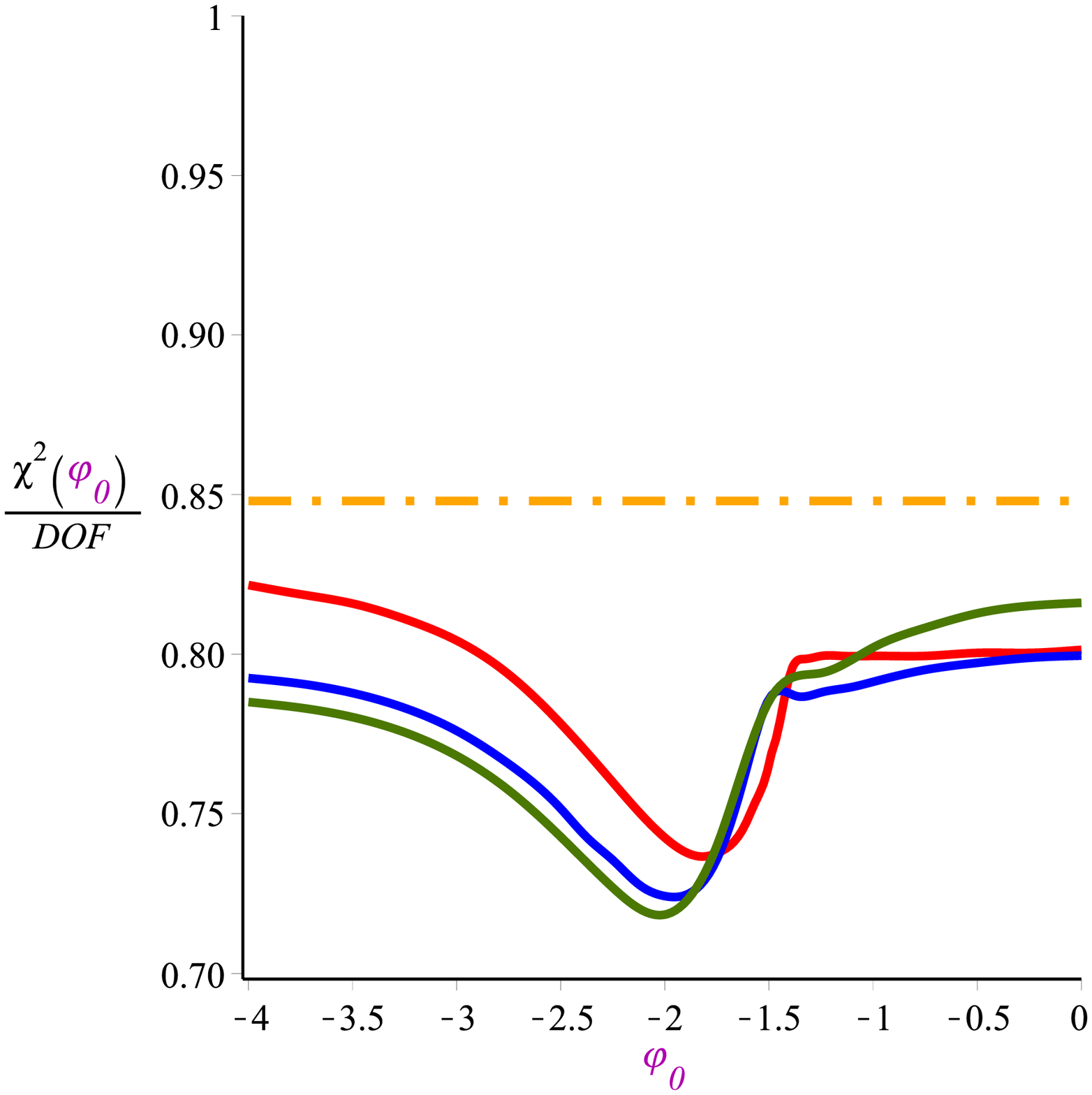, height=1.7in, width=1.7in}
\end{array}$
\end{center}
\caption{\small Comparisons between WMAP9 raw data and the angular power spectra predicted by the two--exponential potential of eq.~\eqref{potwoexp}, in point form (left) and in spline form (right), for $\gamma=0.08$ (red), $\gamma=0.04$ (blue) and $\gamma=0.02$ (green), and by the attractor curve (orange). The minima in the three cases are $\chi^2_{\rm min}/{DOF}=0.737$, $0.724$ and $0.718$.}
\label{fig:angular_fit_double}
\end{figure}
%

\begin{table}[h]
\begin{center}
\caption{Some values of $(\delta,\chi^2)$ in fig.~\ref{fig:angular_fit_double}}
\begin{tabular}{r c c c}
\hline
\hline
$\varphi_0$ & $\gamma=0.08$ & $\gamma=0.04$ & $\gamma=0.02$\\
\hline
\hline
\\
0.0 & $(-0.32,23.233)$ &  $(-0.75,23.187)$ & $(-0.90,23.667)$ \\
-0.5 & $(-0.21,23.203)$ & $(-0.70,23.121)$ & $(-0.85,23.571)$ \\
-1.0 & $(\ \ 0.00,23.191)$ &  $(-0.60,22.949)$ & $(-0.75,23.261)$ \\
-1.5 & $(+1.04,22.212)$ &  $(-0.22,22.793)$ & $(-0.50,22.775)$ \\
-2.0 & $(+1.03,21.537)$ &  $(+1.00,21.040)$ & $(+0.96,20.834)$ \\
-2.5 & $(+1.04,22.574)$ &  $(+1.00,21.796)$ & $(+0.96,21.540)$ \\
-3.0 & $(+1.05,23.326)$ &  $(+1.00,22.504)$ & $(+0.96,22.277)$ \\
-3.5 & $(+1.06,23.651)$ &  $(+1.00,22.847)$ & $(+0.96,22.628)$ \\
-4.0 & $(+1.08,23.821)$ &  $(+1.00,22.983)$ & $(+0.96,22.765)$ \\
\\
\hline
\hline
\end{tabular}
\label{table:norm_delta}
\end{center}
\end{table}
\begin{figure}[h]
\begin{center}$
\begin{array}{ccc}
\epsfig{file=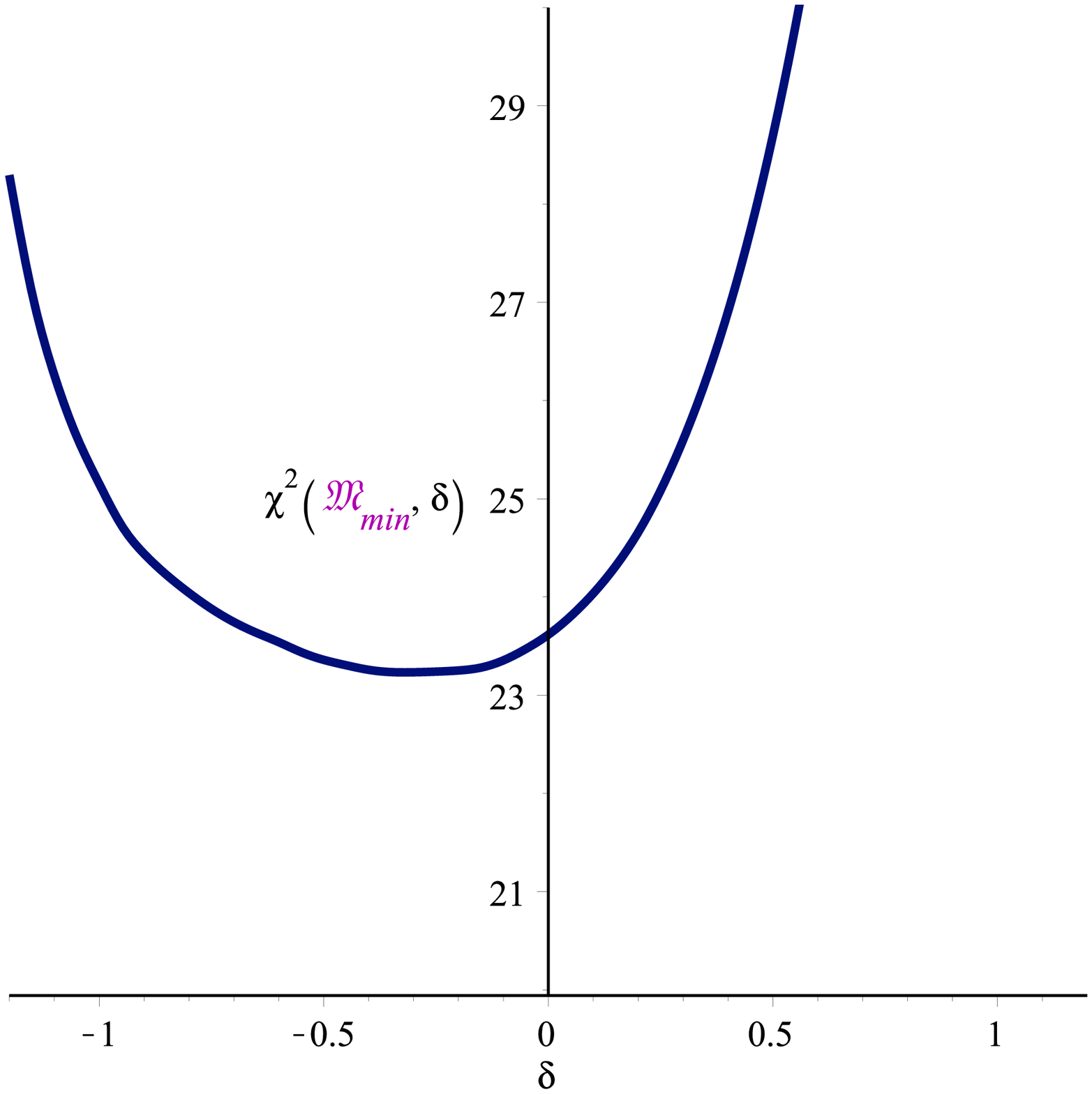, height=1.5in, width=1.5in}& \qquad\quad
\epsfig{file=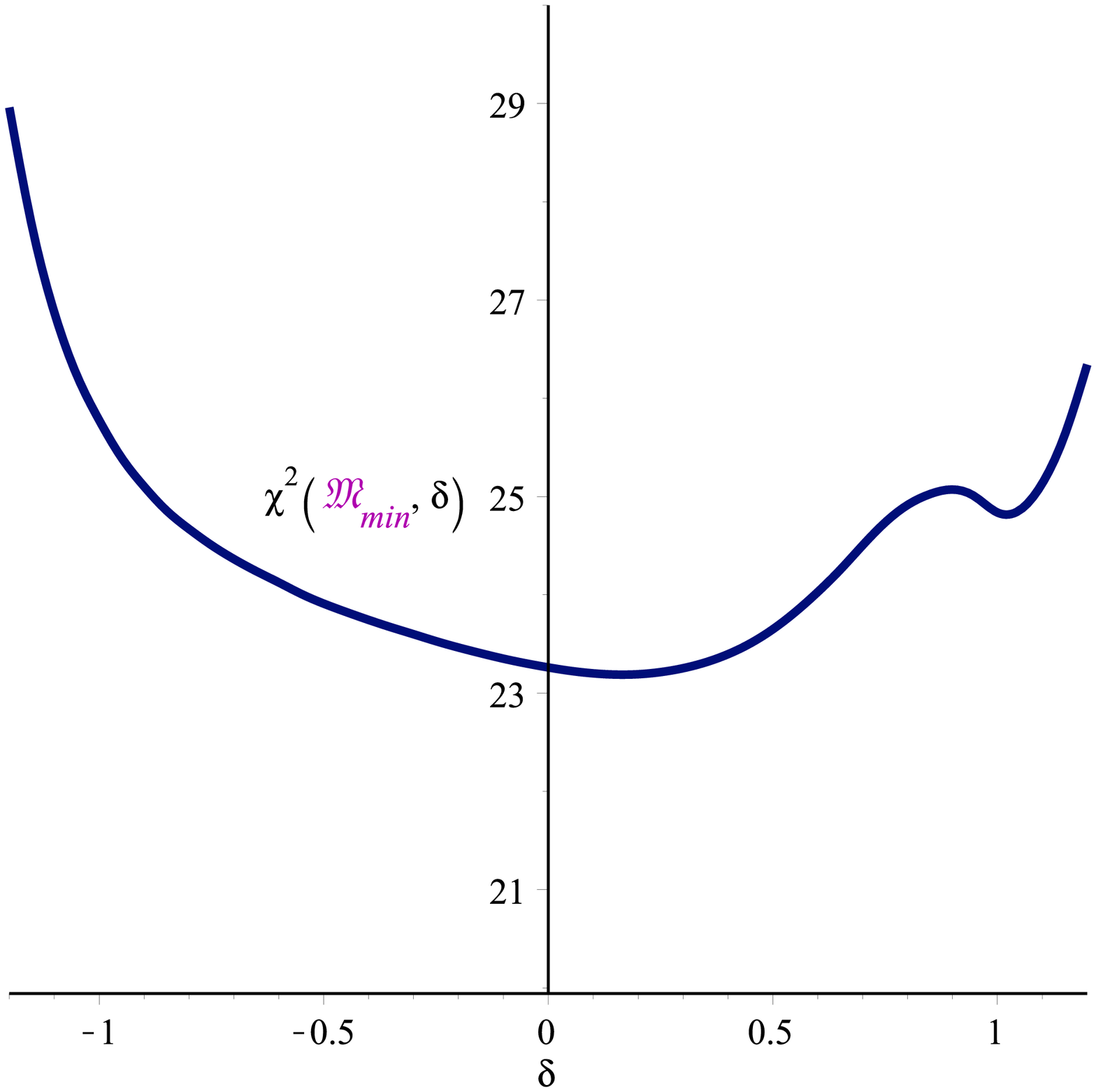, height=1.5in, width=1.5in}& \qquad\quad
\epsfig{file=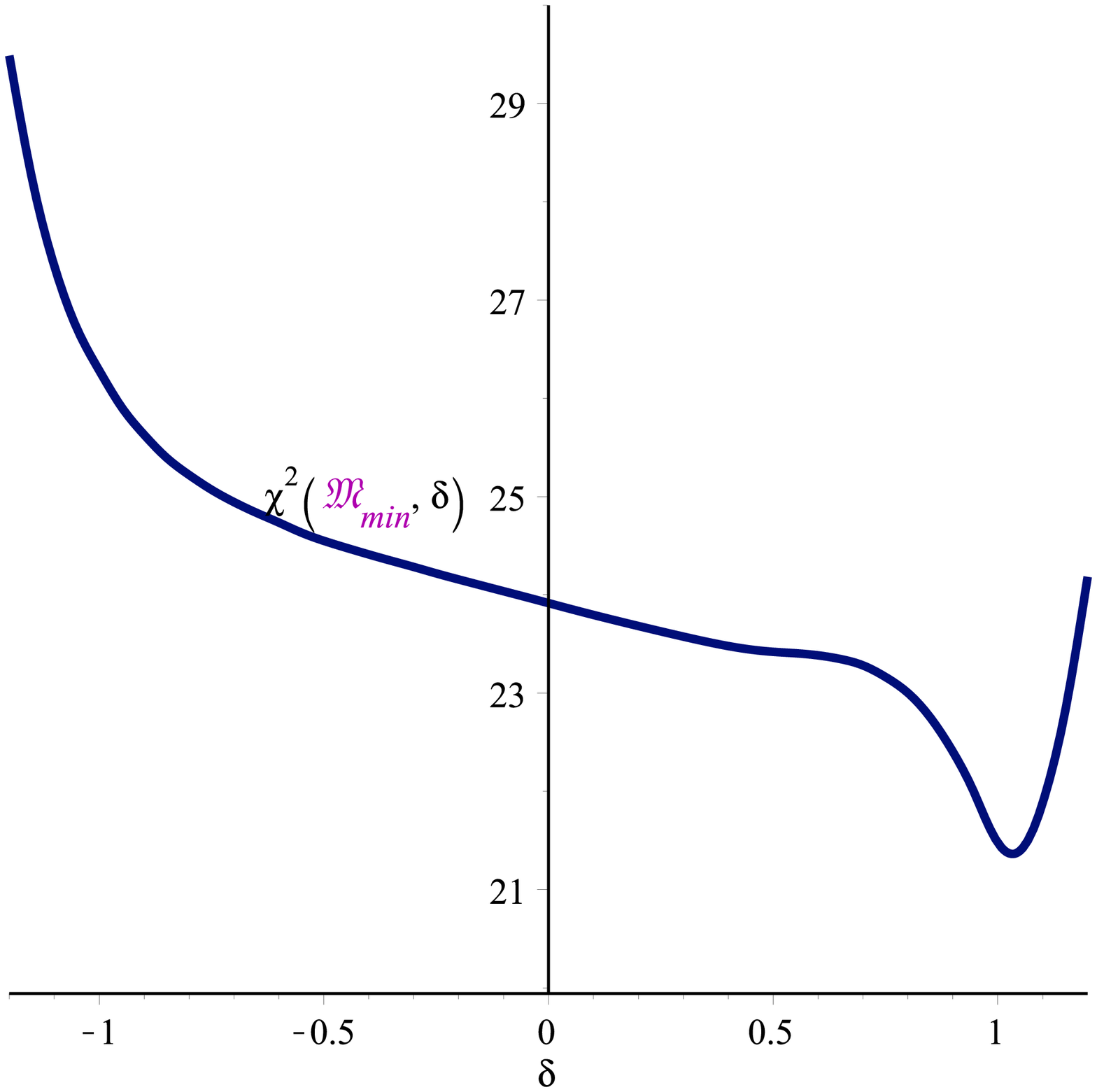, height=1.5in, width=1.5in}
\end{array}$
\end{center}
\caption{\small After optimizing the normalization ${\cal M}$, for $\varphi_0$ close to $0$ the $\chi^2$--fit is driven by the low CMB quadrupole and there is single minimum for $\delta<0$ (left). For intermediate values of $\varphi_0$ a second extremum emerges for $\delta>0$ (center), which readily becomes the overall minimum as the fit becomes eventually dominated by the pre--inflationary peak. The examples refer to $\gamma=0.08$, but the results for $\gamma=0.04$ and for $\gamma=0.02$ are qualitatively similar.}
\label{fig:chi_squared_delta}
\end{figure}

Our best--fit analysis entailed the comparison between spectra for climbing scalars in two--exponential systems and raw WMAP9 data for a number of choices of $\varphi_0$ that follow closely the evolution of the power spectrum from the slow growth that was elaborated upon in \cite{dkps} to the typical peak that reflects conventional transitions from fast to slow roll and was nicely identified in \cite{destri}. As we have said, in all cases we minimized $\chi^2$ of eq.~\eqref{chi_squared}, first analytically with respect to ${\cal M}$ and then, numerically, with respect to $\delta$, exploring to this end more than 60 values in the interval $[-1.2,1.2]$. This process was repeated for more than 20 values of $\varphi_0$ belonging to the interval $[-4,0]$ and including $-4,-3.75,..,-0.25,0$. This rather rich discrete set sufficed to capture clearly the effects of the transition that, as we described in Section \ref{sec:powerspectrum}, accompanies the emergence of the pre--inflationary peak in fig.~\ref{fig:double}.

As can be seen in fig.~\ref{fig:angular_fit_double}, starting from $\varphi_0=-4$ and proceeding toward larger values an initial \emph{decrease} of $\chi^2$ down to a minimum corresponding to  $\chi^2_{min}/DOF \simeq 0.737$ is followed by a more rapid \emph{increase} and then essentially by a plateau that extends up to $\varphi_0=0$. This interesting behavior accompanies the transition from a typical pre--inflationary peak terminating on the attractor spectrum of \cite{destri}, to the intermediate pre--inflationary peak of Section \ref{sec:powerspectrum}, and finally to a region where this peak disappears altogether, as we have seen, leaving only the wide infrared depression elaborated upon in \cite{dkps}. We are inclined to regard this rather rich behavior as a \emph{noticeable, if slight, preference of WMAP9 raw data for an infrared depression followed by a pre--inflationary peak}.

The transitional behavior finds a neat rationale in the $\delta$--dependence, for any given $\varphi_0$, of the values obtained minimizing $\chi^2$ with respect to ${\cal M}$, as can be seen in fig.~\ref{fig:chi_squared_delta}. Briefly stated, when the pre--inflationary peak is not visible or is too small to play a significant role, this function exhibits wide depressions centered around negative values of $\delta \in (-1,0)$. These clearly signal a tendency to link the slow growth of the corresponding power spectra to the CMB quadrupole. However, as $\varphi_0$ is reduced a second local depression by 2--3 units, quite narrow this time, emerges for values of $\delta$ that are now ${\cal O}(1)$. It becomes lower than the other for $\varphi_0 \simeq -1.5$, and clearly reflects a tendency to link the pre--inflationary peak to an oscillation that appears to be present in the CMB angular power spectrum for $\ell \simeq 5$. This discussion has somehow the flavor of Mean Field Theory so that, borrowing some terminology, one could say that the ``order parameter'' $\delta$ undergoes a first--order transition when the pre--inflationary peak becomes sizable.

We have also repeated the analysis for $\gamma = 0.04$ and for $\gamma = 0.02$. These lower values, as we have stated, are meant to simulate the departure from the ``hard'' exponential of concave potentials with lower tensor--to--scalar ratios, as those in eq.~\eqref{starobinsky}. Some of the preferred choices for $\delta$ and the corresponding $\chi^2$ are collected in Table \ref{table:norm_delta}. Notice that \emph{reducing the slope of the mild exponential in eq.~\eqref{potwoexp} leads to slightly improved fits that, as expected, are optimized for slightly lower values of $\varphi_0$}. This tendency affords a simple explanation: with lower values of $\gamma$ the scalar typically approaches the ``hard wall'' more easily, so that comparable conditions obtain only if $\varphi_0$ is correspondingly lowered.
\begin{figure}[h]
\begin{center}
\epsfig{file=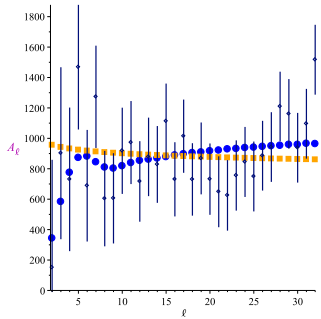, height=1.7in, width=1.7in}

\end{center}
\caption{\small The optimal $A_\ell$ for $\gamma=0.08$ (blue), the attractor $A_\ell$ (orange) and the WMAP9 raw data.}
\label{fig:attractor-optimalCl}
\end{figure}

The differences among the various cases, and the overall preferred status of the two--exponential system, are clearly eye--catching even if, admittedly, they are of limited statistical significance. The values of $\chi^2/DOF$ for the one--exponential potential displayed in fig.~\ref{fig:angular_fit_single} lie indeed rather close to the attractor, while the corresponding values for the two--exponential potential displayed in fig.~\ref{fig:angular_fit_double} lie appreciably farther, especially in the intermediate region. Interestingly, the lowest value of $\chi^2/DOF$ for $\gamma=0.08$, about $0.736$, obtains for $\varphi_0 \simeq -1.8$, and can thus be associated to the perturbative region for String Theory, since as we have stressed $\varphi_0$ sets an upper bound on the string coupling. Moreover, as we have seen lower values of $\gamma$ bring about slightly lower minima that are reached for slightly smaller values of $\varphi_0$. For example, for $\gamma=0.02$ the minimum is about $0.718$ and obtains for $\varphi_0 \simeq -2$. The optimal angular power spectrum for $\gamma=0.08$ is displayed in fig.~\ref{fig:attractor-optimalCl}, together with the optimal attractor angular power spectrum and the corresponding raw WMAP9 data. Notice that the optimal curves for the two--exponential system and for the attractor come together for $\ell \simeq 15$, a value that could be regarded as defining a ``COBE--like'' normalization point for the model.
\begin{figure}[h]
\begin{center}$
\begin{array}{cc}
\epsfig{file=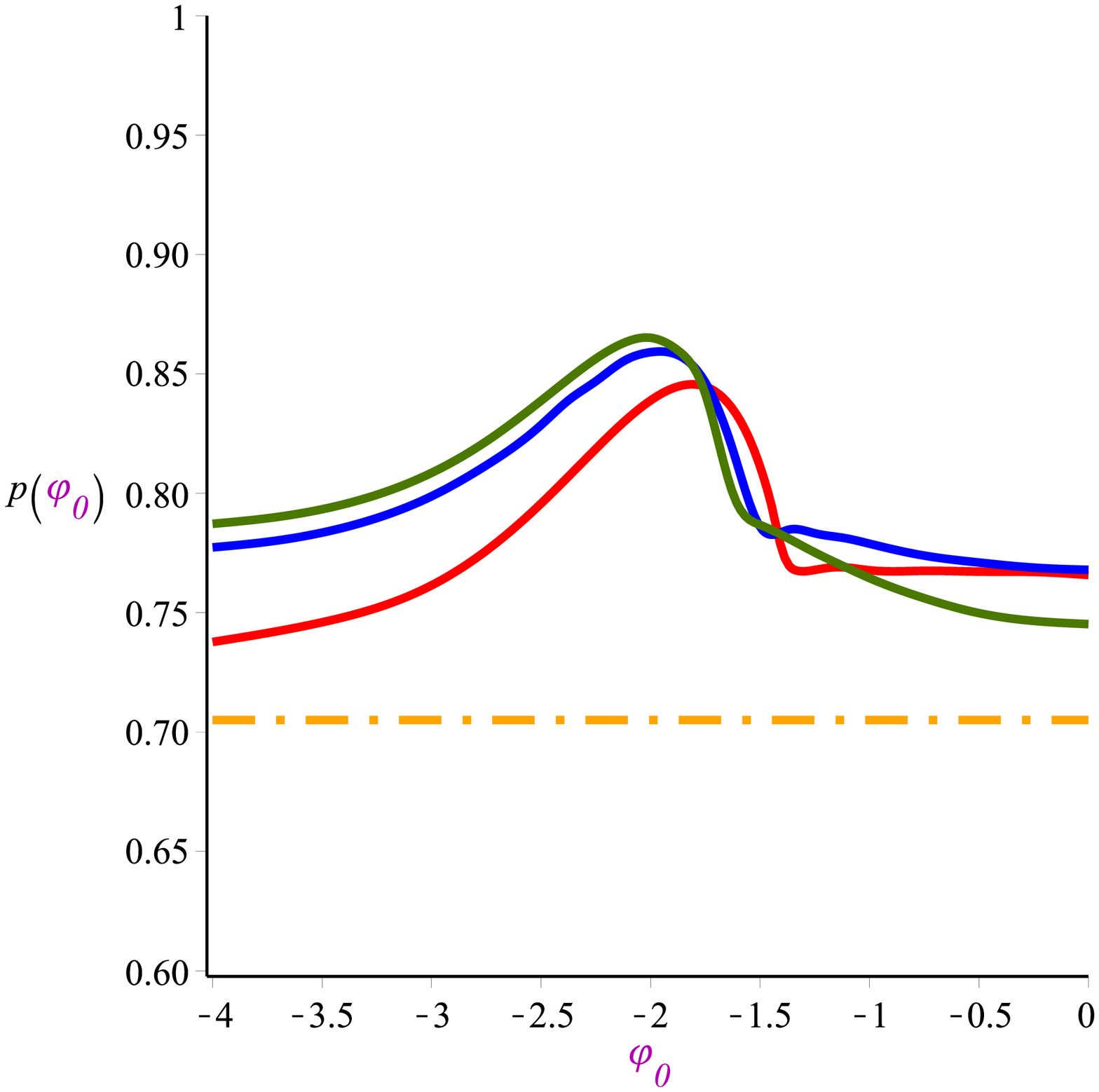, height=1.7in, width=1.7in}& \qquad\qquad
\epsfig{file=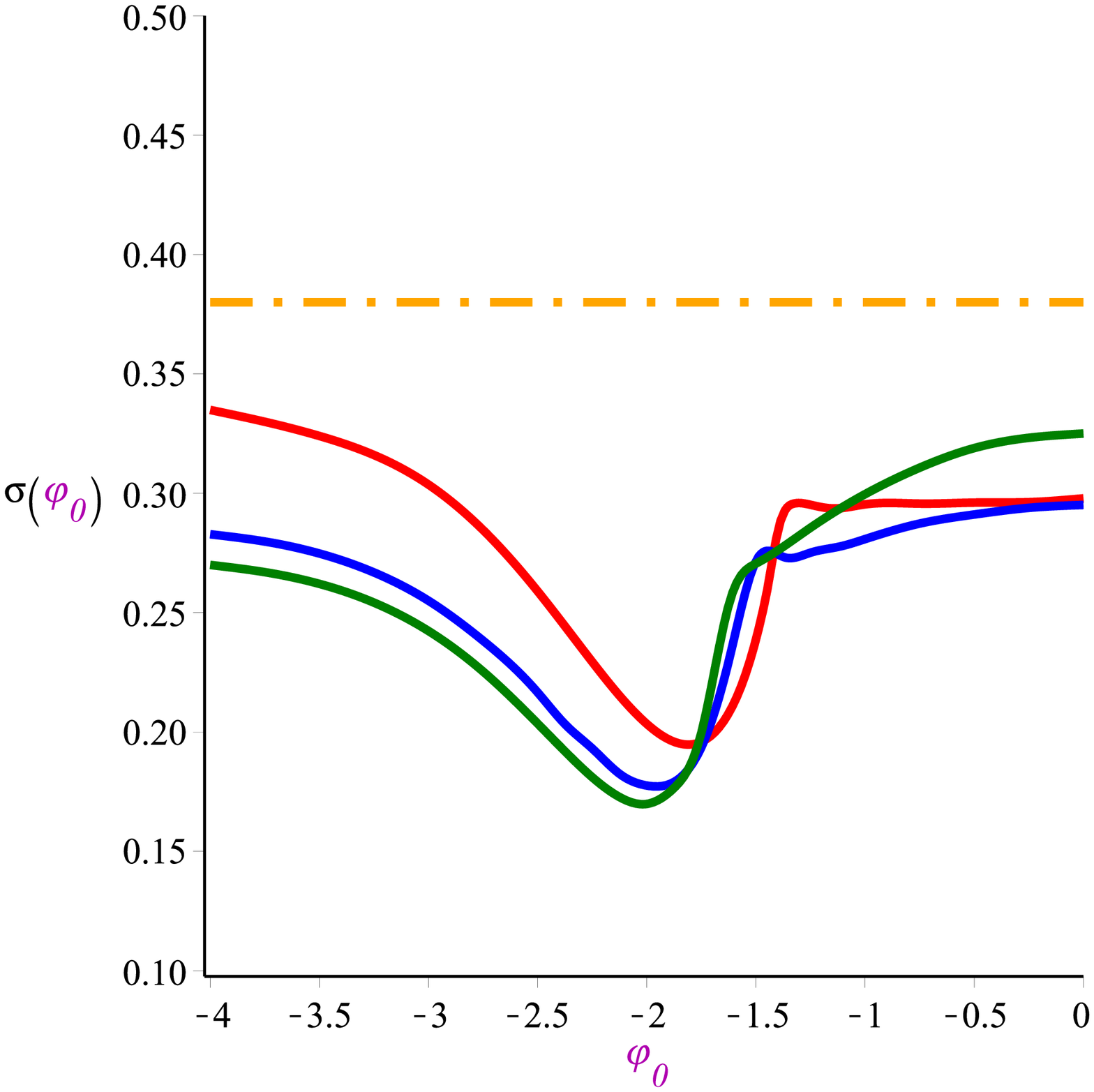, height=1.7in, width=1.7in}
\end{array}$
\end{center}
\caption{\small $p$--values for the two--exponential potential \eqref{potwoexp} with $\gamma=0.08$ (red), for $\gamma=0.04$ (blue) and for $\gamma=0.02$ (green) and for the attractor (orange), and corresponding $\sigma$--values.}
\label{fig:p_and sigma values}
\end{figure}

Fig.~\ref{fig:p_and sigma values} collects, for the three double--exponential models with $\gamma=0.08$, $\gamma=0.04$ and $\gamma=0.02$, another characterization of the fits via the $p$--values, which are determined according to
\be
p\left(\chi^2_{\rm min}\ ,\,n\right) \ = \ \frac{1}{2^\frac{n}{2} \ \Gamma\left(\frac{n}{2} \right)}\ \int_{\chi^2_{\rm min}}^\infty dx \ x^{\frac{n}{2}-1} \ e^{\,-\,\frac{x}{2}} \ ,
\ee
where $n$ is the effective number of degrees of freedom (it was called $DOF$ above, and equals 29 in those three fits and 30 in the attractor fit, as we have explained). Larger $p$--values make an effect related to the model more plausible, and the optimal choices for the double--exponential system clearly result in non--trivial, if still not fully significant values of $p$ that lie between $0.8$ and $0.9$.

One can also recast these considerations in the gaussian setting. To this end, the starting point is provided by a normalized gaussian distribution
\be
f(x) \ = \ \frac{1}{\sqrt{2\pi}} \ e^{\,-\,x^2} \ ,
\ee
since the $p$--values can be mapped into corresponding $\sigma$--values inverting the relation
\be
p \ = \ 2 \, \int_\sigma^\infty f(x)\, dx \ .
\ee
In this fashion, a $p$--value of about $6 \times 10^{-7}$ would translate into five--$\sigma$, while a $p$--value of about $3 \times 10^{-3}$ would translate into three--$\sigma$, just to quote a couple of familiar instances. In our case the $p$--value plots of fig.~\ref{fig:p_and sigma values} can be recast into corresponding plots for $\sigma$, also displayed in fig.~\ref{fig:p_and sigma values}.

All in all, none of the different models is statistically excluded, since they are lie within one--$\sigma$ of the raw WMAP9 data, with an eye--catching preference for the double--exponential models, and especially so for those with lower values of $\gamma$, in regions close to $\varphi_0=-2$ where the pre--inflationary peak is visible and yet lies well apart from the attractor curve. While we cannot claim to be discovering an evident link between the first peak in the CMB angular power spectrum and the climbing phenomenon, we find it hard to dismiss the feeling that something is going on here, insofar as the currently available data are concerned.
\vskip 24pt

\section{\sc  Conclusions}\label{sec:conclusions}
\vskip 12pt

This work builds on two main inputs. The first, drawn from String Theory \cite{strings}, is the existence of a class of orientifold vacua \cite{orientifolds} with ``brane supersymmetry breaking'' \cite{bsb}. In these models, which admit no maximally symmetric vacuum geometries, supersymmetry is broken at the string scale and is non-linearly realized in the low--energy supergravity \cite{10d_bsb_couplings}, which includes a ``hard'' exponential potential, but no tachyon excitations are present at tree level. The second input is drawn from a striking feature of the spatially flat cosmologies allowed by the corresponding low--energy supergravities. These involve a scalar field that emerges from an initial singularity with no other option than climbing up the steep exponential potential \cite{exp_sol,dks}. The process comes to an end at a turning point, while other branes of String Theory \cite{branescan,branesugimoto} can give rise, in principle, to milder exponentials \cite{fss,as13} that can force an inflationary phase during the ensuing descent. Therefore, in this setting the breaking of supersymmetry at high scales can provide a rationale for inflation to begin.

Intuitively, as the climbing phase ends, in reverting its motion the scalar ought to bring along a spurt of exponential expansion for the Universe and a corresponding peak in the power spectrum of scalar perturbations. Here we showed that this is indeed the case for scalar perturbations, while tensor perturbations, which depend essentially on the scale factor alone, do not exhibit a similar phenomenon. It took some effort to quantify this expectation, but we have provided ample evidence for it here, and we have also started to compare these findings with the low--$\ell$ tail of the CMB angular power spectrum. \emph{Leaving aside cosmic variance, a fair summary of our findings is that the low--$\ell$ WMAP9 raw data tend to favor slightly scenarios of this type with respect to the attractor power spectrum underlying the standard $\Lambda$CDM setup}. The most interesting aspect of the whole setting, however, is perhaps the main assumption on which the comparison rests. This posits an essentially direct correspondence between the largest wavelengths entering the cosmic horizon at the present epoch and those that exited at the onset of inflation. If true, it would translate into the enticing perspective of drawing from the low--$\ell$ tail of the CMB power spectrum some information on the very early Universe.
\begin{figure}[h]
\begin{center}$
\begin{array}{cccc}
\epsfig{file=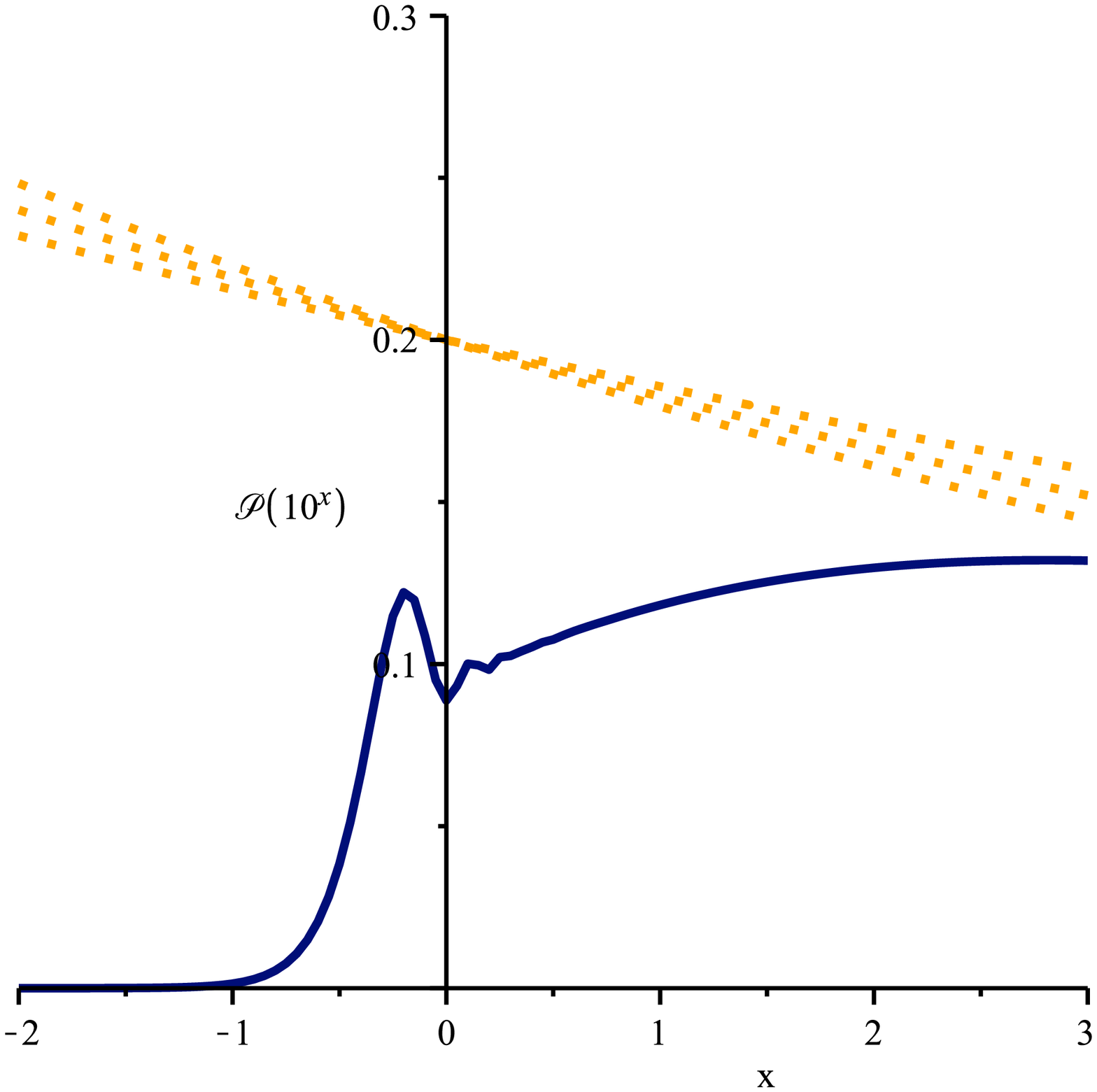, height=1.2in, width=1.2in}& \qquad
\epsfig{file=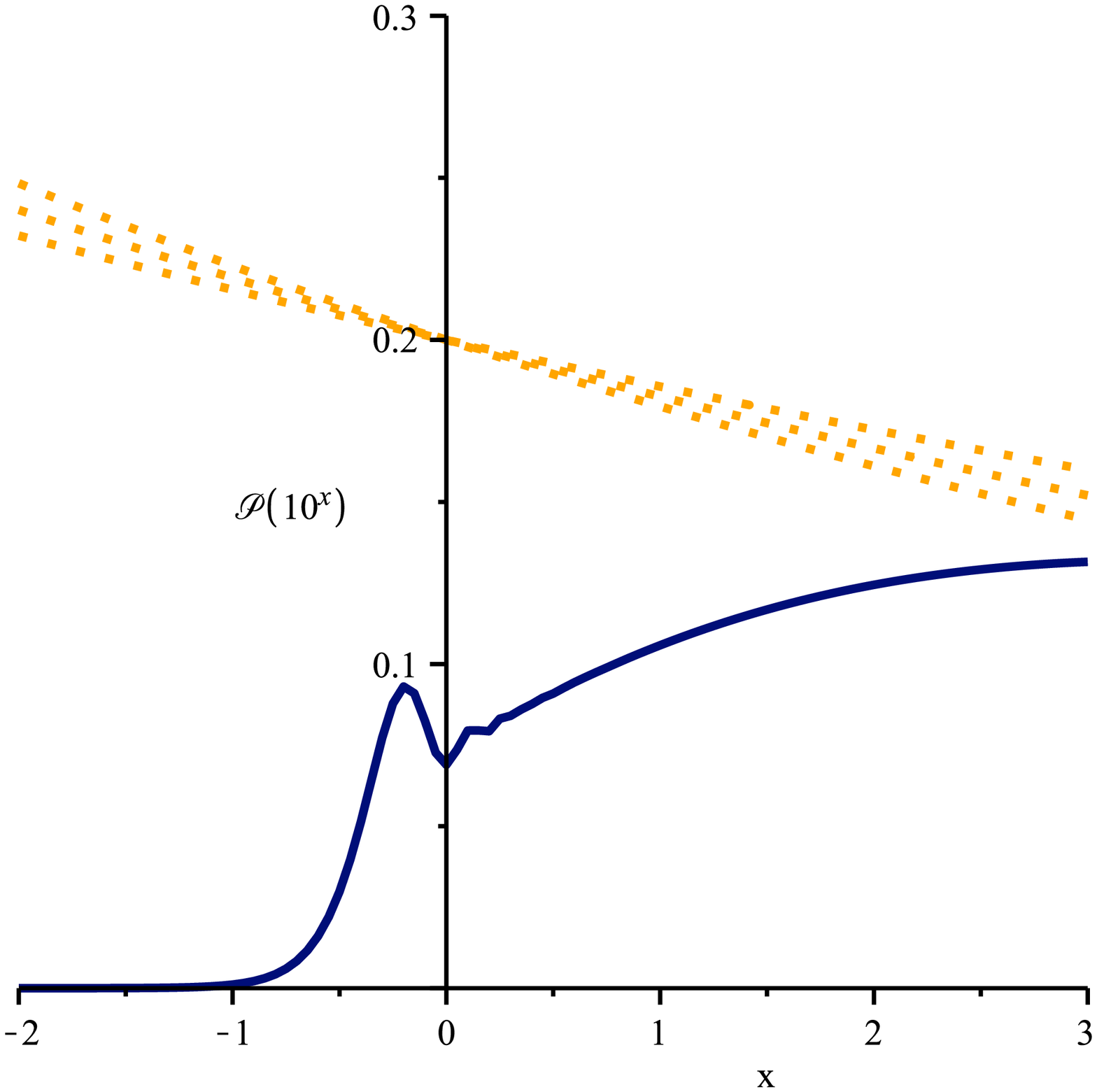, height=1.2in, width=1.2in}& \qquad
\epsfig{file=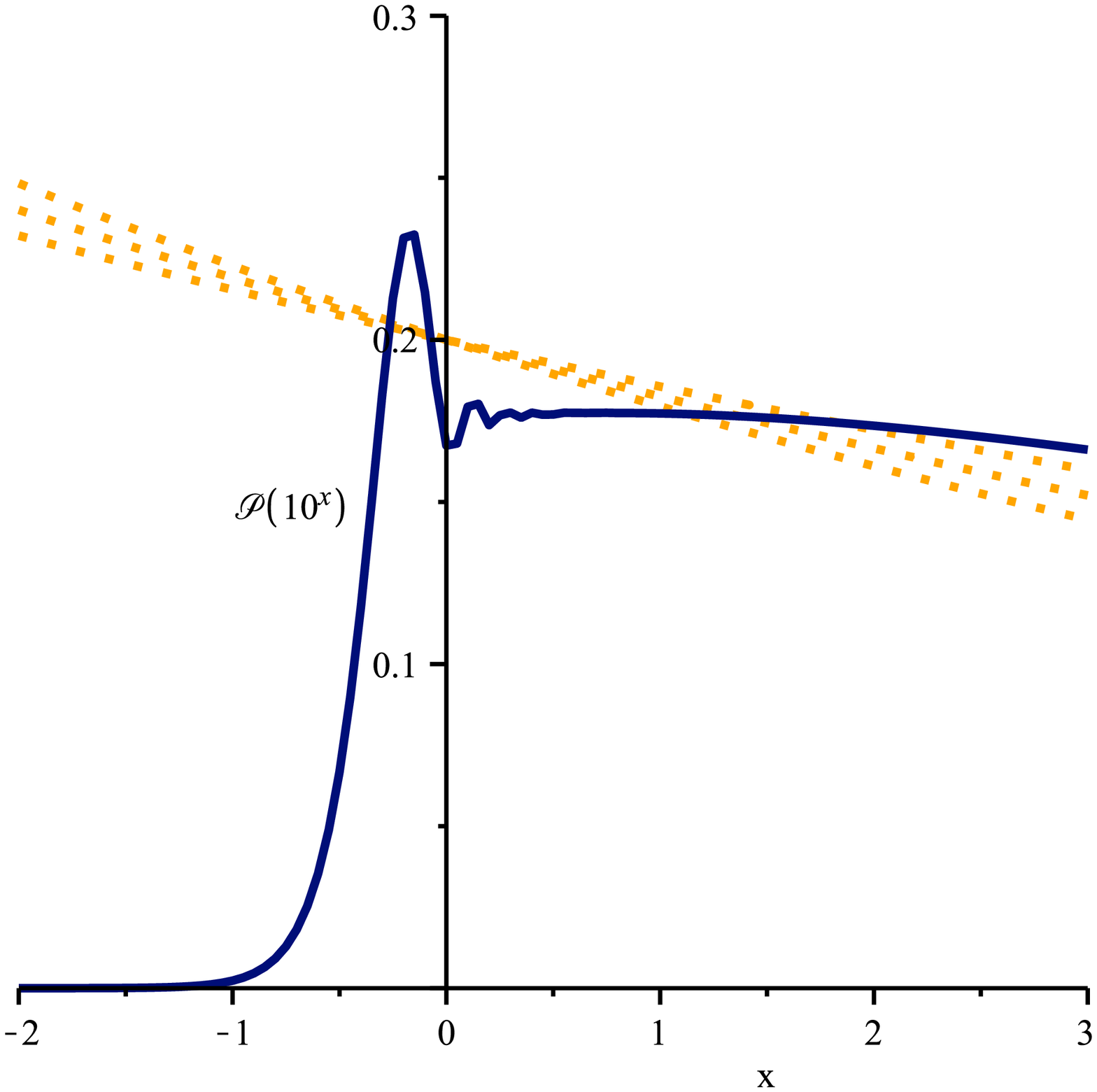, height=1.2in, width=1.2in}& \qquad
\epsfig{file=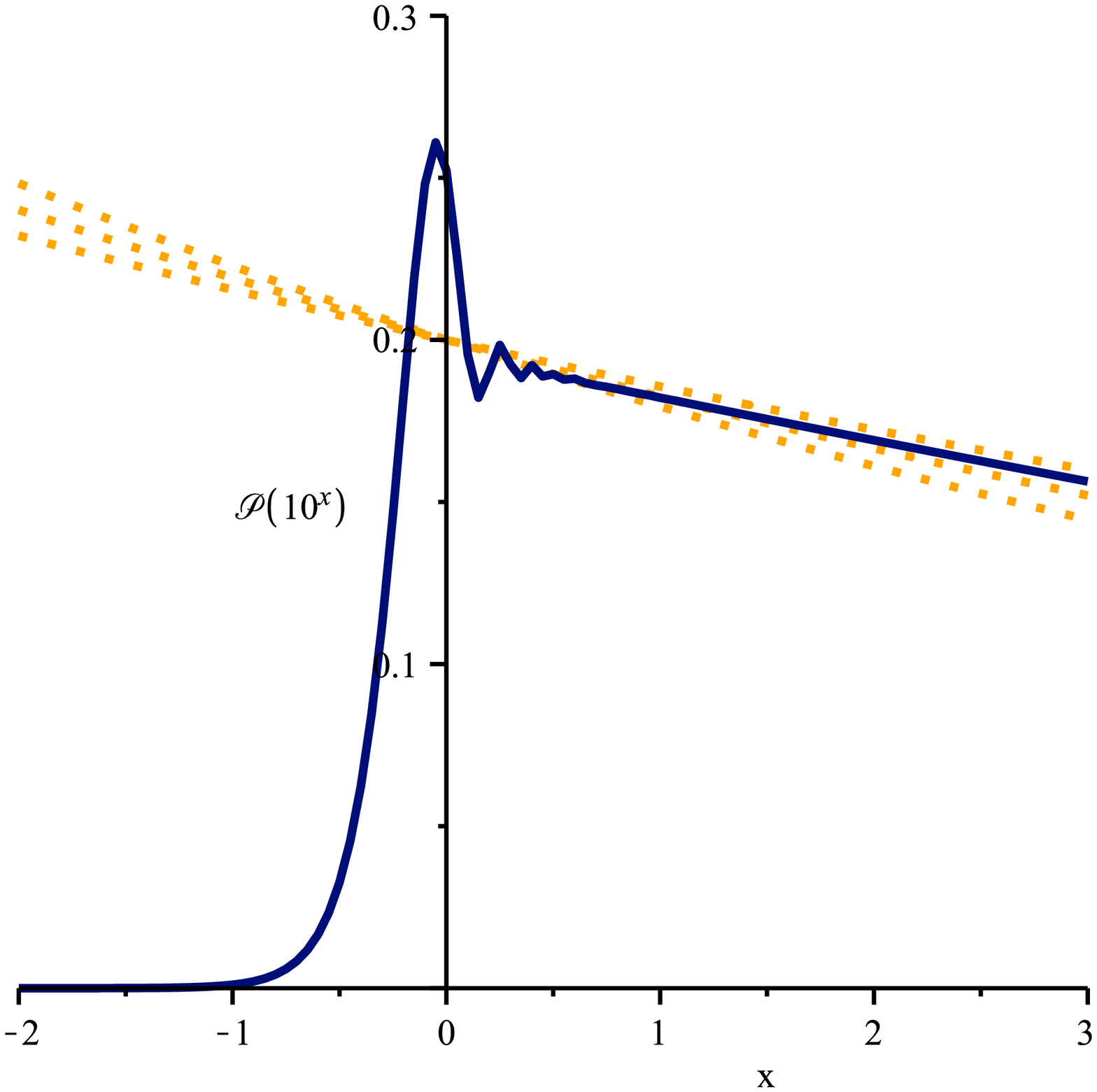, height=1.2in, width=1.2in}
\end{array}$
\end{center}
\caption{\small Power spectra of scalar perturbations computed, in cosmic time, for the Starobinsky--like potentials of eq.~\eqref{starobinsky}, shown here with an arbitrary normalization but with the parameters adjusted in order to guarantee about 60 $e$--folds of inflation and a fair portion of them with $n_s \simeq 0.96$. The slopes of the dotted lines reflect the range of values for $n_s$ that are consistent with current observations.}
\label{fig:starobinsky_power}
\end{figure}

All we have shown so far rests on the two--exponential potentials of eq.~\eqref{potwoexp}, which are relatively simple to analyze but are clearly incomplete in several respects. As we have stressed, however, the pre--inflationary peaks reflect the local behavior in the region where the scalar reverts its motion, and we have also seen that the agreement with WMAP9 data improves slightly as the parameter $\gamma$ of the two--exponential models is reduced below 0.08, consistently with the expected preference for concave potentials. We can actually conclude with a brief discussion of some results obtained directly for the Starobinsky--like potentials of eq.~\eqref{starobinsky}, with parameters adjusted so as to grant about 60 $e$--folds of inflation and a fair portion of them with $n_s \simeq 0.96$. The computation was more subtle since, for one matter, these potentials are negligibly small near the minimum so that we could not resort to the convenient gauge \eqref{gauge}, and moreover the initial values were to be set very close to the singularity. Still, the ``lsode'' method available in Maple worked rather efficiently, and the power spectra displayed in fig.~\ref{fig:starobinsky_power} clearly vindicate our claims. As in the two--exponential models, tuning the impact with the hard exponential a slow growth leaves way to the birth of a pre--inflationary peak and to its eventual merge with an approximately scale--invariant profile. The curves collected in the figure are indeed very similar to those displayed in Section \ref{sec:powerspectrum}, but turn more gradually toward the slope dictated by the observed spectral index $n_s \simeq 0.96$, which we used as an input. We did not perform a systematic analysis as for the two--exponential case, but for example the power spectra displayed in fig.~\ref{fig:starobinsky_power} yield for $\chi^2_{\rm min}/DOF$ the four values 0.738, 0.724, 0.783 and 0.816. These results are along the lines of those displayed in fig.~\ref{fig:angular_fit_double}, and in particular the second reaches the minimum of the curve for $\gamma=0.04$. Let us stress, to conclude, that only a small portion of the power spectra centered around the pre--inflationary peak played a role in our low--$\ell$ analysis via eq.~\eqref{bessel}. There is thus room, in principle, for power spectra of this type to be compatible with the $\Lambda$CDM analysis of higher multipoles, while additional small features in the potential could well account for the other oscillations that are apparently present for $\ell \lesssim 35$ in fig.~\ref{fig:attractor-optimalCl}.

Two very legitimate objections could be raised, within String Theory, against our analysis. The first concerns the values of the string coupling that accompany these phenomena, and in this case we could provide an encouraging answer, since the scenarios that are apparently preferred in the comparison with the CMB rest on relatively small values of the string coupling. There is a second objection, however, on which we have nothing definite to say, now as in the past. It has to do with curvature corrections, which are ubiquitous in String Theory and are expected to dominate near an initial singularity, casting doubts on a low--energy analysis of the climbing phenomenon. They were examined, insofar as possible, at low orders in \cite{cd}, but it is fair to state that at present we do not understand if and how they could be under control. Time and more detailed studies will tell whether these considerations can find a more rigorous origin in String Theory, and whether the PLANCK data to be soon released and a more refined data analysis taking into account wider portions of the CMB power spectrum \cite{gnks} will confirm some encouraging clues that have emerged from the present work.

\vskip 24pt
\section*{Acknowledgments}

\vskip 12pt

We are grateful to E.~Dudas and S.P.~Patil for collaboration at earlier stages of this research, to A.~Gruppuso and P.~Natoli for an ongoing collaboration on detailed likelihood tests and to E.~Akhmedov, P.~Fr\'e, H.~Kodama, K.~Kohri, G.~Rolandi and A.S.~Sorin for discussions. We are also grateful to R.~Barbieri, E.~Dudas and S.~Ferrara for discussions on supersymmetry breaking, to M.~Vietri for his interest in this work and to L.~Sabbatino for computer assistance. This work was supported in part by the ERC Advanced Grants n. 226455 (SUPERFIELDS) and n. 226371 (MassTeV), by Scuola Normale Superiore, by the Tokyo Metropolitan University, by INFN (I.S. ST\&FI), by the MIUR-PRIN contract 2009-KHZKRX and by Grant-in-Aid for Scientific Research on Innovative Areas (\# 24104505) from MEXT Japan. The authors would like to thank Scuola Normale Superiore, the \'Ecole Polytechnique and the Tokyo Metropolitan University for the kind hospitality extended to them while this work was in progress.
\vskip 24pt

\vskip 24pt
\newpage
\appendix
\section{On two--exponential potentials and String Theory}

In this Appendix we would like to sketch some semi--quantitative links between the two--exponential potentials of eq.~\eqref{potwoexp} and String Theory. To begin with, we associated the ``hard'' term to the uncanceled tension present in the Sugimoto model in \cite{bsb} and originating from $\overline{D9}$--branes and $O9_+$--planes. Therefore, one should begin by considering the ten--dimensional contribution
\be
S_{10,\, D9} \ = \ - \ \frac{64}{(2\pi)^9\, (\alpha^\prime)^5} \ \int d^{10}x \, \sqrt{-\tilde{g}} \ e^{\,-\, \phi_{10}} \ , \label{app1}
\ee
here written in the string frame following \cite{strings}, where the overall 64 accounts for the total contribution of the extended objects and $\phi_{10}$ is the ten--dimensional dilaton.

This expression can give rise to a four-dimensional potential that can be retrieved first performing the Weyl rescaling
\be
\widetilde{g}_{MN} \ = \ g_{MN} \ e^{\,\frac{\phi_{10}}{2}} \label{app2}
\ee
of the ten--dimensional metric and then effecting the compactification on a metric of the form \cite{witten85}
\be
ds^2 \ \equiv \ {g}_{MN} \, dx^M\,dx^N \ = \ e^{\,-\, 3\, \sigma} \ g_{\mu\nu}\, dx^\mu\, dx^\nu \ - \ e^{\,\sigma} \ \delta_{ij}\, dx^i \, dx^j \ . \label{app3}
\ee
For present purposes, it will suffice to consider a collection of circle--like dimensions characterized by a radius $R$. As discussed in \cite{fss,as13}, but reverting to the present notation, one can then define the two canonically normalized combinations
\be
\sqrt{2}\, k_N\, \phi \ = \  \frac{\sqrt{3}}{2}\ \phi_{10}\ - \  \sqrt{3} \ \sigma \ , \qquad
\sqrt{2}\, k_N\, \Phi_s \ = \  \frac{1}{2}\ \phi_{10} \ + \ 3 \, \sigma   \label{app4}
\ee
of the breathing mode $\sigma$ and the dilaton, with the end result
\be
S_{4,\, D9} \ = \ - \ \frac{64}{(2\pi)^3\, (\alpha^\prime)^2}\ \left( \frac{R^{\,2}}{\alpha^\prime}\right)^3 \ \int d^{10}x \, \sqrt{-\tilde{g}} \ e^{\,\sqrt{6}\, k_N\, \phi} \ . \label{app5}
\ee

A comparison with eq.~\eqref{redef} shows that this term, when expressed in terms of $\varphi$, is indeed a ``hard'' exponential potential with $\gamma=1$. It will result directly in the climbing phenomenon provided the other combination $\Phi_s$ in eq.~\eqref{app4} is somehow stabilized, which we shall assume to be the case with $\Phi_s \simeq 0$.

As an illustration, let us also sketch how an NS fivebrane with one of its dimensions wrapped on a small internal circle of radius $R^{\,\prime} \ll R$ could give rise, in principle, to the mild term in eq.~\eqref{potwoexp} with $\gamma=1/12$ \cite{fss,as13}. As stressed in \cite{fss,as13}, its eventual decay could even bring inflation to an end, but the process tends to occur too rapidly to allow for a reasonable number of $e$--folds. Still, it could drive a bounce on the hard potential, giving rise to the first peak in fig.~\ref{fig:attractor-optimalCl}, but leaving way quickly to other dynamics. We can only leave aside, at the present time, very important issues related to the effects on matter of the primordial perturbations accompanying the climbing phenomenon. Some related issues are reviewed in detail in \cite{recent_review_PP}.

At any rate, proceeding as above one can identify the additional contribution
\be
S_{4,\,NS5} \ = \ - \ \frac{1}{(2\pi)^3\, (\alpha^\prime)^2} \ \frac{R\, R^\prime}{\alpha^\prime} \ \int d^{10}x \, \sqrt{-\tilde{g}} \, e^{\,\frac{\sqrt{6}}{12} \, k_N\, \phi} \ , \label{app6}
\ee
and the next issue is turning the two terms of eqs.~\eqref{app5} and \eqref{app6} into the form of eq.~\eqref{potwoexp}. This can be attained via a variant of eq.~\eqref{redef}, a redefinition that includes a shift $\Delta$,
\be
\sqrt{2}\, k_N\, \phi \ = \ \frac{2}{\sqrt{3}} \ \left( \varphi \ + \ \Delta \right) \ ,
\ee
with
\be
e^{\,\Delta} \ = \ \left[ \frac{1}{64}\ \left(\frac{R^{\,\prime}}{R}\right)\, \left(\frac{ \alpha^\prime}{R^{\,2}}\right)^2 \right]^\frac{6}{11} \ ,
\ee
which also determines the overall scale of eq.~\eqref{potwoexp},
\be
V_0 \ = \ \frac{1}{(2\pi)^3\, (\alpha^\prime)^2}\, \left(\frac{R^{\,2}}{ \alpha^\prime}\right)^\frac{9}{11}\,  \left(\frac{R^{\,\prime\,2}}{2\, R^{\,2}}\right)^\frac{6}{11} \ .
\ee
The reader can verify that $\varphi \ + \ \Delta$ coincides with the ten--dimensional dilaton in this case, while $\Delta$ can be naturally negative for $R^\prime \ll R$, in the spirit of a cosmological evolution within the perturbative regime.

Let us add that, in our numerical investigation of two--exponential models, we identified a recurrent link between the overall scale $V_0$ of the potential and typical values of Hubble parameter during the inflationary phase,
\be
V_0 \ \simeq \ 50 \, \left(\frac{H_\star}{k_N}\right)^2 \ ,
\ee
while in this type of compactification String Theory leads to the relation
\be
\left(\alpha^\prime\right)^4 \ = \ \frac{k_N^{\,2}\, R^{\,6}}{\pi\, g_s^{\,3}} \ ,
\ee
so that one can also present $V_0$ in the form
\be
V_0 \ = \ \frac{g_s^{\,6}}{8\,\pi\, k_N^{\,4}}\, \left(\frac{ \alpha^\prime}{R^{\,2}}\right)^\frac{57}{11}\,  \left(\frac{R^{\,\prime\,2}}{2\, R^{\,2}}\right)^\frac{6}{11} \ .
\ee

Combining these results finally yields
\be
{k_N}\, {H_\star} \ \simeq \ \frac{g_s^{\,3}}{20\, \sqrt{\pi}} \ \left( \frac{\alpha^\prime}{R^{\,2}} \right)^\frac{57}{22} \ \left( \frac{R^{\,\prime\, 2}}{2\, R^{\,2}}\right)^\frac{3}{11} \ ,
\ee
an expression that is potentially compatible with the lower end of the allowed range
\be
10^{\,-\,12} \ \lesssim \ {k_N}\, {H_\star} \ \lesssim \ 10^{\,-\,5}
\ee
for the Hubble scale during the inflationary phase  \cite{Hrange,concave}.

\end{document}